\documentclass[11pt,aps,prf,final,superscriptaddress,longbibliography]{revtex4-1}
\usepackage{hyperref}
\hypersetup{
    colorlinks,
    linkcolor={red},
    citecolor={blue},
    urlcolor={blue}
}
\usepackage{rotating}
\usepackage{mwe}
\usepackage{graphicx}
\usepackage{amsmath}
\usepackage{epstopdf, epsfig}
\usepackage{fancyhdr}
\usepackage{graphicx}
\usepackage{caption}
\usepackage{subcaption}
\usepackage{mathrsfs}
\usepackage{enumerate}
\usepackage{wasysym}
\usepackage{multirow}
\usepackage{soul}
\usepackage{natbib}

\usepackage[usenames, dvipsnames]{color}
\definecolor{mygray}{gray}{0.8}

\begin{document}


\title{Local velocity variations for a drop moving through an
orifice: effects of edge geometry and surface wettability}


\author{Ankur D. Bordoloi}
\email[]{ankur@lanl.gov}
\affiliation{Physics Division, Los Alamos National Laboratory, Los Alamos, New Mexico}

\author{Ellen K. Longmire}
\affiliation{Dept. of Aerospace Engineering \& Mechanics, University of Minnesota, USA}


\date{\today}

\begin{abstract}
We investigate velocity variations inside of and surrounding a gravity driven drop impacting on and moving through a confining orifice, wherein the effects of edge geometry  (round- vs. sharp-edged) and surface wettability (hydrophobic vs. hydrophilic) of the orifice are considered. Using refractive index matching and time-resolved PIV, we quantify the redistribution of energy in the drop and the surrounding fluid during the drop's impact and motion through a round-edged orifice. The measurements show the importance of a) drop kinetic energy transferred to and dissipated within the surrounding liquid, and b) the drop kinetic energy due to internal deformation and rotation during impact and passage through the orifice. While a rounded orifice edge prevents contact between the drop and orifice surface, a sharp edge promotes contact immediately upon impact, changing the near surface flow field as well as the drop passage dynamics. For a sharp-edged hydrophobic orifice, the contact lines remain localized near the orifice edge, but slipping and pinning strongly affect the drop propagation and outcome. For a sharp-edged hydrophilic orifice, on the other hand, the contact lines propagate away from the orifice edge, and their motion is coupled with the global velocity fields in the drop and the surrounding fluid.  By examining the contact line propagation over a hydrophilic orifice surface with minimal drop penetration, we characterize two stages of drop spreading that exhibit power-law dependence with variable exponent. In the first stage, the contact line propagates under the influence of impact inertia and gravity. In the second stage, inertial influence subsides, and the contact line propagates mainly due to wettability.
\end{abstract}


\maketitle

\section{Introduction}
\label{sec01}
Interaction of a moving drop with a solid confining boundary is of interest to many industrial applications, such as the displacement of supercritical $\mathrm{CO_2}$ through porous rock, underground petroleum recovery, food separation, and the design of pharmaceutical devices. In these applications, the drop propagates through a porous matrix and displaces an immiscible liquid. The presence of a solid boundary introduces many complexities to the local dynamics of these systems via contact and wettability effects. For example, if the drop fluid makes contact with the solid phase, the wettability of the surface affects the dynamics through complex physiochemical interactions which are challenging to observe locally or to model numerically.

Considering the simplified scenario investigated in \citet{Bordoloi14}, a drop moving under gravity that encounters a narrow opening has to deform significantly and displace a second fluid.  In the absence of wettability effects, the entry and passage of a drop through an opening is an outcome influenced by the interplay between the hydrodynamic resistance in the form of surface tension and viscosity, and the driving potential such as inertia or gravity. Many of the previous investigations on drops entering a narrow opening were focused on {parameters that separate} capture from passage of the bulk of a drop through various opening geometries, such as a capillary tube, a toroidal ring or an orifice \citep{Lorenceau03, Zinchenko06, Delbos10, Ratcliffe10, Bordoloi14}. When a drop approaches an orifice under gravity, the interfacial tension at the opening entrance acts against gravity, and a Bond number ($Bo = \Delta\rho gD^2/\sigma$, where $\Delta\rho$ is the density difference between the two fluids and $g$ is the acceleration due to gravity), helps characterize the flow \citep{Ratcliffe10, Bordoloi14}. \citet{Ratcliffe10} used an axisymmetric boundary integral method to predict the capture and release of a drop through a smooth toroidal opening. They found that, in addition to the Bond number, the capture/release transition depended on opening-to-drop diameter ratio ($d/D$). Using visualization experiments on drops moving through an orifice in a liquid/liquid system, \citet{Bordoloi14} showed that the transition between the two outcomes could be described by a combination of these two parameters in the form:  $\mathcal{R} = \sqrt{Bo(d/D)^3 } \approx 0.9$. This criterion was shown to be valid for cases where the drop fluid was separated from the orifice surface by a thin layer of the surrounding fluid. This film was also predicted in earlier numerical studies conducted on various confining geometries \citep{Tsai94, Cachile96, Guido10, Protiere10, Zinchenko06, Ratcliffe10}. An overview of existing literature on this subject can be found in a recent review article by \citet{Zinchenko2017}.

The direct interaction of a drop with a solid surface, wherein the wettability and other surface effects influence the flow, is more complex \citep{Delbos10, Ding12, Bordoloi14}. In their experiments, \citet{Bordoloi14} employed sharp-edged hydrophobic and hydrophilic orifices to force a contact between the drop and the orifice surface.  Under these circumstances, the contact line motion as well as the contact line pinning were shown to alter the parameter limits separating drop capture from drop release as well as the eventual drop topology.  

As they were based on visualization methods, the studies discussed above were focused almost exclusively on the bulk motion of the drop. A few previous studies used refractive index matched liquids to perform Particle Image Velocimetry (PIV) and reveal the local motions interior and exterior to a drop impacting and coalescing with a liquid/liquid interface \citep{Mohamed03, Mohamed04, Ortiz10, Bordoloi12}. Recently, \citet{Kumar17} used a ray-tracing algorithm and PIV to characterize velocity variations within a drop falling through air onto a solid surface. The local motions within the drop and in the displaced surrounding liquid during the liquid/liquid/solid interactions discussed above are important to have a complete physical understanding of these processes and yet have not been thoroughly explored. 

The present work is motivated by the lack of understanding of the local velocity fields in a liquid/liquid system during a drop\rq{}s passage through a narrow opening under different {contact and} wettability conditions. Here we extend the work of \citet{Bordoloi14} by examining the velocity variations inside and outside of a drop and their associations with the drop dynamics following the impact upon a given orifice.

\section{Experimental setup and methods}
\label{sec02}
\subsection{Setup}
\label{sec2.1}
\begin{figure}[h]
  \centering
	\includegraphics[scale=0.8]{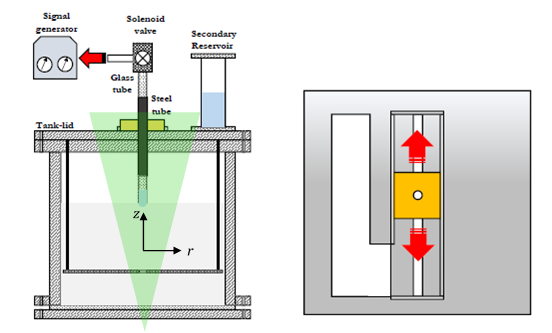}
  \caption{(Colour online) Schematic of the experimental setup and the drop generating mechanism:  side view (left) and top view (right). }
\label{fig01}
\end{figure}

The experiments were performed in the same rectangular acrylic tank (with cross-section 255 mm $\times$ 255 mm and height of 280 mm) described in \citet{Bordoloi14} (see figure \ref{fig01}). The tank was supported on a steel frame and filled with silicone oil to a height of 160 mm. A rectangular plate (254 mm $\times$ 230 mm $\times$ 2mm) with a circular orifice at its center was suspended horizontally at a height 120 mm below the free surface. Drops of water/glycerin were released into the pool of oil so that they reached terminal velocity before encountering the plate. The width of the tank was large enough (about 20 times the drop diameter) that any effect of the surrounding walls on the dynamics of the drop motion was negligible.

The properties of the drop and surrounding fluids are summarized in Table \ref{tab01}. We used silicone oil (Polydimethylsiloxane, Dow Corning Fluid) with viscosity 50 cSt as the continuous fluid and an aqueous solution of glycerin as the drop fluid. The percentage of glycerin in the mixture was set at 48\% by volume such that the refractive indices of the two fluids were matched. The resulting ratio of drop viscosity to surrounding fluid viscosity was $\lambda$=0.14.

\begin{table}
  \begin{center}
\def~{\hphantom{0}}
\renewcommand{\arraystretch}{1.15}
  \begin{tabular}{lcc}
\hline
						              & \textbf{Drop fluid}   	                  &    \textbf{Surrounding fluid} \\
	                                                                   & (\textit{48 \% glycerin}                &  (\textit{Silicone}\\
							   & \textit{in water})	                 & \textit{oil})\\
\hline
	    Density ($\rho$), $\mathrm{g/cm^3}$ & 1.131     & 0.96 \\
	   Dynamic viscosity ($\mu$ ), g/(cm-s)      & 0.07      & 0.48 \\
             Surface tension ($\sigma$), mN/m           &        \multicolumn{2}{c}{29.5}\\
\hline

  \end{tabular}
  \caption{Properties of glycerin/water solution and silicone oil (Dow Corning fluid). The subscripts $d$ and $s$ are used to denote the individual properties of the drop and surrounding fluids respectively. }
  \label{tab01}
  \end{center}
\end{table}

Drops were generated using a solenoid-valve-driven mechanism as shown in figure \ref{fig01}. For each test, the drop fluid (water/glycerin mixture) was extracted from a separate reservoir fixed above the tank using a long glass tube (6.5 mm internal diameter). The tube, carrying the drop fluid, was passed through a second metallic tube rigidly fixed inside a rectangular plastic piece (see figure \ref{fig01}). The rectangular piece was confined to a slot cut across the tank lid, so that the glass tube could be held fixed inside the tank. The bottom end of the tube was placed at the oil/air interface. The top end of the release tube was connected via a rubber hose to a closed solenoid valve that sealed the drop fluid inside the tube. To release the drop, the valve was opened using the signal from the leading edge of a positive square wave with amplitude of 12 V. The pressure difference created due to the sealing of the tube at the top end, and the interfacial tension between the drop and the silicone oil at the bottom end, held the drop fluid inside the tube until the valve was triggered open.  The wider slot on the left of the tank-lid (see top-view in figure \ref{fig01}) was made to allow accessibility inside the tank during experiments.

Orifice plates composed of two materials, commercial acrylic (hydrophobic, HPB) and glass (hydrophilic, HPL), were considered. Two orifice geometries were considered as in \citet{Bordoloi14}. In one case, the orifice edges were rounded (with edge radius approximately 0.5 mm) on both sides of the plate. In the other case, the orifice edges were sharp (edge radius $\approx$ 100 microns). Utmost care was taken in fabricating both round and sharp edged orifices to minimize any irregularity near the edge, particularly with the glass plates. Before each experiment, a given plate was washed with distilled water before being dried and carefully cleaned with ethyl alcohol and acetone in sequence to remove any oil residue.

\subsection{Simultaneous PIV and interface tracking}
\label{sec2.2}
For the PIV experiments, the drop-orifice plane of symmetry was illuminated from below using a Photonix Nd:YLF laser with pulse energy 30 mJ and pulse frequency 1 kHz. The beam was formed into a sheet of 1 mm thickness. A Photron Fastcam Ultima APX camera (10 bit CMOS sensor, 1024 $\times$ 1024 resolution) with a 105 mm and a 200 mm Nikkor lens was synchronized with the laser to image the flow. A very small quantity of Rhodamine B was added to the drop fluid in order to distinguish between the two fluids. Addition of Rhodamine and tracer particles has no significant effect on the physical properties of a drop \citep{Bordoloi12}.

We used silver coated hollow glass spheres of average diameter 15$\mu$m and density 1.6 $\mathrm{g/cm^3}$ to seed both the drop fluid and the surrounding oil. Particles were first introduced into a secondary volume ($\approx$ 100 ml) of each fluid. To avoid particle clumping and eventually achieve a uniform distribution throughout a given volume, beads and a small amount of fluid were first squashed using a mortar and pestle, and subsequently mixed with the remaining volume in the respective beaker. A strainer screen was used to remove remaining clumps from the fluid. At this time, the particle number density in the secondary volume was extremely high. These fluids were then added to the primary test fluids in small installments until the particle density needed for PIV measurements was obtained.

\begin{table}
  \begin{center}
\def~{\hphantom{0}}
\renewcommand{\arraystretch}{1.15}
  \begin{tabular}{cccccc}
\hline
Test & Orifice & $Bo$ & $d/D$ & $We$ & Re\\
\hline
RHPB1 & round-edged hydrophobic & 3.7 & 0.62 & 2.6 & 13.2\\
SHPB1 & sharp-edged hydrophobic & 3.7 & 0.62 & 2.6 & 13.2\\
SHPB2 & sharp-edged hydrophobic & 5.3 & 0.62 & 3.4 & 18.6\\
SHPL1 & sharp-edged hydrophilic   & 5.3 & 0.62 & 3.4 & 18.6\\
SHPL2 & sharp-edged hydrophilic   & 4.8 & 0.44 & 3.2 & 17.5\\
\hline

  \end{tabular}
  \caption{Experimental parameters for PIV test cases. {The drop Weber number, $We=\frac{\rho_dU_t^2D}{\sigma}$ and Reynolds number, $Re=\frac{\rho_sU_t D}{\mu_s}$ are calculated based on the drop terminal velocity, $U_t$.}}
  \label{tab02}
  \end{center}
\end{table}

The experimental parameters for the representative cases considered for the PIV analysis are summarized in Table \ref{tab02}. The PIV data acquisition was carried out using Davis 7.2 software from LaVision, whereas {Davis 8.3} was used for processing. The field of view, starting and finishing interrogation areas for the multi-pass cross-correlation algorithm, and vector spacing for all cases considered are summarized in Table \ref{tab03}. Typically, a velocity field yielded 99\% valid vectors. Vorticity and strain rate were determined from the velocity fields with a central difference scheme.

\begin{table}
  \begin{center}
\def~{\hphantom{0}}
\renewcommand{\arraystretch}{1.15}
  \begin{tabular}{cccccc}
\hline
Test & FOV ($\mathrm{mm^2}$) & initial and final IWS ($\mathrm{pixel^2}$) & vector spacing\\
\hline
RHPB1 & 32$\times$32 & 64$\times$64; 32$\times$32 with 50\% overlap & 0.5 mm\\
SHPB1 & 32$\times$32 & 64$\times$64; 32$\times$32 with 50\% overlap & 0.5 mm\\
SHPB2 & 32$\times$32 & 48$\times$48; 24$\times$24 with 50\% overlap & 0.38 mm\\
SHPL1 & 32$\times$32 & 48$\times$48; 24$\times$24 with 50\% overlap & 0.38 mm\\
SHPL2 & 20.5$\times$20.5 & 48$\times$48; 24$\times$24 with 50\% overlap & 0.24 mm\\
\hline

  \end{tabular}
  \caption{Processing parameters for PIV test cases considered. Acronyms FOV: field of view and IWS: interrogation window size.}
  \label{tab03}
  \end{center}
\end{table}

\begin{figure}[h]
 \centering
        \begin{subfigure}[b]{0.2\textwidth}
                \centering
                \includegraphics[width=\textwidth]{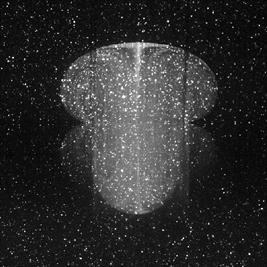}
                \caption{}
                \label{fig2a}
        \end{subfigure}
        \begin{subfigure}[b]{0.2\textwidth}
                \centering
                \includegraphics[width=\textwidth]{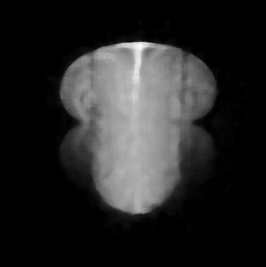}
                \caption{}
                \label{fig2b}
        \end{subfigure}
        \begin{subfigure}[b]{0.2\textwidth}
                \centering
                \includegraphics[width=\textwidth]{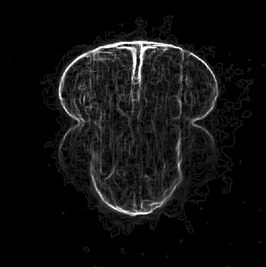}
                \caption{}
                \label{fig2c}
        \end{subfigure}\\

        \begin{subfigure}[b]{0.2\textwidth}
                \centering
                \includegraphics[width=\textwidth]{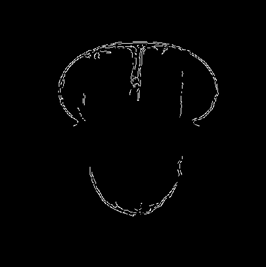}
                \caption{}
                \label{fig2d}
        \end{subfigure}
        \begin{subfigure}[b]{0.2\textwidth}
                \centering
                \includegraphics[width=\textwidth]{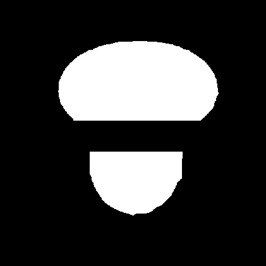}
                \caption{}
                \label{fig2e}
        \end{subfigure}
        \begin{subfigure}[b]{0.2\textwidth}
                \centering
                \includegraphics[width=\textwidth]{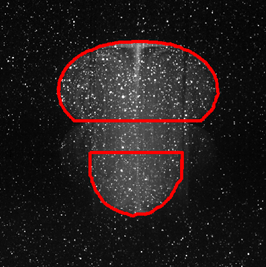}
                \caption{}
                \label{fig2f}
        \end{subfigure}\\        \caption{Image processing steps: (a) raw PIV image, (a)-(b) median filter, (b)-(c) standard deviation filter, (c) to (d) \lq{}Canny\rq{} edge detection, (d)-(e) morphological closing and filling and (f) raw PIV image with detected interface.}\label{fig02}
\end{figure}

\begin{figure}[h]
  \centering
	\includegraphics[scale=0.8]{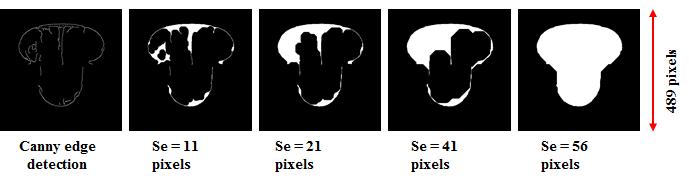}
  \caption{Iterative morphological closing and filling operations performed on a pre-processed image. In each step, the size of the structuring element (Se) was incremented until the entire drop was filled.}
\label{fig03}
\end{figure}

To isolate the drop interface and estimate various parameters, PIV images were processed with a custom MATLAB program. Figure \ref{fig2a} shows a sample raw PIV image for a drop that has impacted on the orifice. Figures \ref{fig2b}-\ref{fig2f} depict the image file after various processing steps were applied. If the image contained high intensity noise due to reflections near the orifice, an image without the drop was subtracted from the test image. Then median (figure \ref{fig2a}-\ref{fig2b}) and standard deviation filters (figure \ref{fig2b}-\ref{fig2c}) were applied on the resulting image. The median filter removed tracer particles, and the standard deviation filter enhanced the edges of the drop. Next, a ‘Canny’ edge detection scheme (figure \ref{fig2c}-\ref{fig2d}) was applied to detect the edges of the drop. The Canny edge detection is a multistep process. In the first step, the image is smoothed by a Gaussian convolution operator. Then, a simple 2D first derivative operator highlights the regions with high first order spatial derivative. The algorithm then tracks the edges by scanning the entire image using non-maximum suppression, wherein the pixels not associated with high first order gradient are set to zero. Other classical edge detection algorithms, such as Sobel and Prewitt were also considered. However, in regions with more noise, Canny edge detection worked better.

Although the Canny method detected edges of the drop fairly accurately, the edges were sometimes disconnected in specific regions. Therefore, the disconnected boundaries were closed using a series of iterative morphological operations (see figures \ref{fig2d}-\ref{fig2e}). In each step of the iterative process, the size of the structuring element was increased incrementally by 5 pixels until the entire drop was filled. The stepwise iterative operation is shown separately in figure \ref{fig03}. After closing the entire drop, the location of the orifice plate was masked. The final drop boundary superposed on a raw PIV image is shown in figure \ref{fig2f}  (see also \citet{Bordoloi14b}).

\begin{figure}[h]
  \centering
	\includegraphics[scale=0.6]{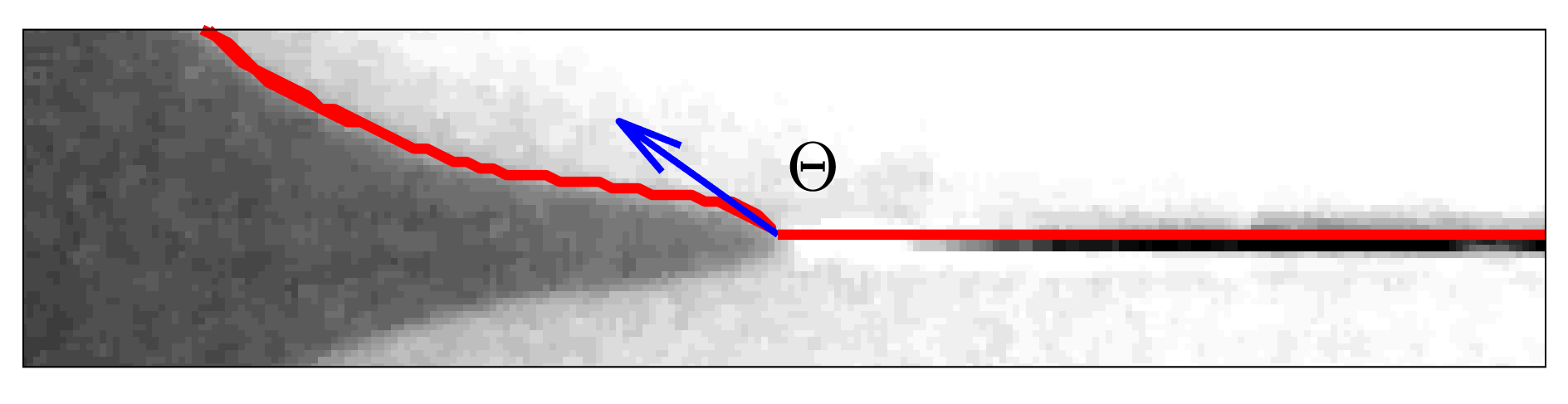}
  \caption{Sample visualization image of a drop contact line at the hydrophilic orifice. The arrow indicates the contact angle obtained via image processing.}
\label{fig_CL}
\end{figure}

A similar image processing method was employed on backlit visualization images to identify the contact line for the analysis in Section \ref{dcontact}. For this specific analysis, the drop fluid was mixed with a very small amount of food-dye instead of Rhodamine. First a region of interest extending 150 pixels above the plate was selected, such that the image intensity inside the drop was fairly uniform. Due to better signal quality, the image processing in the visualization images was easier than in the PIV images. One difference between a PIV raw image and a visualization image was that, in case of PIV, the high-intensity pixels within the drop were surrounded by a dark background. In the visualization image on the other hand, it was exactly the opposite because of the nature of backlit silhouetting technique. Therefore, these images were inverted before performing any of the image processing steps described above. Secondly, since no seeding particles were visible in the visualization images, no median filter was applied. With this method, the maximum uncertainty in contact line location was $\pm$5-7 pixel (0.1-0.14 mm). The contact line for a representative image is shown in  figure \ref{fig_CL}). To measure the contact angle (shown in figure \ref{fig_CL}), the identified contact line extending about 300 microns above the contact point was linearly fitted using a least-squares method.

\subsection{Energy calculation}
\label{sec02c}
\begin{figure}[h]
  \centering
	\includegraphics[scale=0.6]{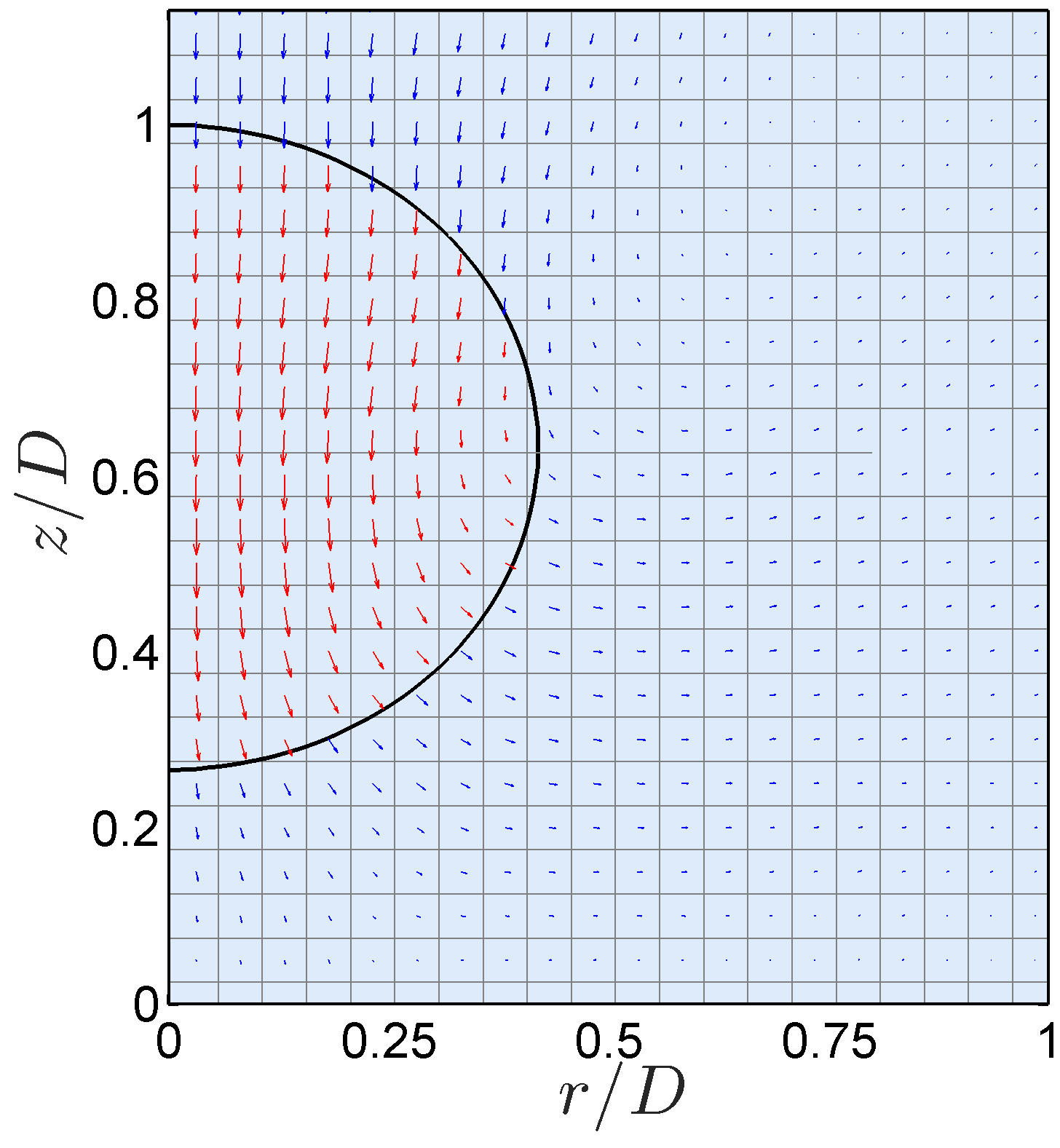}
  \caption{(Color online) Isolation of velocity fields inside a drop (red) and in the surrounding fluid (blue) based on interface location}
\label{fig04}
\end{figure}

The redistribution and dissipation of mechanical energy within a dispersed liquid/liquid system is complex and difficult to predict without knowledge of the local velocity fields. To estimate the total kinetic energy in the drop ($KE_d$) and in the surrounding oil ($KE_s$) at a given time, we first isolated the velocity fields in each phase according to the interface location (see figure \ref{fig04}). The kinetic energy for each phase was then calculated by integrating the local kinetic energy   $ke(i,j) = \frac{1}{2}(u_r^2(i,j) + u_z^2(i,j))$ within each grid cell over our measurement region (19.2 mm $\times$ 19.2 mm) as:
\begin{equation}
  \centering
KE_{d/s} = 2\pi\sum_{i=1}^{m}\sum_{j=1}^{n}r(i,j)ke(i,j)\Delta r\Delta z
\end{equation}
\noindent
where $r(i,j)$ is the radial distance of a cell from the drop axis.  In estimating  $ke(i,j)$, the the azimuthal component of velocity ($u_\theta$) was assumed to be negligible. Before performing the integration, velocity vectors corresponding to locations within the orifice (unresolved by PIV) were interpolated from the neighboring vector values above and below the orifice. The translational kinetic energy of the drop ($KE_{d,T}$) was determined based on its centroidal displacement between time steps.  This value was subtracted from the total kinetic energy $KE_{d}$ to determine the drop kinetic energy associated with local strain and rotation ($KE_{d,D}$). The surface deformation energy of the drop, representing the energy required to deform it from a spherical shape, was estimated as
\begin{equation}
  \centering
E_{\sigma}(t) = \sigma(A-A_0).
\end{equation}
\noindent
Here, $A$ is the instantaneous surface area of the drop estimated assuming symmetry of the drop about the vertical axis, and $A_0$ is the surface area of a volume equivalent sphere.

\section{Results and discussion}
\label{sec03}
\subsection{General dynamics}
\label{sec3.1}
\begin{figure}
 \centering
        \begin{subfigure}[b]{0.4\textwidth}
                \centering
                \includegraphics[width=\textwidth]{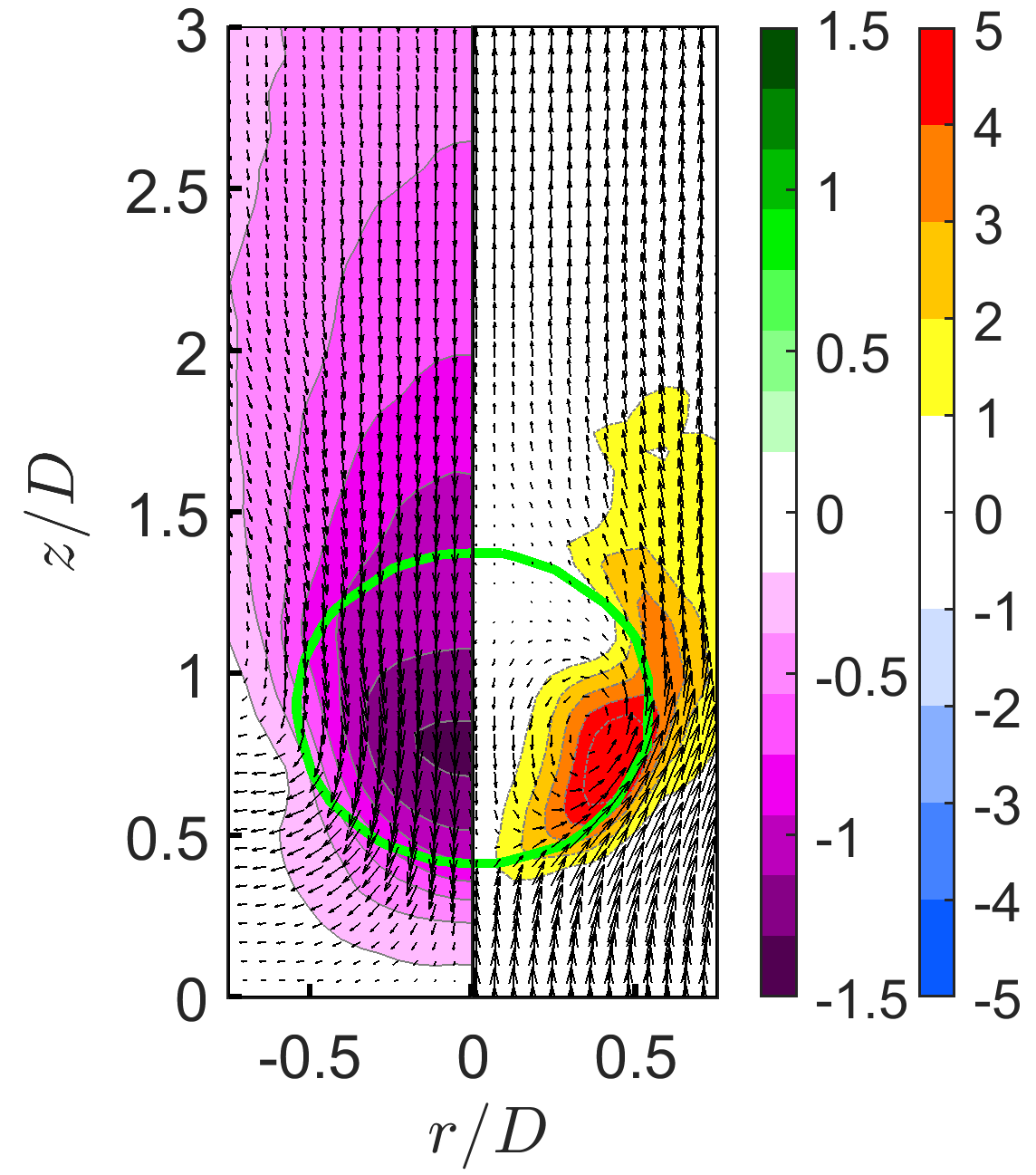}
                \caption{}
                \label{fig5a}
        \end{subfigure}
        \begin{subfigure}[b]{0.45\textwidth}
                \centering
                \includegraphics[width=\textwidth]{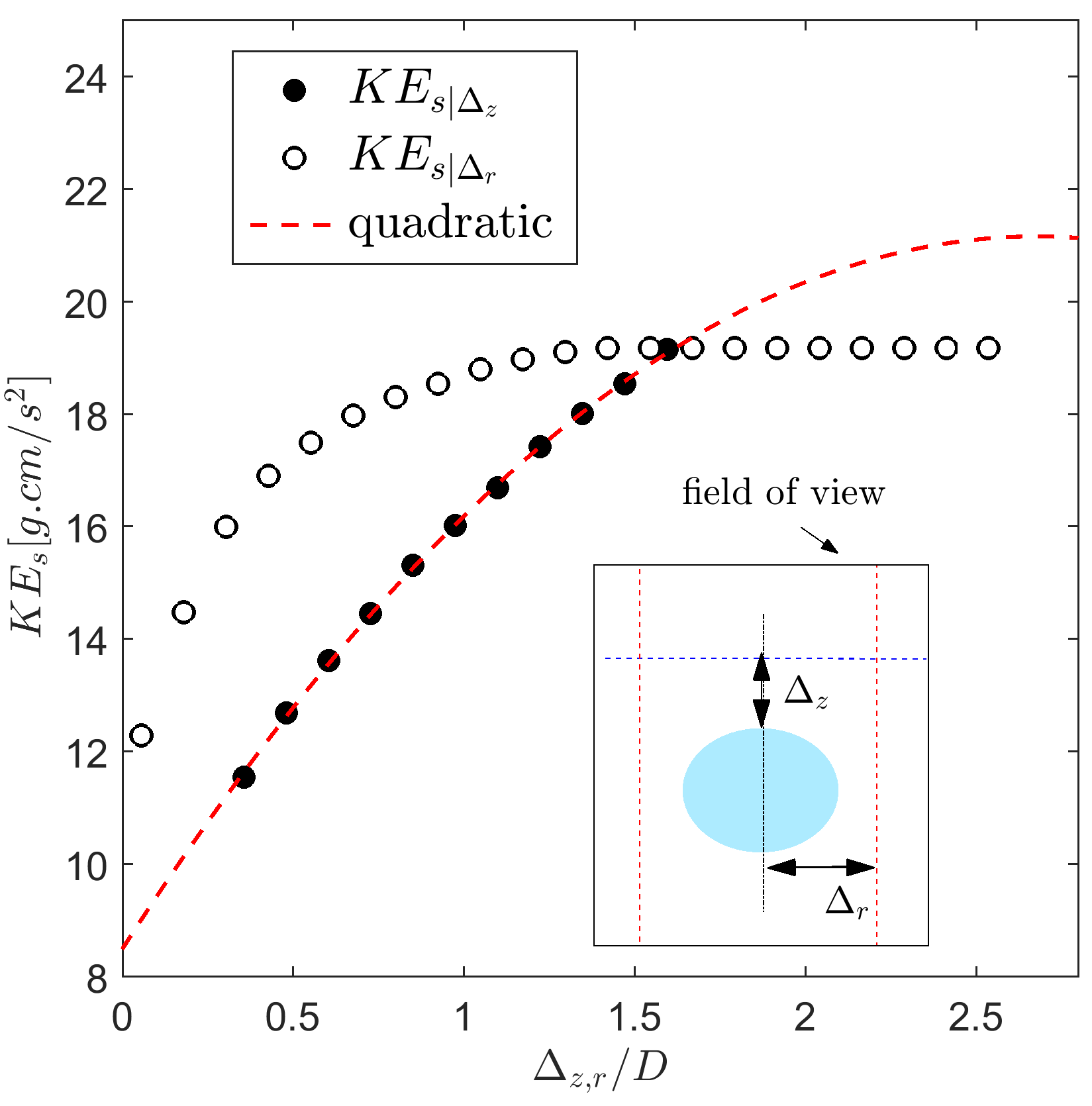}
                \caption{}
                \label{fig5b}
        \end{subfigure}
     
        \caption{a) Axial velocity ($u_z/U_t$: left panel) and out-of-plane vorticity ($\omega_{\theta}D/U_t$: right panel) contours superposed with velocity vectors. Contour levels and vectors on the left panel are shown with respect to the lab reference frame. Velocity vectors on the right panel are shown with respect to the drop reference frame. b) Kinetic energy of the surrounding fluid estimated over test windows of increasing size in $z$ ($KE_{s|\Delta z}$) and $r$ ($KE_{s|\Delta r}$) directions. }\label{fig05}
\end{figure}

To describe the general dynamics of a drop approaching and moving through an orifice, we first choose an example case, where a drop with $Bo$ = 3.7 impacts a round-edged hydrophobic orifice with $d/D$ = 0.62 (RHPB1 in Table \ref{tab02}). The corresponding Reynolds and Weber numbers of the falling drop are 13.2 and 2.6, respectively. A release height of 120 mm above the orifice plate allowed the drop to reach its terminal speed ($U_t$ = 8.2 cm/s) before impacting on the orifice. The terminal state of the drop is captured at $t=-0.47t_i^*$ in figure \ref{fig5a} when its leading interface is located 0.33$D$ above the plate surface. Here, $t_i^*= D^3/d^2U_i$ is a passage time-scale through the orifice based on the drop velocity $U_i$ immediately before the impact and the fact that the drop must be deformed into an elongated shape during its passage \citep{Bordoloi14}. The drop shape is symmetric about {the} central vertical axis, and it approaches the orifice with an oblate spheroidal shape (width-to-height ratio = 1.1). The drop interior carries a steady vortex ring developed as a combination of shearing and baroclinity generated during and after the drop’s release from the tube. The core of the vortex ring is identified based on the vectors seen from the drop reference frame and the accompanying vorticity contours.  The right-hand side of figure \ref{fig5a} shows that the contours of anticlockwise (positive) vorticity are focused near the interface and inside of the drop because the interior viscosity is lower than the exterior viscosity. The velocity induced by the vortex ring along the centerline of the drop increases the maximum axial velocity at the core of the drop to  about 30\% larger than $U_t$  (see left hand side of figure \ref{fig5a}). The drop kinetic energy ($KE_d$) at the terminal state ($t=-0.45t_i^*$) was estimated to be 12 $\mathrm{g.cm/s^2}$. The motion in the fluid downstream of the drop is small and significant only within 0.3$D$ downstream (see figure \ref{fig5a}). On the other hand, a wake of trailing fluid extends upward behind the drop beyond the field of view.  

The kinetic energy of the surrounding fluid ($KE_s$) was estimated across a field of view of 2.4$D$ along the $z$-axis and $\pm$2.5$D$ along the $r$-axis. The validity of this estimate was tested by examining the kinetic energy distribution in the z and r directions. The kinetic energy (labeled as $KE_{s|\Delta_z}$  and $KE_{s|\Delta_r}$ ) was calculated over test windows of increasing size (see inset in figure \ref{fig5b}) in $z$ and $r$ independently through increments  $\Delta_z$ and $\Delta_r$, respectively. While varying the increment in one direction, the field in the other direction included the complete span of the measurement window. Results show that  $KE_{s|\Delta_r}$ converges at 1.5$D$ from the $z$-axis such that the added increments outside this radius become negligible. The curve for $KE_{s|\Delta_z}$  does not converge completely within the field of view. However, based on a quadratic fit to the data, the extent of the upstream wake in $z$ is estimated as 2.5$D$ beyond the trailing interface of the drop. Within the field of view at this time, the kinetic energy of the surrounding fluid ($KE_s$) was estimated as 19.3 $\mathrm{g.cm/s^2}$. The fit in figure \ref{fig5b} also suggests that the kinetic energy of the surrounding fluid at this time was underestimated by about 8\%.  The estimated $KE_s$ is only 4\% of the gravitational potential lost during the drop's descent from the release height, suggesting strong dissipation during the free fall.

\begin{figure}[h]
  \centering
	\includegraphics[scale=4]{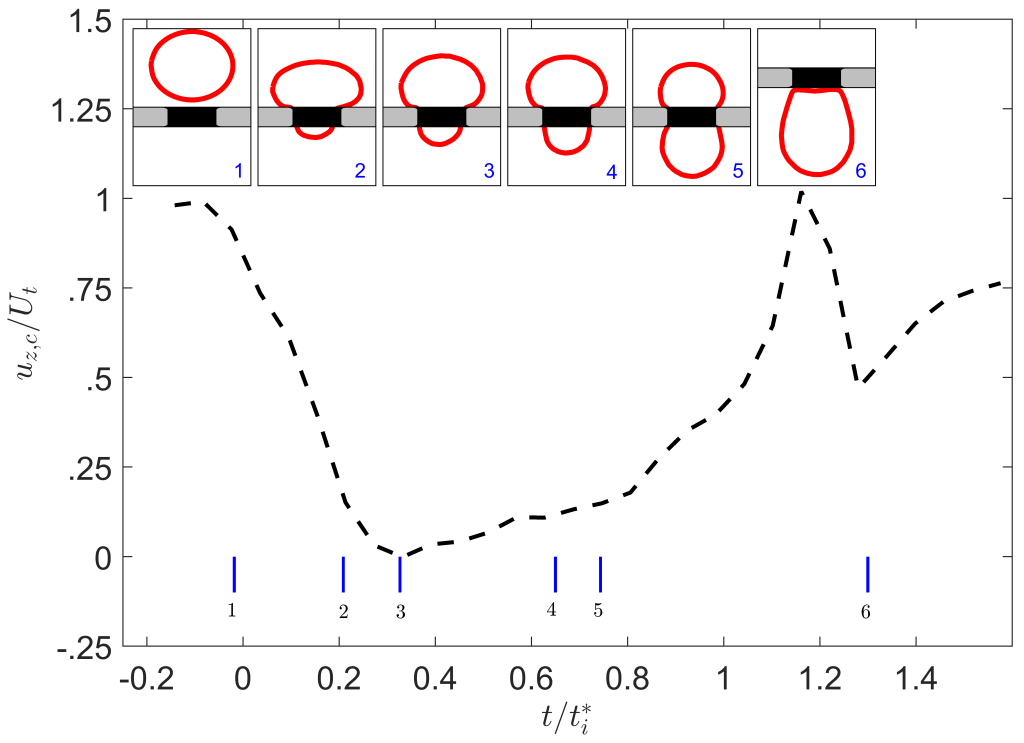}
  \caption{(Color online) Variation of drop centroid velocity ($u_{z,c}/U_t$) for a drop with $Bo$ = 3.7  moving through a round-edged orifice of $d/D$ = 0.62 (RHPB1 in Table \ref{tab02}), and outlines of the drop interface at six specified times.}
\label{fig06}
\end{figure}

For a drop passing through a round-edged orifice, the dynamics are independent of the surface properties of the plate since a thin film of oil separates the drop fluid from the orifice surface \citep{Bordoloi14}. As the drop impacts on and moves through the orifice, it decelerates and deforms before eventually re-accelerating toward its terminal velocity. Figure \ref{fig06} shows the non-dimensional centroid velocity ($u_{z,c}/U_t$) with respect to normalized time ($t/t_i^*$), accompanied by the drop shape and position at specified times. The reference time $t$ = 0 occurs when the leading drop interface reaches the height of the upper plate surface. The descending drop begins decelerating from its terminal velocity as it approaches the plate, and reaches approximately 0.8$U_t$ by the time of impact (frame 1). Post impact, the drop undergoes a series of deformations, first by flattening radially outward above and across the plate (frame 2) and then contracting inward (frame 3) at $t = 0.33t_i^*$. During this period, the drop decelerates in the vertical direction until it momentarily stagnates above the plate. After $t = 0.33t_i^*$, the drop begins to accelerate, first slowly (frame 3-frame 5) as the fluid above the plate retracts and starts penetrating into the orifice under gravity, and then more strongly (after frame 5) when the plate no longer blocks any of the drop fluid. An additional short deceleration occurs between frames 5 and 6, when the trailing interface of the drop drags the more viscous trailing fluid into and through the orifice. Beyond frame 6, the drop rearranges into a spheroidal shape and begins accelerating toward its terminal speed. We discuss below in detail the unsteady nature of the flow field inside and outside of the drop during its motion through the orifice.

\begin{figure}[h]
 \centering
        \begin{subfigure}[b]{0.3\textwidth}
                \centering
                \includegraphics[width=\textwidth]{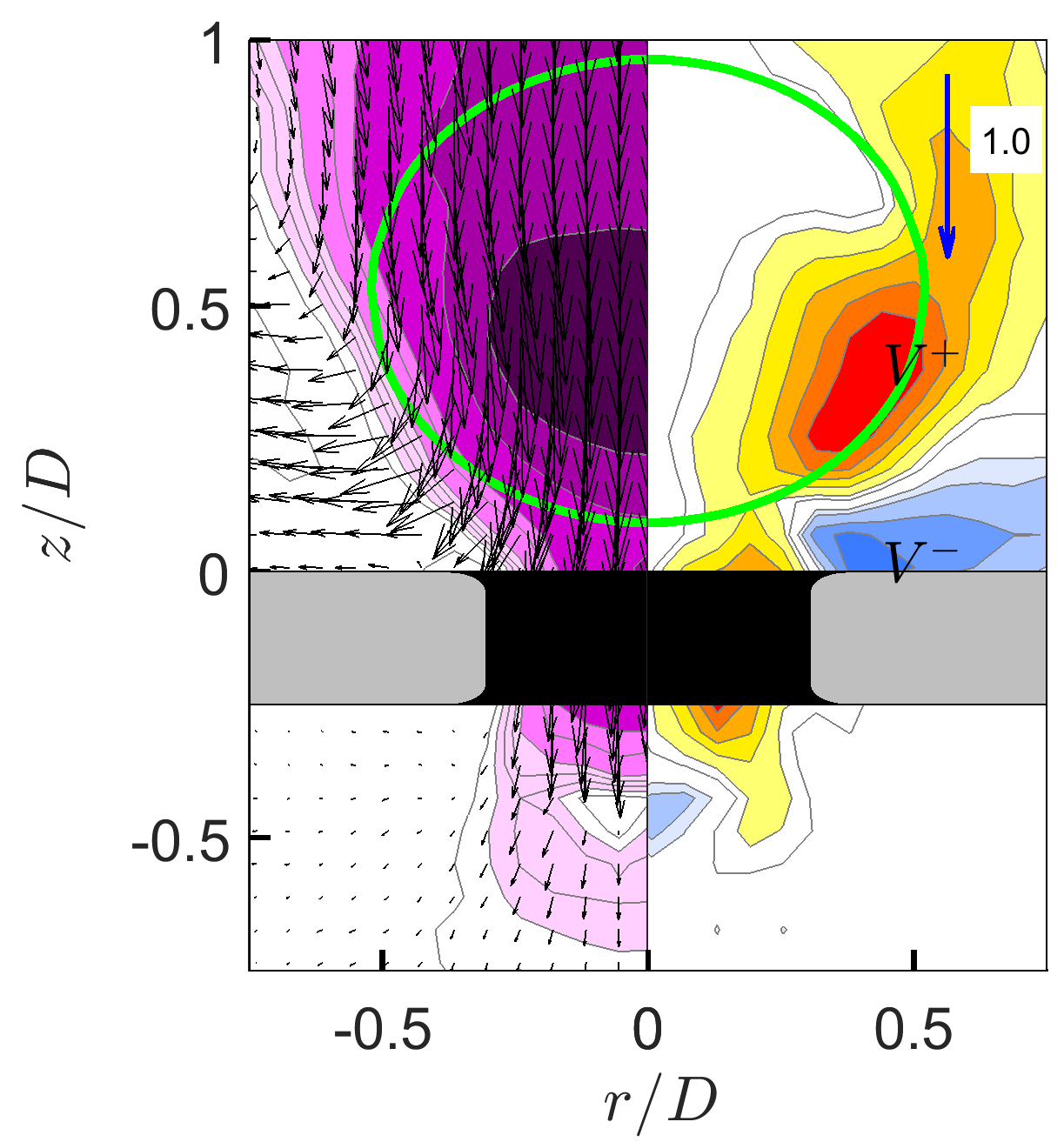}
                \caption{}
                \label{fig7a}
        \end{subfigure}
        \begin{subfigure}[b]{0.3\textwidth}
                \centering
                \includegraphics[width=\textwidth]{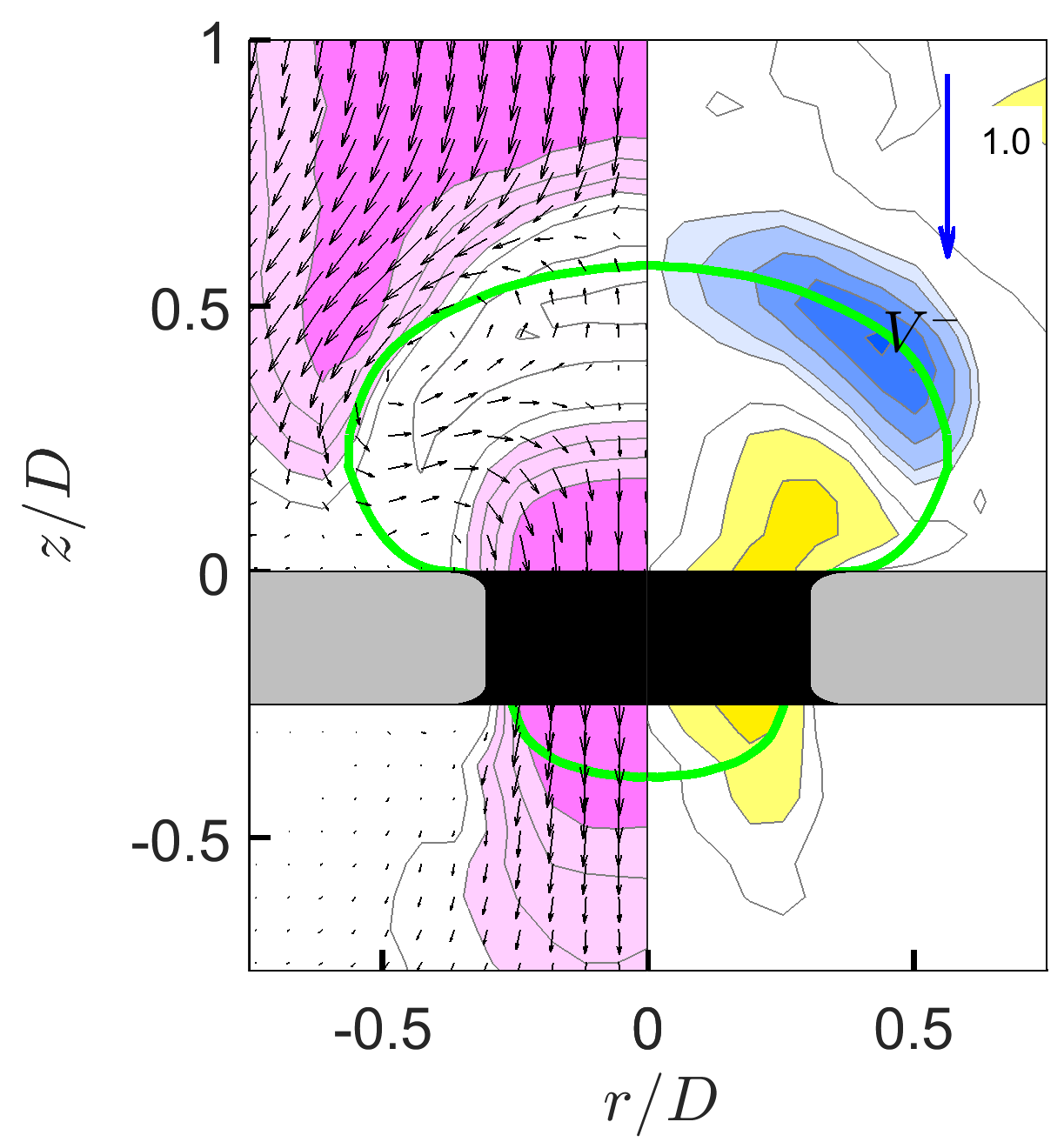}
                \caption{}
                \label{fig7b}
        \end{subfigure}
        \begin{subfigure}[b]{0.374\textwidth}
                \centering
                \includegraphics[width=\textwidth]{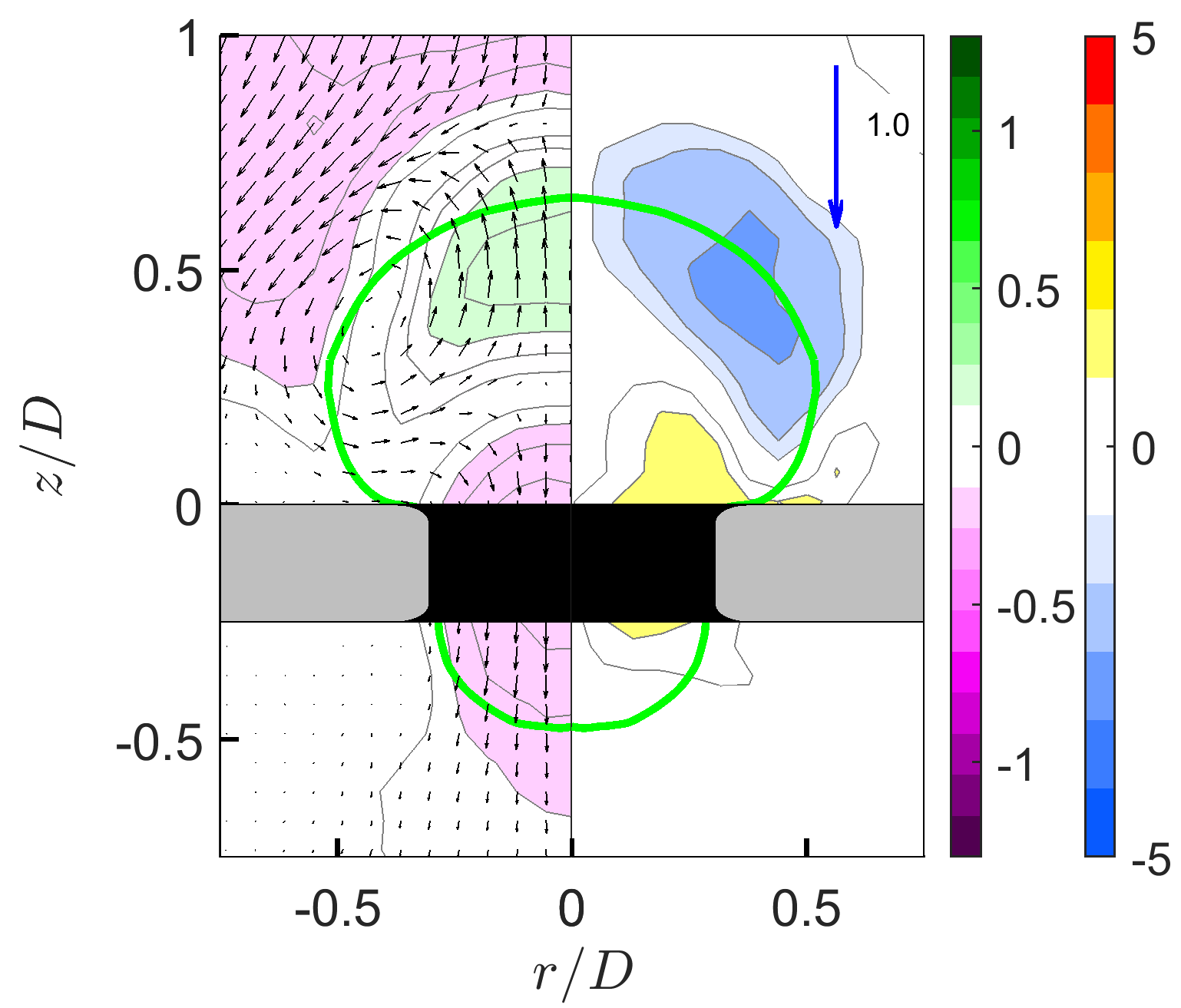}
                \caption{}
                \label{fig7c}
        \end{subfigure}
          
        \caption{(Color online) Contours of axial velocity ($u_z/U_t$: left panel) and out of plane vorticity ($\omega_{\theta}D/U_t$: right panel) for a drop with $Bo$ = 3.7 impacting on a round-edged orifice with $d/D$ = 0.62 during the initial deceleration phase at times, $t =$ a) -0.02$t_i^*$, b) 0.22$t_i^*$ and c) 0.33$t_i^*$.  }\label{fig07}
\end{figure}

The initial deceleration phase of the drop is shown in Figure \ref{fig07} via the normalized axial velocity contours ($u_z/U_t$) superposed with velocity vectors (on the left), and the out of plane normalized vorticity ($\omega_\theta D/U_t$) contours (on the right) at three specified times. Immediately before the impact, at $t=-0.02t_i^*$ the oil downstream of the drop moves through the orifice with axial velocity, $u_z \approx 0.5U_t$.   Although the mean velocity of the drop before impact is reduced to about 0.8$U_t$, the maximum downward velocity, located near the drop center, is about 1.2$U_t$ due to the internal circulation. The maximum counterclockwise drop vorticity resolved by the PIV is 5$D/U_t$. This positive vorticity region is labeled as $V^+$ for future reference. On the other hand, the radially outward motion of the viscous surrounding oil above the plate produces opposing vorticity ($V^-$) as a boundary layer develops close to the surface. 

The oscillation caused by the impact on the drop fluid above the plate is captured in  figures \ref{fig7b} (expansion) and \ref{fig7c} (retraction). The opposing capillary force within the orifice continues to decelerate the drop front reducing the axial velocity of the drop fluid below the orifice plate to 0.4$U_t$ at $0.22t_i^*$ and 0.2$U_t$ at $0.33t_i^*$. In addition, the radial contraction above the plate causes an upward flow within the drop which also impedes its downward motion. The positive vorticity within the drop fluid above the plate decreases continuously to a local maximum of 2$D/U_t$ at $t=0.22t_i^*$ and becomes negligible by $t = 0.4t_i^*$ (not shown here).  Meanwhile, as the bulk of the drop fluid nearly stagnates above the plate in the axial direction ($t=0.2-0.4t_i^*$), the surrounding fluid in the wake continues to impinge on the drop and shear against the upper drop surface.  The lateral shear caused by the outwardly moving wake fluid generates a region of clockwise vorticity (labeled as $V^-$ at $0.22t_i^*$ in figure \ref{fig7b}). The subsequent radial contraction of the drop rotates the interface in a similar sense, also influencing the local vorticity field. At $t = 0.33t_i^*$, the combination of these two effects has resulted in a secondary vortex ring. A similar secondary vortex was also observed for a drop impacting on a planar liquid/liquid interface \citep{Mohamed03}, and on a flat solid surface in a liquid/air system \citep{Kumar17}. The viscosity ratio between the two fluids influences the location of the secondary vortex ring. In the liquid/air system with large viscosity ratio between the drop and the surrounding air considered in  \citet{Kumar17}, the secondary vortex ring is located inside the drop away from the drop interface. In our case, the surrounding liquid has higher viscosity than the drop, and the vortex ring straddles the interface.

The secondary vortex motion within the upper drop fluid pushes the trailing interface upward so that gravitational potential of the fluid above the plate increases. The increased gravitational potential, on the other hand, opposes rotation in the secondary vortex ring. Consequently, vorticity in $V^-$ begins to decrease after $t = 0.33t_i^*$. Between $t=0.33-0.74t_i^*$, the fluid above the plate retracts laterally inward towards the orifice, while gravity induces slow axial motion in the gradually increasing volume of the drop fluid below the orifice.

\begin{figure}[h]
 \centering
        \begin{subfigure}[b]{0.29\textwidth}
                \centering
                \includegraphics[width=\textwidth]{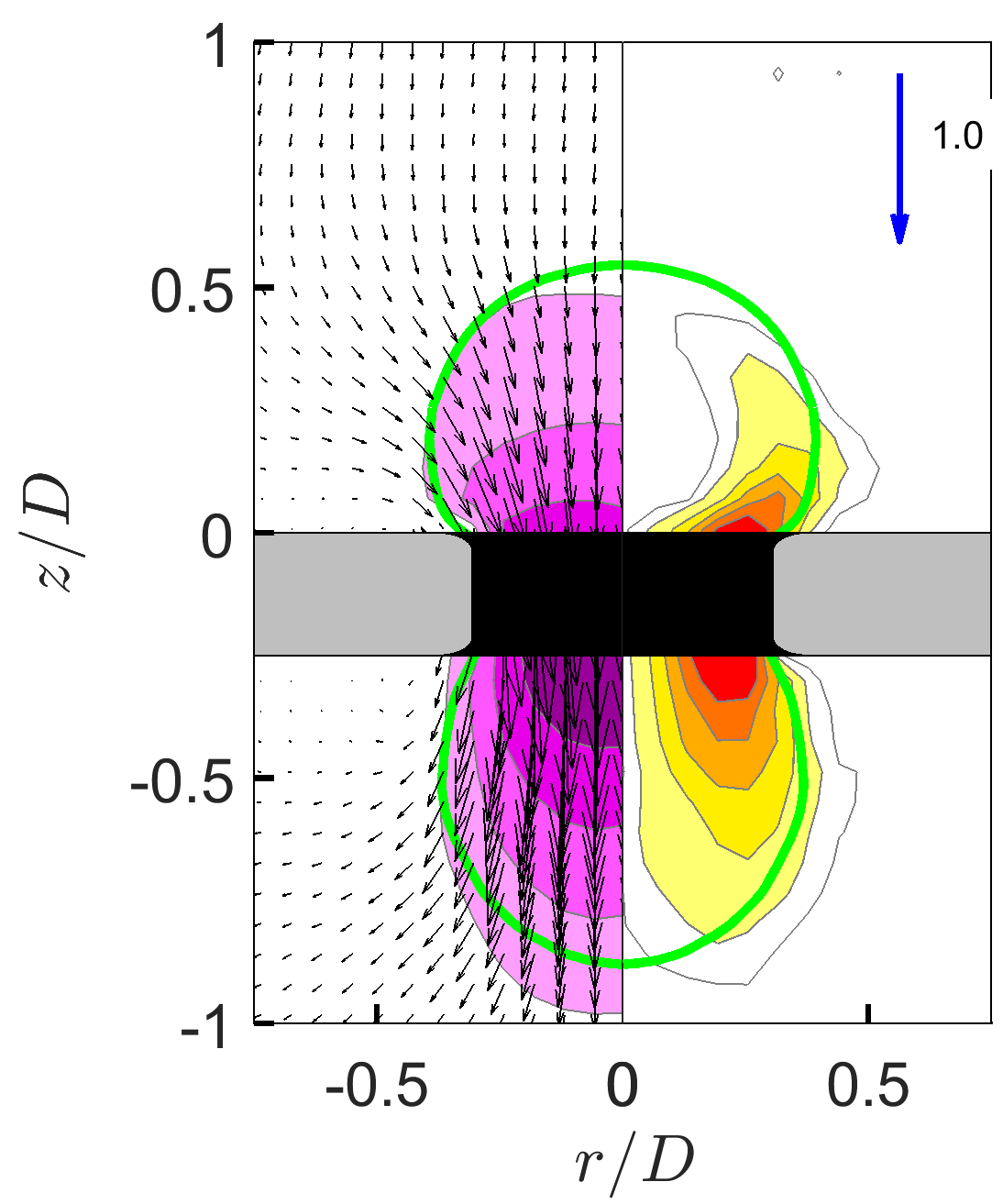}
                \caption{}
                \label{fig8a}
        \end{subfigure}
        \begin{subfigure}[b]{0.29\textwidth}
                \centering
                \includegraphics[width=\textwidth]{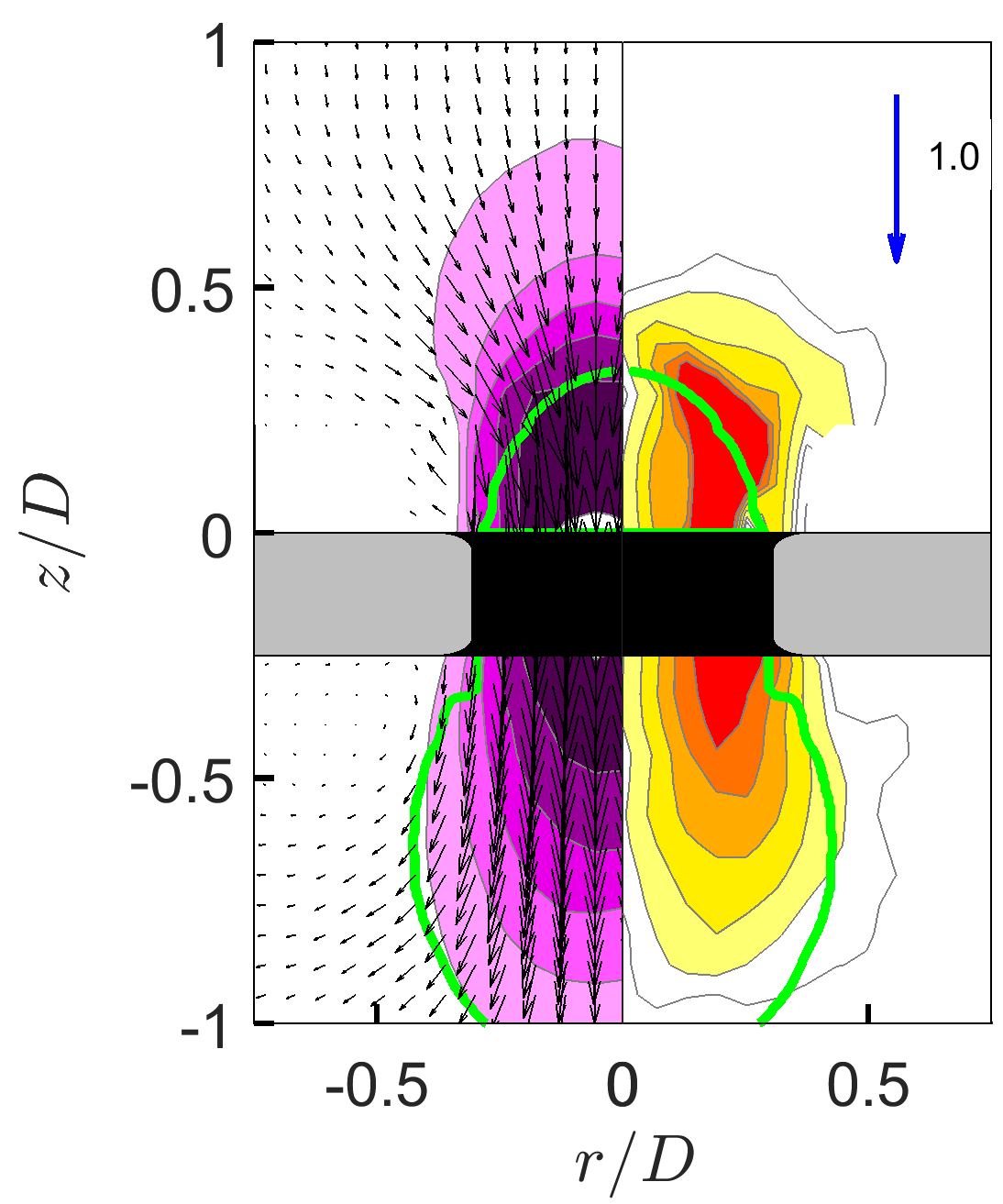}
                \caption{}
                \label{fig8b}
        \end{subfigure}
        \begin{subfigure}[b]{0.397\textwidth}
                \centering
                \includegraphics[width=\textwidth]{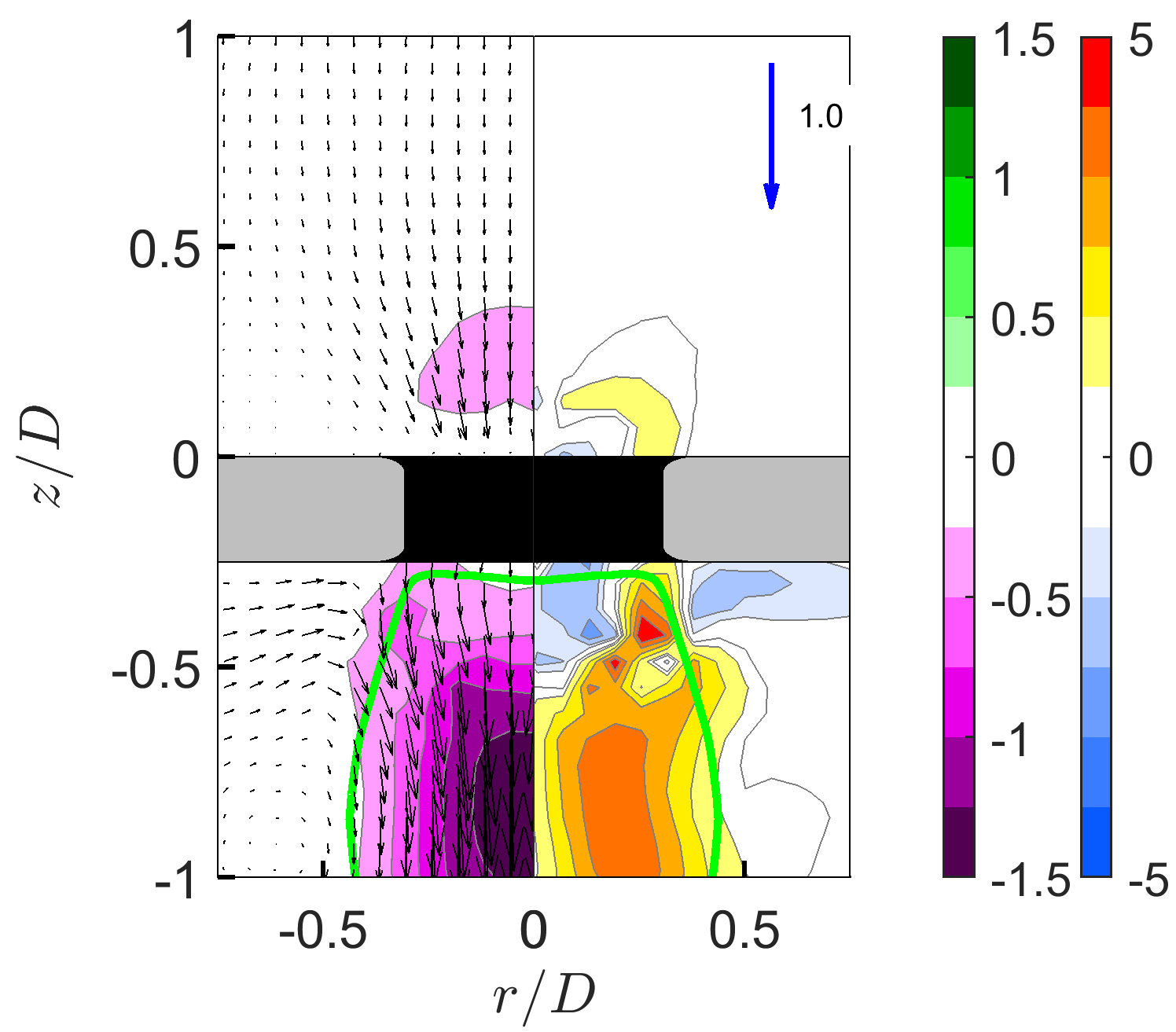}
                \caption{}
                \label{fig8c}
        \end{subfigure}
          
        \caption{(Color online) Contours of axial velocity ($u_z/U_t$:  left panel) and normalized out of plane vorticity ($\omega_{\theta}D/U_t$: right panel) for a drop with $Bo$ = 3.7 impacting on a round-edged orifice with $d/D$ = 0.62 during the re-acceleration phase at times $t=$ a) 0.74$t_i^*$, b) 1.0$t_i^*$ and c) 1.3$t_i^*$.  }\label{fig08}
\end{figure}

After all of the drop fluid above the plate retracts to within $r \approx D/2$ (see $t=0.74t_i^*$ in figure \ref{fig06}), the trailing interface forms a cap-like shape, and due to gravity, the drop internal fluid reaches a maximum axial velocity of 1.2$U_t$. {The axial velocity and out-of-plane vorticity during the acceleration phase of the drop are shown at three specified times in figure} \ref{fig08}. The accelerated penetration with increasing volume of drop fluid passing through the orifice results in regions of strong clockwise vorticity within the drop near the orifice.  Also, the curvature of the trailing interface increases (see $t=1.0t_i^*$).  The enhanced capillary force consequently pushes the remaining fluid downward, adding to the drop acceleration. The maximum velocity inside the drop at $t=1.0t_i^*$ increases to 1.5$U_t$. The acceleration of the drop fluid also influences the velocity field in the trailing surrounding fluid: the axial velocity in the immediate wake of the drop increases from 0.2$U_t$ at 0.75$t_i^*$ to 0.6$U_t$ at 1.0$t_i^*$. {However, as the drop trailing interface exits the orifice, the viscous drag related to the inner orifice surface causes the upstream surrounding fluid and the downstream drop to decelerate (see 1.3$t_i^*$). The internal axial velocity decreases to a maximum of 1.0$U_t$.}

\subsection{Distribution of energy}
\label{sec3.2}
\begin{figure}
 \centering
        \begin{subfigure}[b]{0.46\textwidth}
                \centering
                \includegraphics[width=\textwidth]{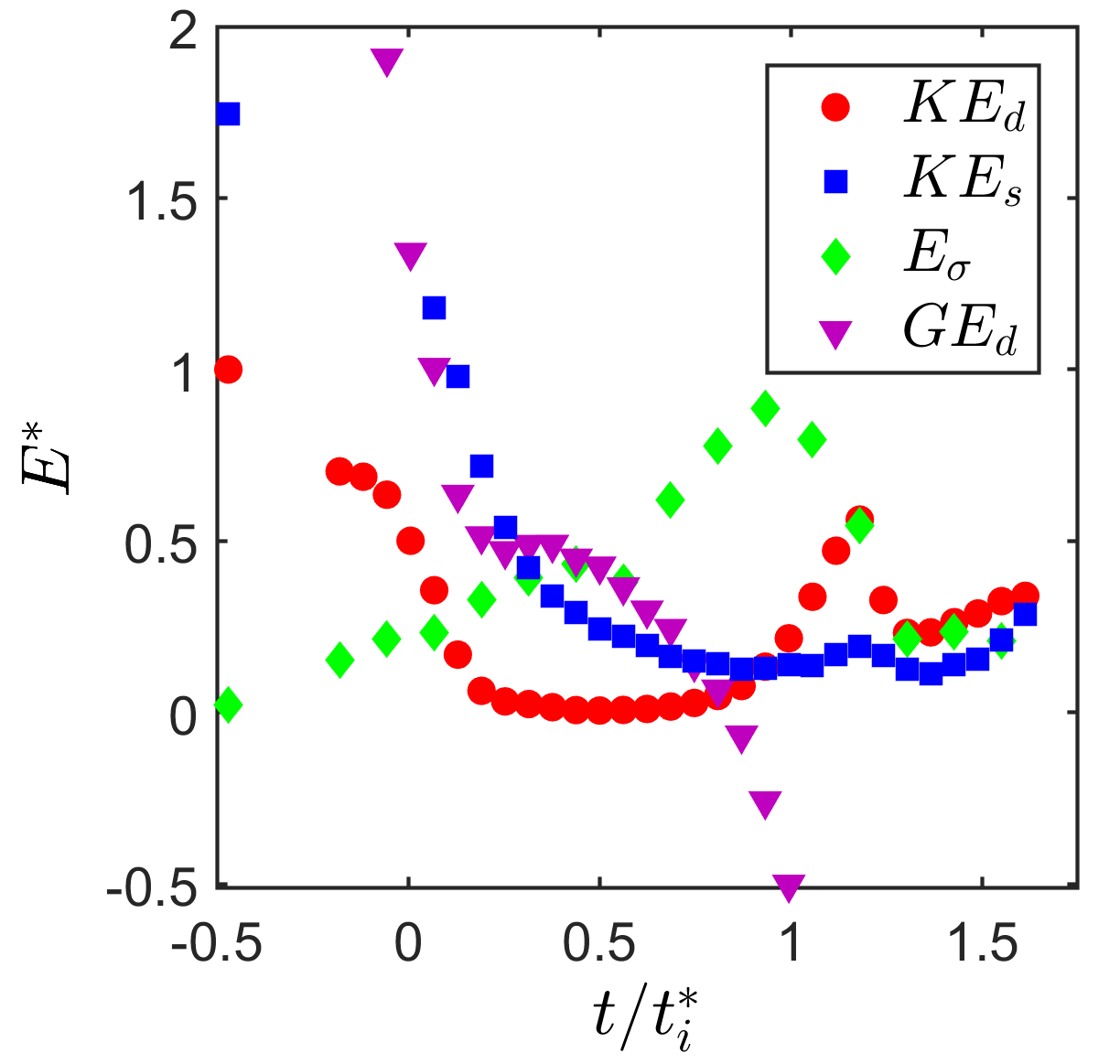}
                \caption{}
                \label{fig9a}
        \end{subfigure}
        \begin{subfigure}[b]{0.45\textwidth}
                \centering
                \includegraphics[width=\textwidth]{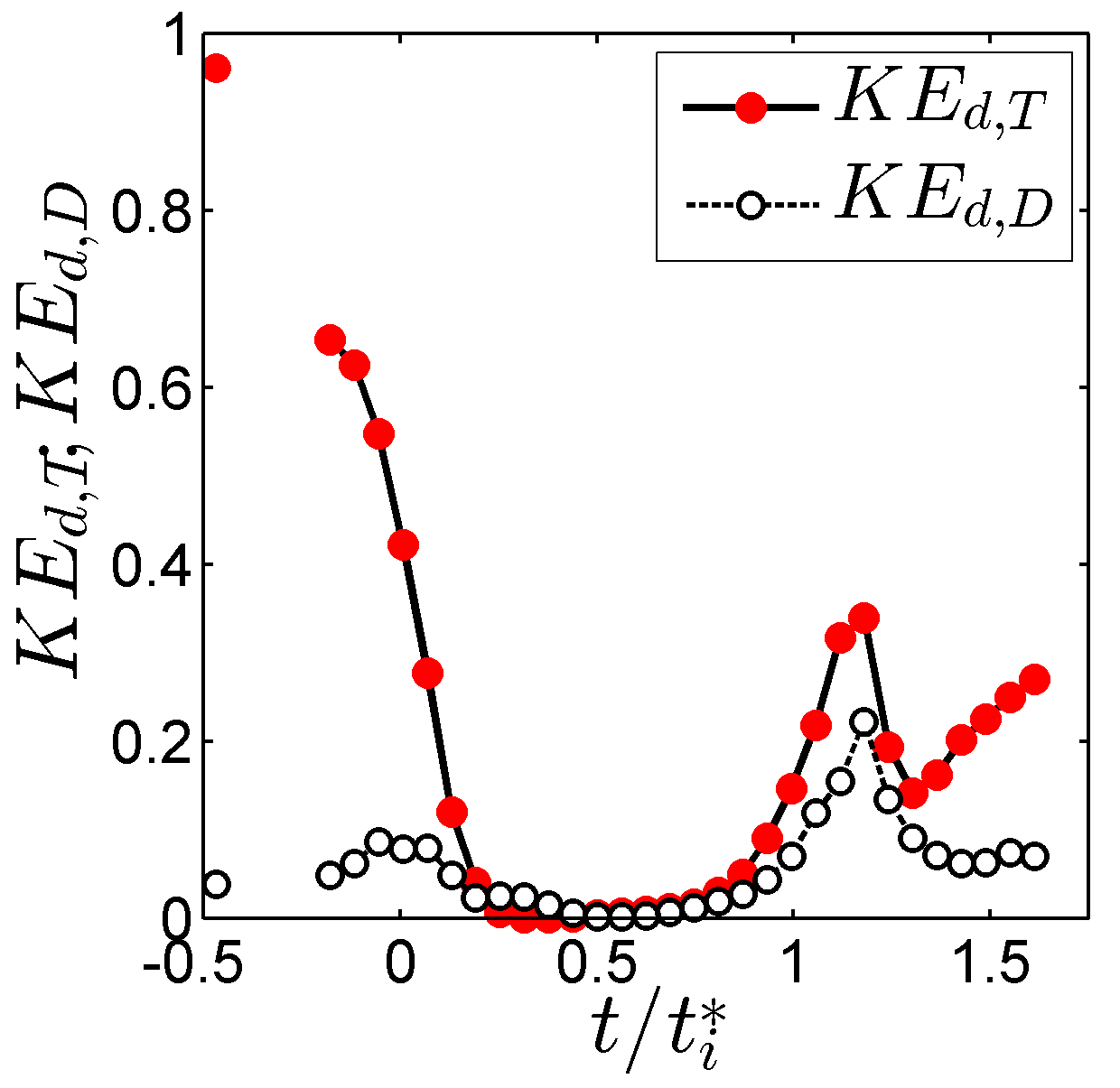}
                \caption{}
                \label{fig9b}
        \end{subfigure}
       
        \caption{(Color online) a) Variation of total kinetic energy inside the drop ($KE_d$), total kinetic energy in surrounding oil ($KE_s$), drop deformation energy ($E_{\sigma}$), {and the gravitational potential energy ($GE_d$) above the orifice plate} for a drop with Bo= 3.7 moving through round-edged orifice with d/D=0.62, and b) decomposition into its translational component ($KE_{d,T}$) and the component due to local deformation and internal circulation ($KE_{d,D}$). Each quantity is normalized by the drop kinetic energy in its free-falling state.}\label{fig09}
\end{figure}

The redistribution of energy during the process described above is captured in figure \ref{fig9a}, which shows the evolution of the total kinetic energy of the drop ($KE_d$) and the surrounding oil ($KE_s$), the drop deformation energy ($E_{\sigma}$), and the drop gravitational energy ($GE_d$). {The underestimate in $KE_s$ due to the limited field of view is corrected at every instant using a quadratic fit as discussed in section} \ref{sec02c} (see figure \ref{fig5b}). The components of drop kinetic energy due to: (i) vertical translation ($KE_{d,T}$) and (ii) internal strain and rotation ($KE_{d,D}$) are shown in figure \ref{fig9b}. Each quantity is normalized by the drop kinetic energy in its free-falling state. During the free-falling stage captured at $t=-0.45t_i^*$, the drop kinetic energy ($KE_d$) is dominated by translational motion ( $\approx$ 96\%), with only a small contribution from internal circulation ( $\approx$ 4\%).  The surrounding oil attains more than 1.5 times the kinetic energy inside the drop as a result of the drop\rq{}s losing gravitational potential and the shear between the two fluids. During free fall, the deformation energy ($E_{\sigma}$) of the drop is negligible. As the drop front approaches the orifice and decelerates due to the resistance of the plate, $KE_d$ and $KE_s$ {decrease to approximately half and two thirds} of their initial values (see figure \ref{fig9a}). A fraction of this energy results in an $E_{\sigma}$ increase of 20\%, but clearly a large portion of the kinetic energy from the drop and the surrounding fluids is lost in viscous dissipation. The decomposition of $KE_d$ in figure \ref{fig9b} shows that the decrease in $KE_d$ prior to impact is only translational. By contrast, $KE_{d,D}$ increases by about 4\% as the drop approaches impact.          

After impact, all kinetic energy parameters continue to decrease through dissipation while $E_\sigma$ increases as shown in figure \ref{fig9a}. The increased deformation in the drop during its expansion and retraction above the plate (until $t=0.33t_i^*$) imparts local strain and circulation in the drop fluid, which results in $KE_{d,D}$ exceeding $KE_{d,T}$ (see figures \ref{fig07} and \ref{fig9b}). Subsequently (0.33-0.74$t_i^*$), $E_\sigma$ continues to increase because of limited axial deformation during the drop’s slow penetration through the orifice. After t=0.74$t_i^*$ however, $E_\sigma$ increases more rapidly as the drop volume beneath the plate elongates under gravity.  During this period, the drop kinetic energy also increases at a faster rate due to both translation ($KE_{d,T}$) under gravity and the non-uniform internal motion ($KE_{d,D}$) caused by axial strain and vorticity development within the drop (see figure \ref{fig08} and \ref{fig9b}). The deformation energy reaches a maximum at $t=0.9t_i^*$ before decreasing as the drop relaxes toward its equilibrium spheroidal shape while exiting through the orifice. 

Although the trend in $KE_s$ variation is similar to that in $KE_d$, the decrease and later increase are delayed in time (see figure \ref{fig9a}).  These delays can be explained based on a scaling analysis. The drop Reynolds number ($Re$) can be expressed as the ratio between a viscous time-scale ($t_\mu \approx \rho_s D^2/\mu_s$) and an inertial time scale ($t_i = D/U_t$). The Reynolds number, $Re\approx 10$, in this example suggests that any change in drop inertia takes longer to diffuse into the surrounding fluid than the inertial time.  This is evident in the velocity fields upstream of the drop both after the impact (see figure \ref{fig07} at 0.33$t_i^*$), and during release (see figure \ref{fig08} at 0.74$t_i^*$). In the first, the surrounding oil shows a wake with significant downward momentum behind a stationary drop. In the second, the wake is relatively weak despite a strong velocity field within the accelerating drop. 

\subsection{Effect of orifice edge}
\label{orifice_edge}
When the orifice edge is sharp, the drop fluid makes contact with the edge immediately upon impact. The subsequent dynamics are influenced {significantly} by both the solid boundary and the wettability of the orifice surface. The contact at the {sharp} edge results in a leading contact line that connects the leading drop fluid with the inner surface of the orifice, and a trailing contact line that connects the trailing drop fluid with the upper plate surface.

\begin{figure}[h]
  \centering
	\includegraphics[scale=0.75]{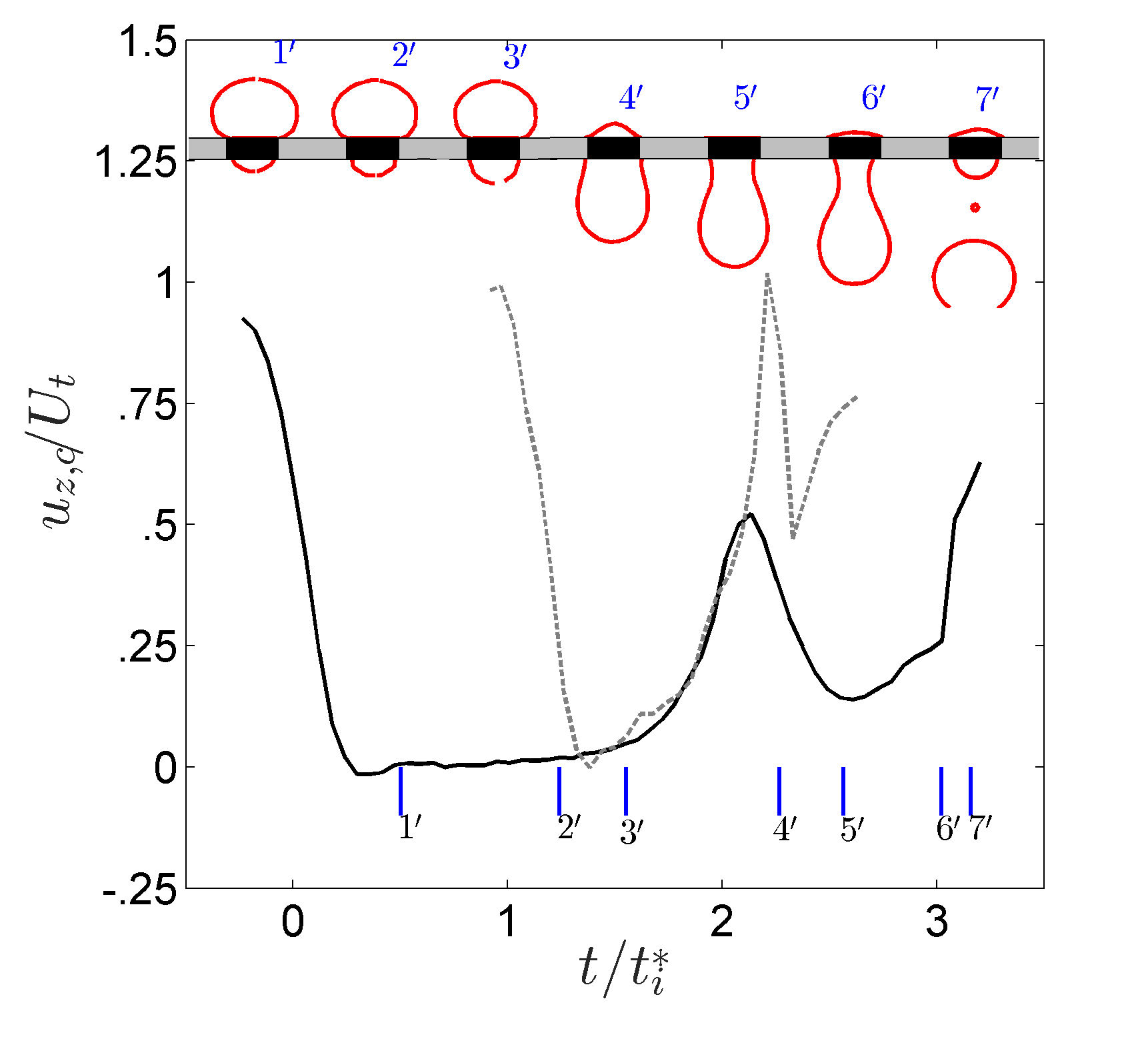}
  \caption{(Color online) Variation of drop centroid velocity ($u_{z,c}/U_t$) for a drop with $Bo$ = 3.7 and $d/D$ = 0.62 moving through a sharp-edged hydrophobic {(SHPB1 in Table} \ref{tab02}) orifice superposed with the outlines of the drop interface at seven specified times. The plot from figure \ref{fig06}, {with time-axis shifted to match the onset of penetration in the {SHPB1} case}, is included for comparison.}
\label{fig10}
\end{figure}

To compare with the results of the round-edged orifice described above, we consider a case (SHPB1 in Table \ref{tab02}) with a sharp-edged hydrophobic orifice for the same initial conditions as in figure \ref{fig05}.  First we present the variation of normalized centroid velocity ($u_{z,c}/U_t$) in combination with interface outlines at specified  {times in figure} \ref{fig10}. The {same velocity variation} for the round-edged orifice case (see figure \ref{fig06}) is included in figure \ref{fig10} with time-axis shifted to match the onset of penetration in the sharp-edged orifice case. Compared to the round-edged orifice, the drop motion through the sharp-edged hydrophobic orifice extends over a much longer period ($3.2t_i^*$). Also the final outcome differs; {here, part of the drop fluid is captured unlike the in the round-edged orifice case where all of the fluid is released below the plate.}

\begin{figure}[h]
 \centering
        \begin{subfigure}[b]{0.29\textwidth}
                \centering
                \includegraphics[width=\textwidth]{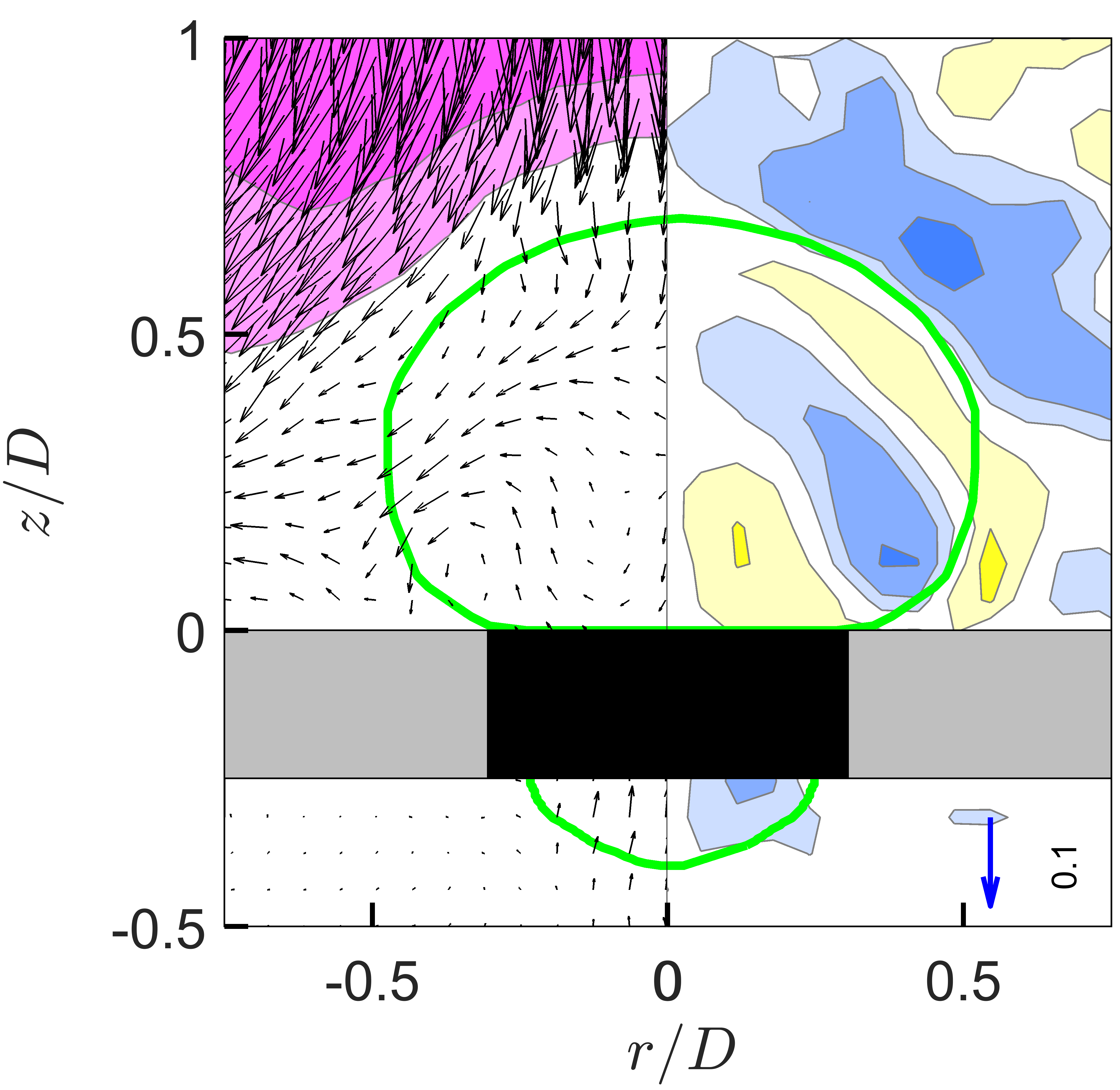}
                \caption{}
                \label{fig11a}
        \end{subfigure}
        \begin{subfigure}[b]{0.29\textwidth}
                \centering
                \includegraphics[width=\textwidth]{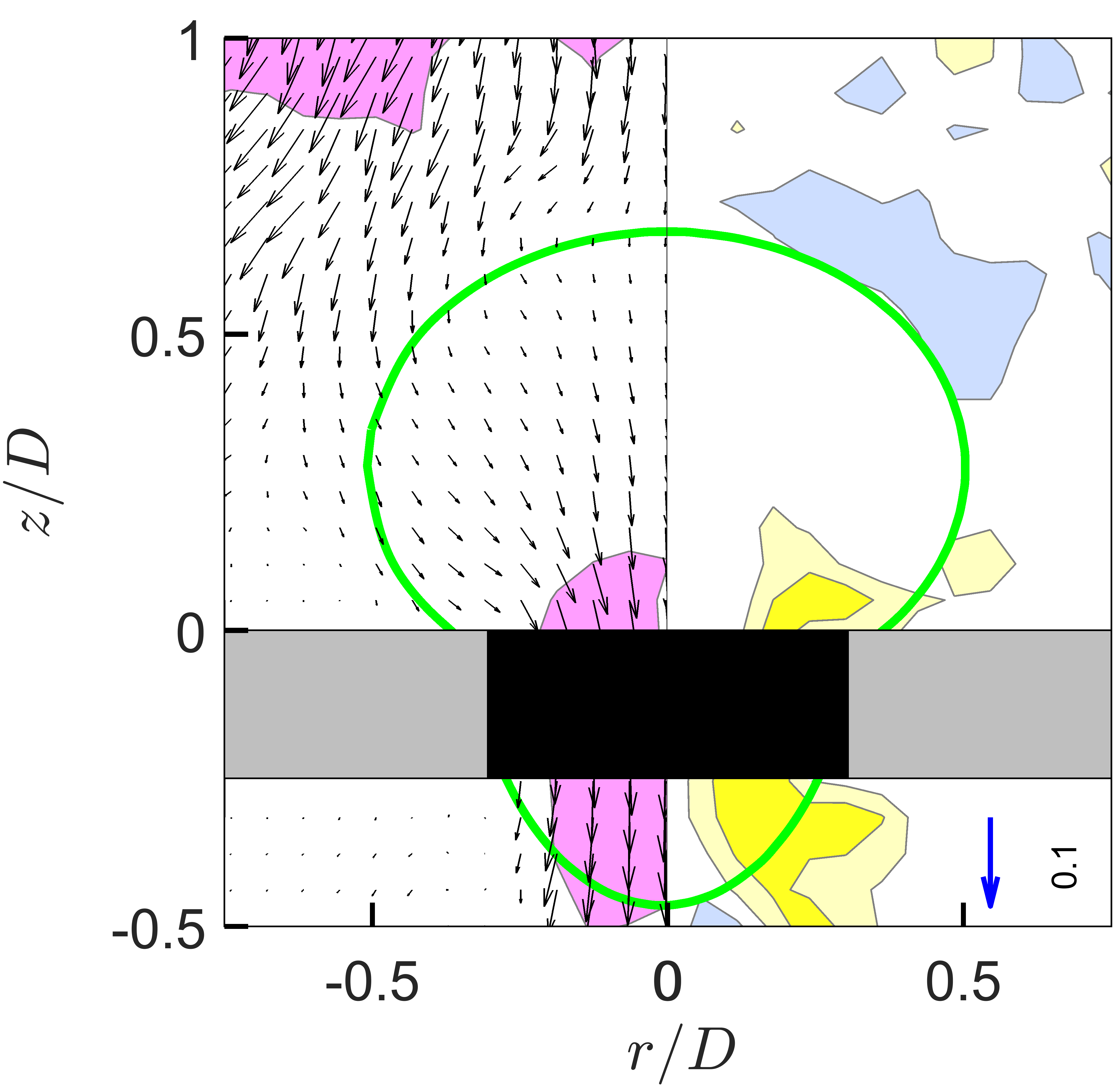}
                \caption{}
                \label{fig11b}
        \end{subfigure}
        \begin{subfigure}[b]{0.38\textwidth}
                \centering
                \includegraphics[width=\textwidth]{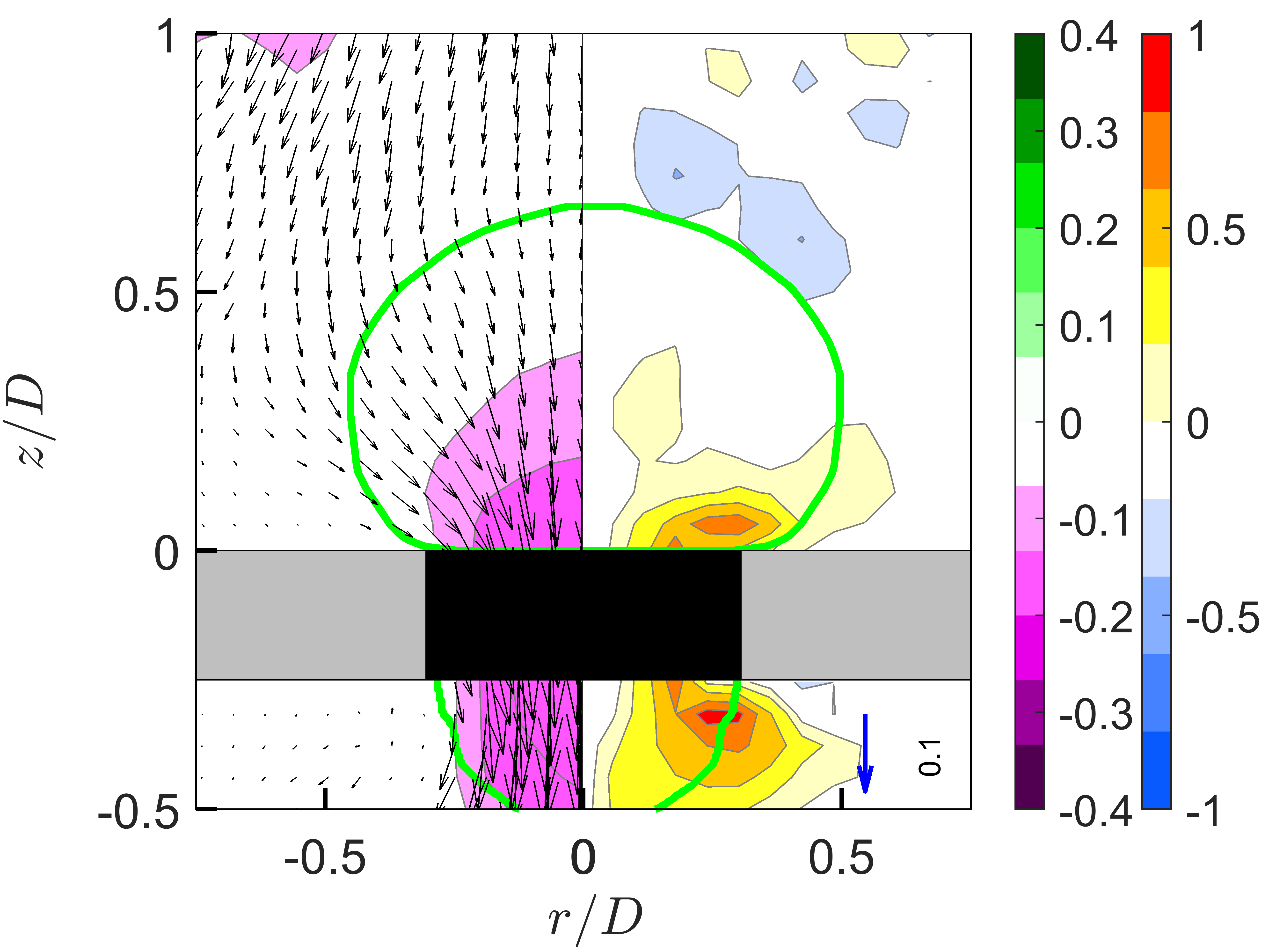}
                \caption{}
                \label{fig11c}
        \end{subfigure}
          
        \caption{(Color online) Contours of axial velocity ($u_z/U_t$: left panel) and out of plane vorticity ($\omega_{\theta}D/U_t$: right panel) for a drop with $Bo$ = 3.7 impacting on a sharp-edged hydrophobic orifice with $d/D$ = 0.62 during the contact line initiated motion {at $t=$ a) 0.5$t_i^*$, b) 1.24$t_i^*$, and c) 1.55$t_i^*$.}}\label{fig11}
\end{figure}

{The drop fluid contacts the orifice edge at t=0.1$t_i^*$ in the sharp-edged case. Figure} \ref{fig10} shows that both drops first decelerate in similar ways until the leading interface penetrates approximately 0.2$D$ beyond the lower plate surface {(see $t=0.25t_i^*$ for the sharp-edged case).} Then, a clear difference arises when the no-slip initiated at the sharp edged orifice stagnates the drop above it.  Normalized local velocity and vorticity fields in this drop and surrounding fluid for the initial deceleration phase are shown at three specified times in figure \ref{fig11}. The stagnated drop with relatively weak internal motion is shown at $t = 0.5t_i^*$. Although the drop is stationary, the trailing fluid continues to impinge from above and diverge radially outward, shearing the drop interface. Consequently, regions of opposing vorticity with magnitude 0.5$U_t/D$ appear in the trailing fluid and the drop fluid. The drop remains stationary for a significant time while the wake weakens (see figure \ref{fig11b}  and \ref{fig11c}). While the macroscopic motion remains minimal, the leading contact line \emph{slips} {and begins to propagate slowly through the orifice at $t=0.75t_i^*$. This action initiates downward motion in the leading portion of the drop as captured at $t=1.24t_i^*$ in figure} \ref{fig11b}. After this downward motion is initiated, the drop begins to accelerate as shown in the frame at $t=1.55t_i^*$.  The trailing contact line, on the other hand, remains pinned near the orifice edge through the entire sequence.

\begin{figure}[h]
 \centering
        \begin{subfigure}[b]{0.35\textwidth}
                \centering
                \includegraphics[width=\textwidth]{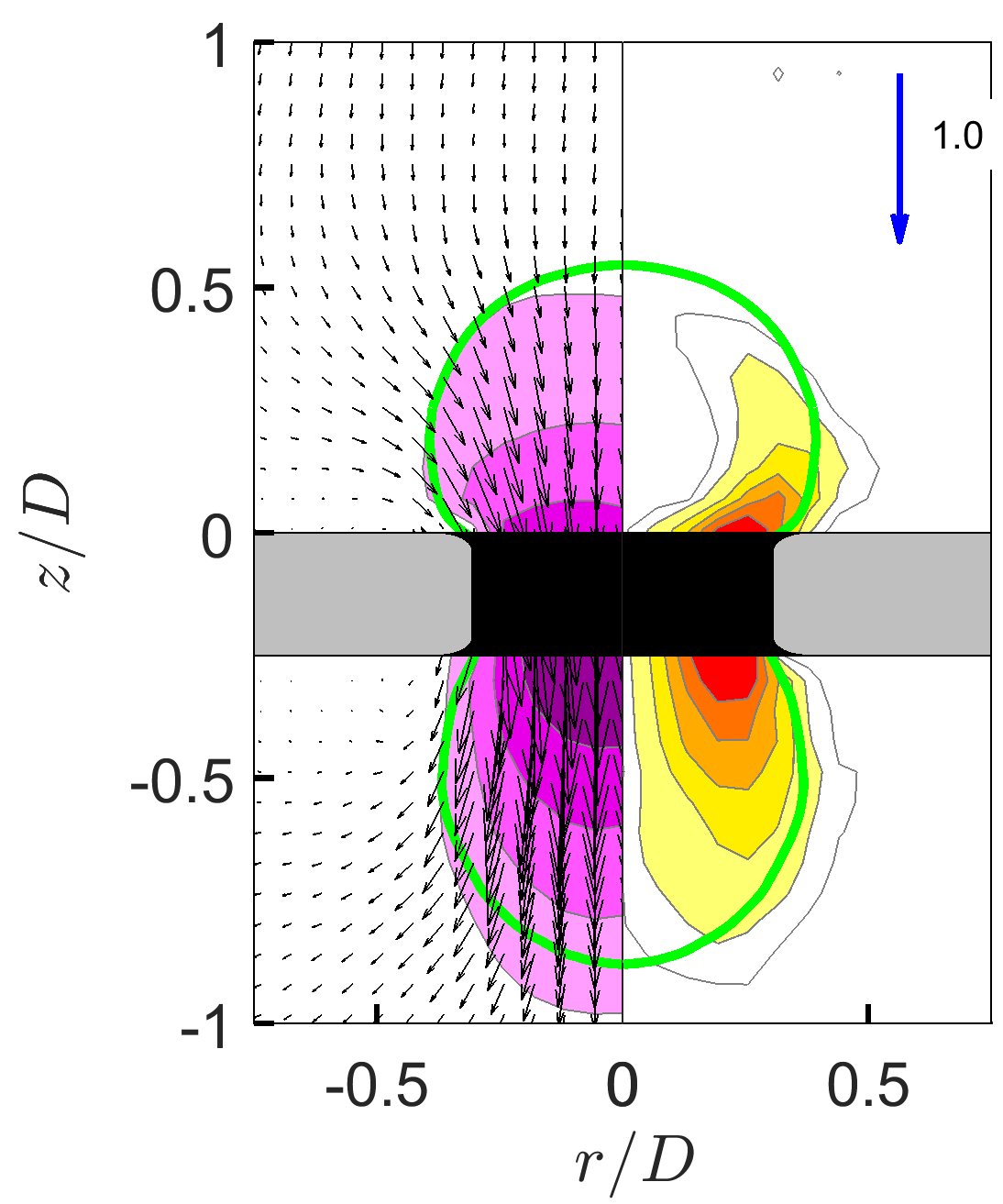}
                \caption{}
                \label{fig02a}
        \end{subfigure}
        \begin{subfigure}[b]{0.45\textwidth}
                \centering
                \includegraphics[width=\textwidth]{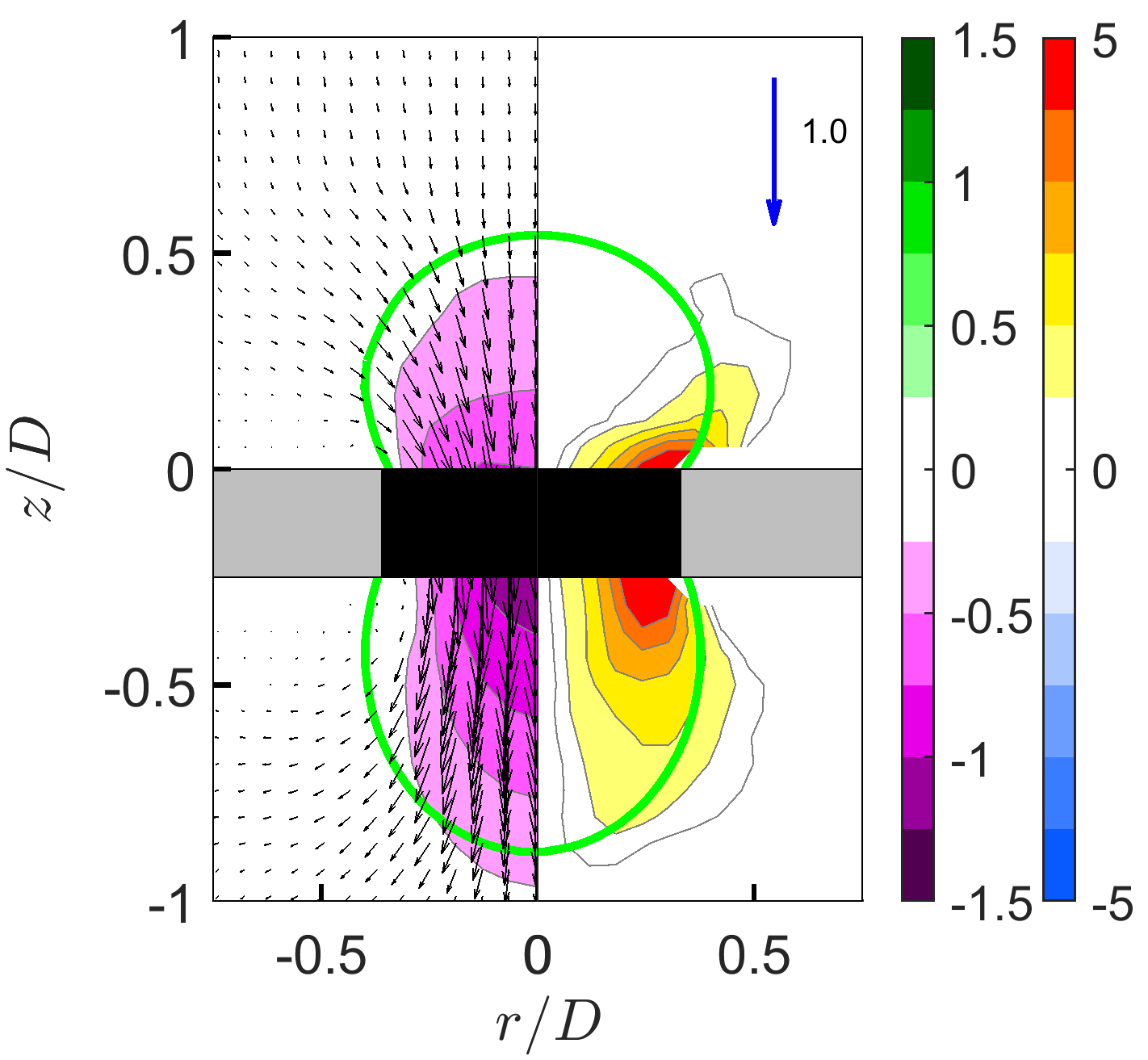}
                \caption{}
                \label{fig02b}
        \end{subfigure}

        \caption{(Color online) Contours of axial velocity ($u_z/U_t$: left panel) and out of plane vorticity ($\omega_{\theta}D/U_t$: right panel) at an equivalent  time ({$t = 1.75t_i^*$} in figure \ref{fig10}) for a drop with $Bo$ = 3.7 impacting on a) round-edged and b) sharp-edged orifice with $d/D$ = 0.62.}\label{fig_compare}
\end{figure}

{Based on visualization images, the initial velocity of the leading contact line was estimated as 0.01$U_t$. At this speed, the contact line would traverse only about one third of the orifice thickness during the rest of the sequence. This scaling, together with observation of the shape of the leading drop interface exiting the orifice, suggests that the leading contact line does not advance very far and that the surrounding fluid continues to coat most of the interior orifice surface. Thus, the viscous resistance within the orifice is similar to that in the round-edged case where a thin film of surrounding fluid separates the drop from the interior orifice surface.  Thus, although the contact and no-slip near the upper orifice edge delay the onset of drop penetration in the sharp-edged case, they do not alter the propagation of the drop fluid through the orifice significantly during the re-acceleration phase. In figure} \ref{fig10}, {the centroidal speed of the drop matches closely for both cases. Also, the local velocity ($u_z/U_t$) and vorticity ($\omega_{\theta}D/U_t$) fields at comparable stages demonstrate remarkable similarity (see figure} \ref{fig_compare} for example).

\begin{figure}[h]
 \centering
        \begin{subfigure}[b]{0.22\textwidth}
                \centering
                \includegraphics[width=\textwidth]{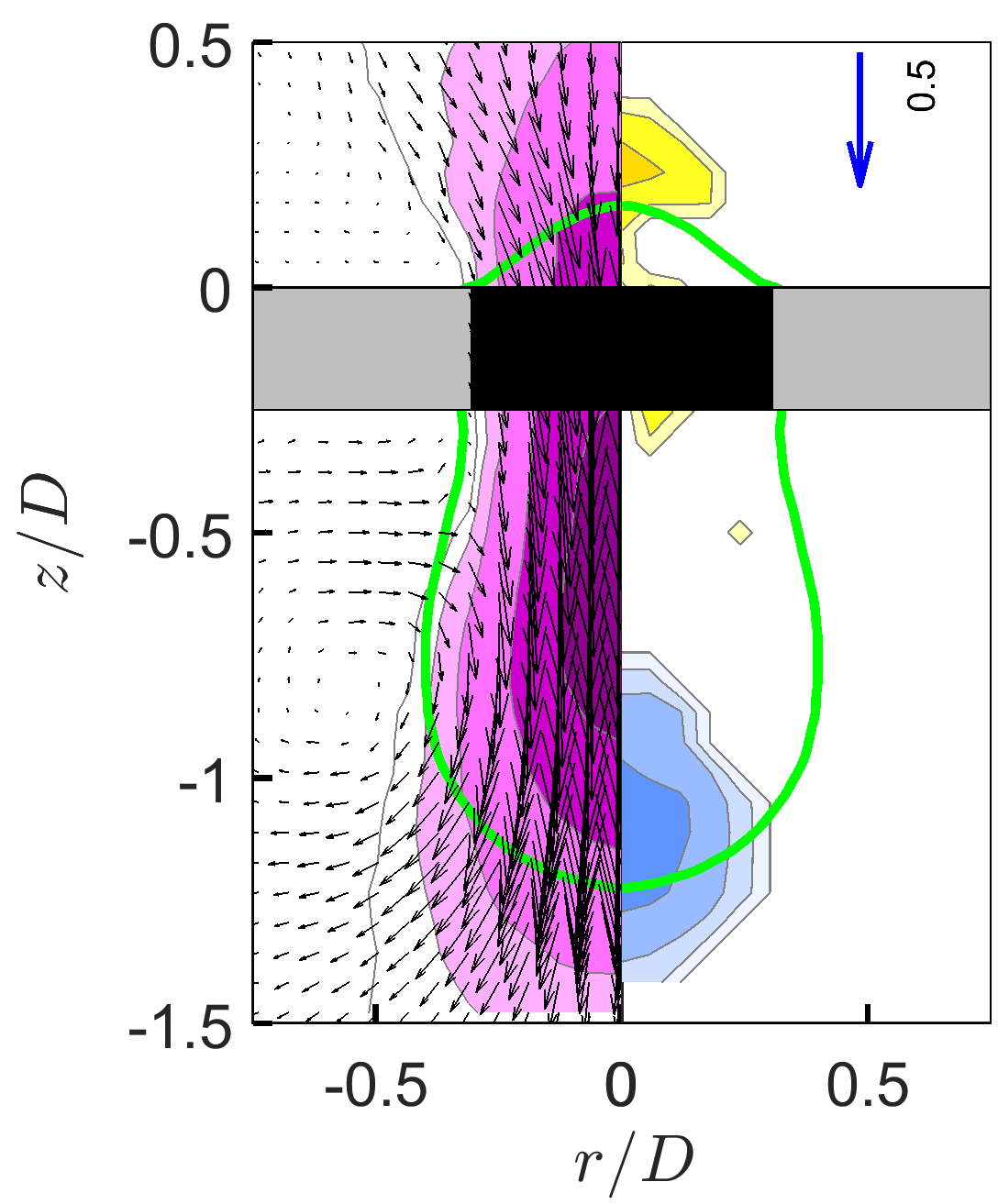}
                \caption{}
                \label{fig12a}
        \end{subfigure}
        \begin{subfigure}[b]{0.22\textwidth}
                \centering
                \includegraphics[width=\textwidth]{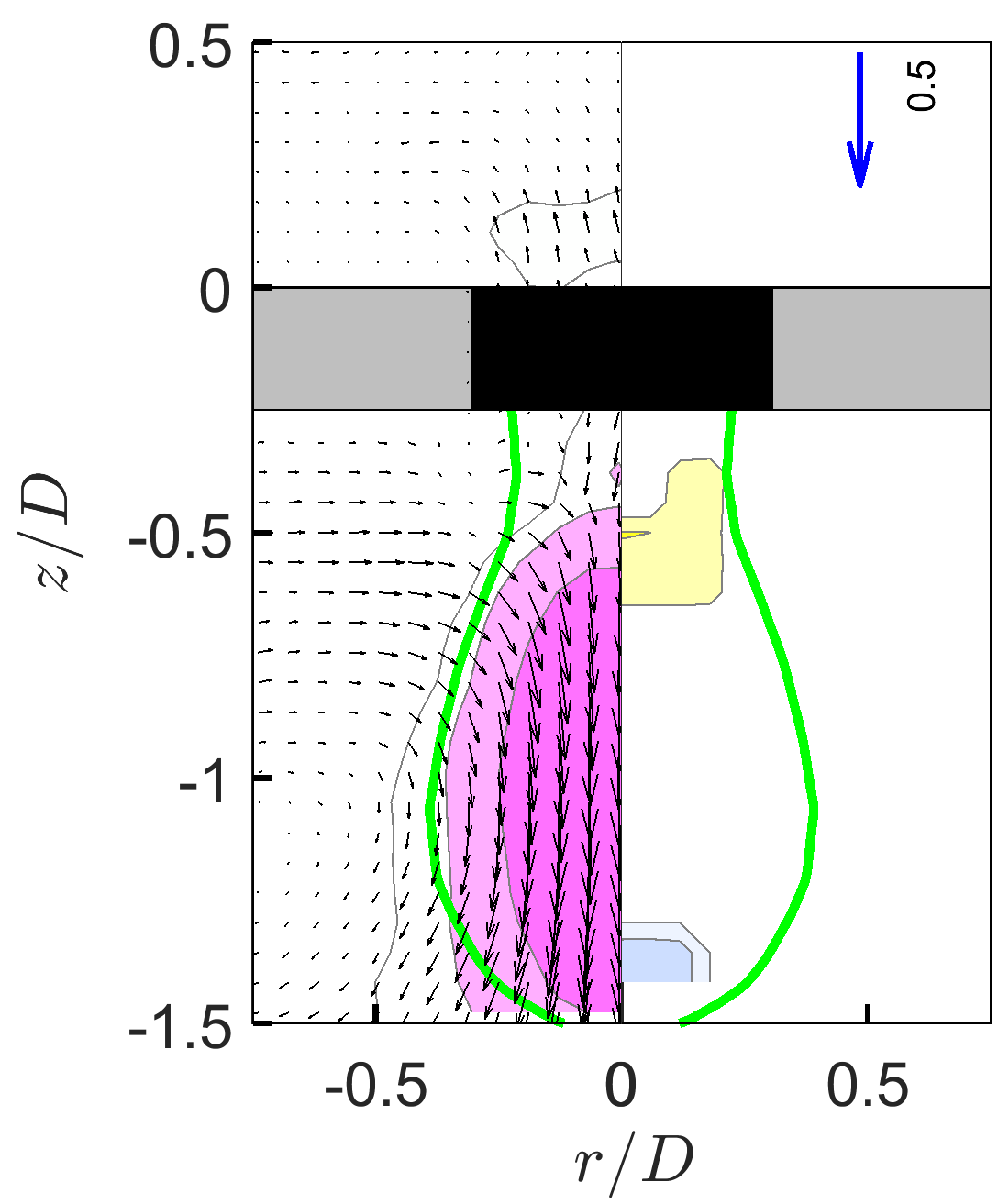}
                \caption{}
                \label{fig12b}
        \end{subfigure}
        \begin{subfigure}[b]{0.22\textwidth}
                \centering
                \includegraphics[width=\textwidth]{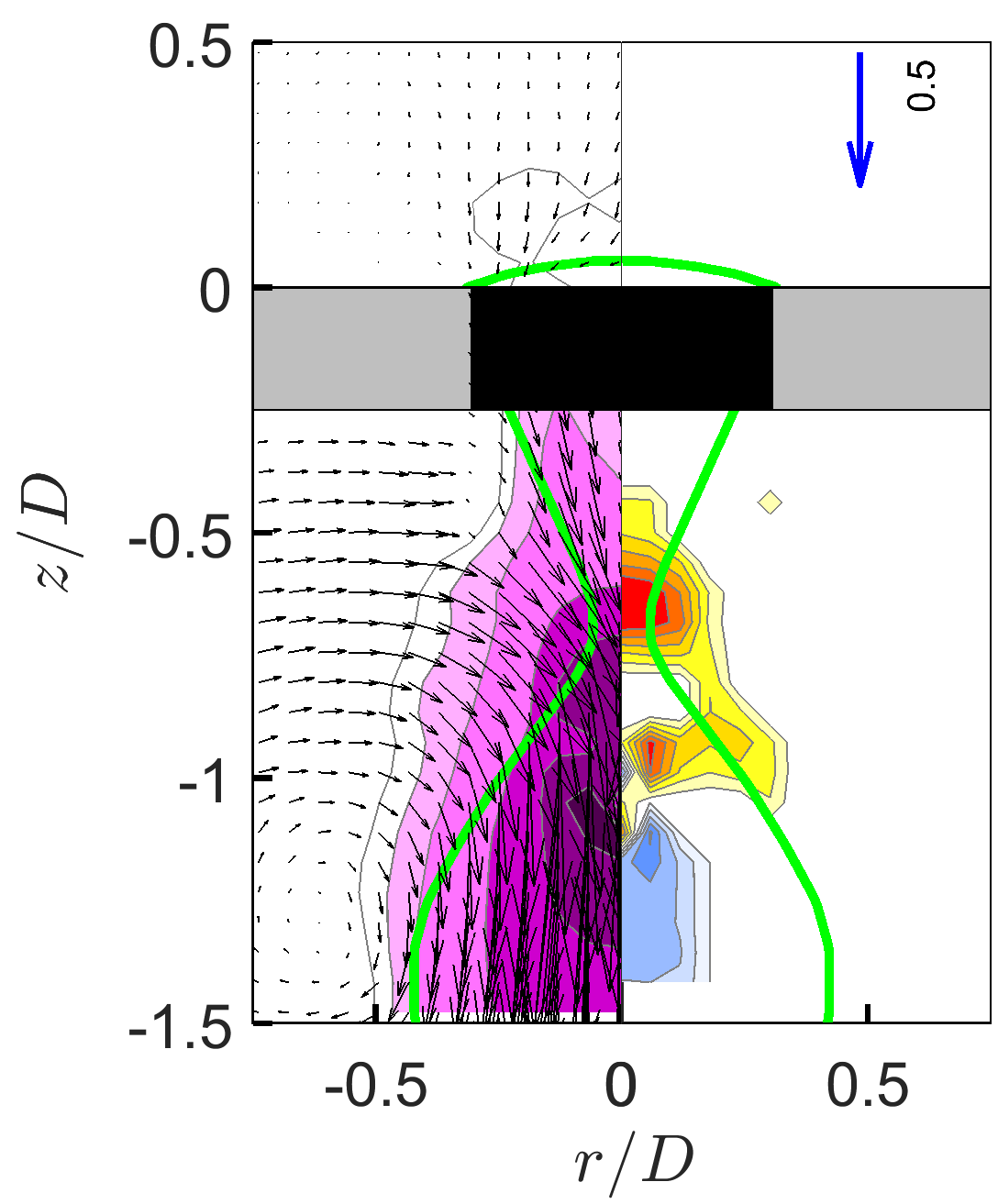}
                \caption{}
                \label{fig12c}
        \end{subfigure}
        \begin{subfigure}[b]{0.29\textwidth}
                \centering
                \includegraphics[width=\textwidth]{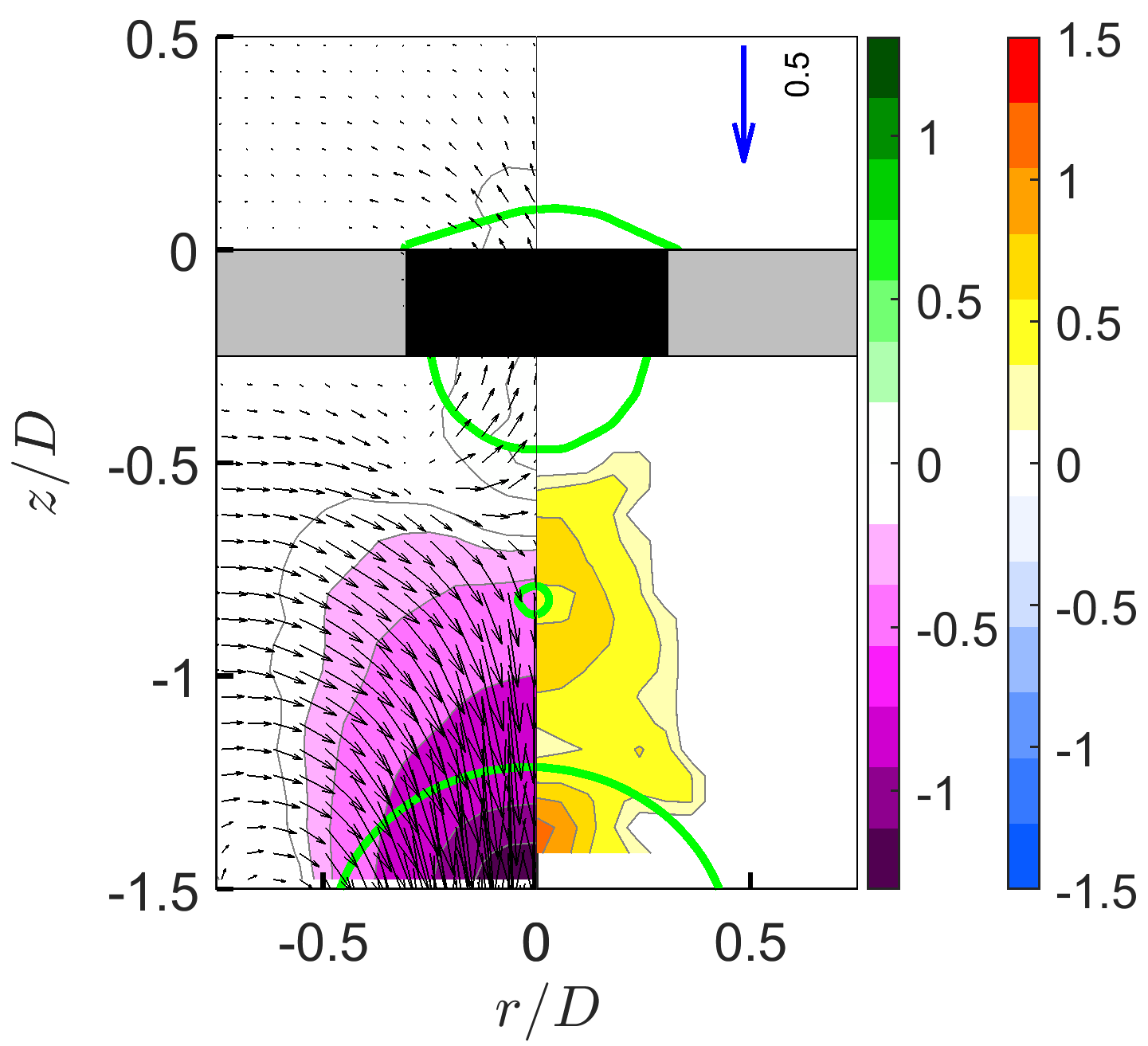}
                \caption{}
                \label{fig12d}
        \end{subfigure}

        \caption{(Color online) Contours of axial velocity ($u_z/U_t$: left panel) and normalized axial strain ($\alpha D/U_t$: right panel) for a drop with $Bo$ = 3.7 impacting and partially releasing through a sharp-edged hydrophobic orifice with $d/D$ = 0.62 {at times $t=$ a) 2.27$t_i^*$, b) 2.57$t_i^*$, c) 3.02$t_i^*$, and d) 3.16$t_i^*$.}}\label{fig12}
\end{figure}

{The pinning of the trailing contact line resists the complete release of the drop from the orifice} \citep{Bordoloi14}. In the example shown here, {the pinning results in strong axial deformation, leading to a breakup and a partial capture of the drop fluid} at the orifice. Based on the local velocity fields (shown in the left hand panels of figure \ref{fig12}), we examine the local axial strain $\alpha = \partial u_z/\partial z$  in the corresponding  right hand panels after the trailing interface enters the orifice. With most of the drop fluid underneath the plate, the gravitational force accelerates the leading portion of the drop such that the maximum drop interior velocity exceeds  1.0$U_t$ (see figure \ref{fig12a}). The trailing drop fluid above the orifice with increased curvature ($\ge$ $1/d$) on the other hand, gains inertia due to the combination of surface tension and continuity, and accelerates to a similar axial velocity.  The gradients in velocity between the drop fluid and the surrounding oil ahead of and behind the drop produce two regions of strong opposing axial strains as shown in figure \ref{fig12a}. 

{At $t=2.57t_i^*$, the pinned interface inside the orifice, which is rebounding upward, causes the drop fluid beneath it to decelerate significantly. The maximum velocity in the drop fluid decreases to 0.4$U_t$. With the drop front moving downward under gravity and the trailing fluid moving upward due to pinning, a neck develops beneath the orifice. The fast-moving fluid below and slow-moving fluid above the neck and related capillary instability cause its diameter to shrink at subsequent times, and surrounding fluid is induced radially inward (see $t=3.02t_i^*$). After the axial strain in the neck reaches a maximum of 1.5$U_t/D$, the leading drop fluid breaks away approximately 0.7D below the lower place surface. A small satellite droplet subsequently breaks off of the leading drop fluid. The fraction of drop fluid captured at the orifice through pinning is approximately 15\% of the impacting drop volume.  }

\begin{figure}[h]
 \centering
                \includegraphics[width=0.5\textwidth]{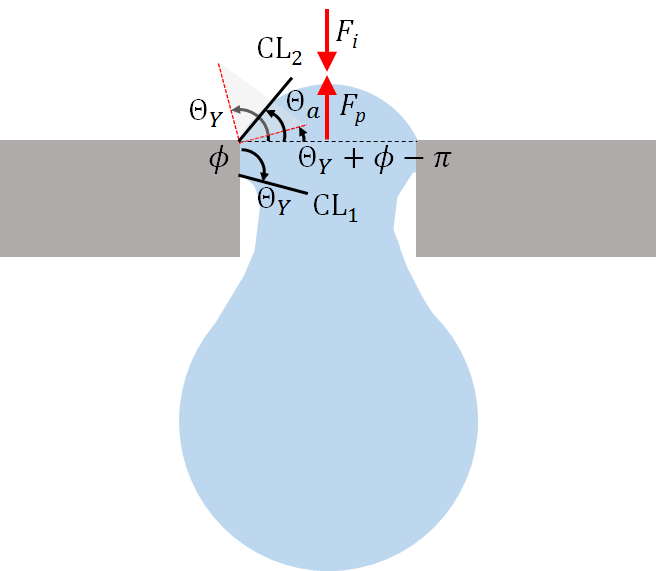}
        \caption{Schematic of drop penetrating a sharp-edged hydrophobic orifice demonstrating pinning force $F_p$ and inertial force $F_i$ acting on the trailing drop fluid, and angles of contact: Young\rq{}s angle of the leading contact line ($CL_1$), the experimentally observed apparent contact angle ($\Theta_a$) of the trailing contact line ($CL_2$), the wedge angle ($\phi$), and the \emph{canthotaxis regime} [$\Theta_Y+\phi-\pi,\Theta_Y$] marked by dashed lines. }\label{fig_pinHPB}
\end{figure}

Based on the \emph{canthotaxis} principle, the apparent contact angle ($\Theta_a$) of  a receding pinned interface must lie within the regime, [$\Theta_Y+\phi-\pi, \Theta_Y$], where $\Theta_Y = 105^\circ$ and $\phi= 90^\circ$ are the equilibrium contact angle and the wedge angle, respectively \citep{PiedNoir2005, Berthier2009, Hensel2014}. {The value of $\Theta_a$ for the captured drop fluid measured on the left-hand corner anticlockwise from the upper orifice plane (see figure} \ref{fig_pinHPB}) is approximately $20^\circ-30^\circ$, which is within this regime [$15^\circ\leq\Theta_a\leq105^\circ$].

{The case presented and the related discussion raise the question: under what condition will the trailing contact line depin from the orifice edge? To address this question, we estimated the pinning force ($F_p)$ of the trailing contact line based on the expression in} \citet{PiedNoir2005} and  \citet{Bodiguel2009}: 
\begin{equation}
F_{p} \approx \sigma d(cos\Theta_Y-cos(\Theta_Y+\phi-\pi)).
\end{equation}
\noindent
The force $F_p$, acting upward around the periphery of the orifice, opposes the force due to inertia ($F_i\approx \rho_d U_t^2d^2$) in the downward-moving drop fluid (see figures \ref{fig12} and \ref{fig_pinHPB}).  For the example discussed above, the ratio $F_p/F_i \approx 1$, so that the two forces are approximately balanced. If the inertia is increased sufficiently, as in the case of SHPB2 in table \ref{tab02} and discussed below in Section \ref{sec3.3}, the inertia overcomes the pinning force ($F_p/F_i=0.6$). The contact line depins, and all of the drop fluid moves through the orifice.

\subsection{Effect of orifice wettability}
\label{sec3.3}
\begin{figure}[h]
 \centering
        \begin{subfigure}[b]{0.27\textwidth}
                \centering
                \includegraphics[width=\textwidth]{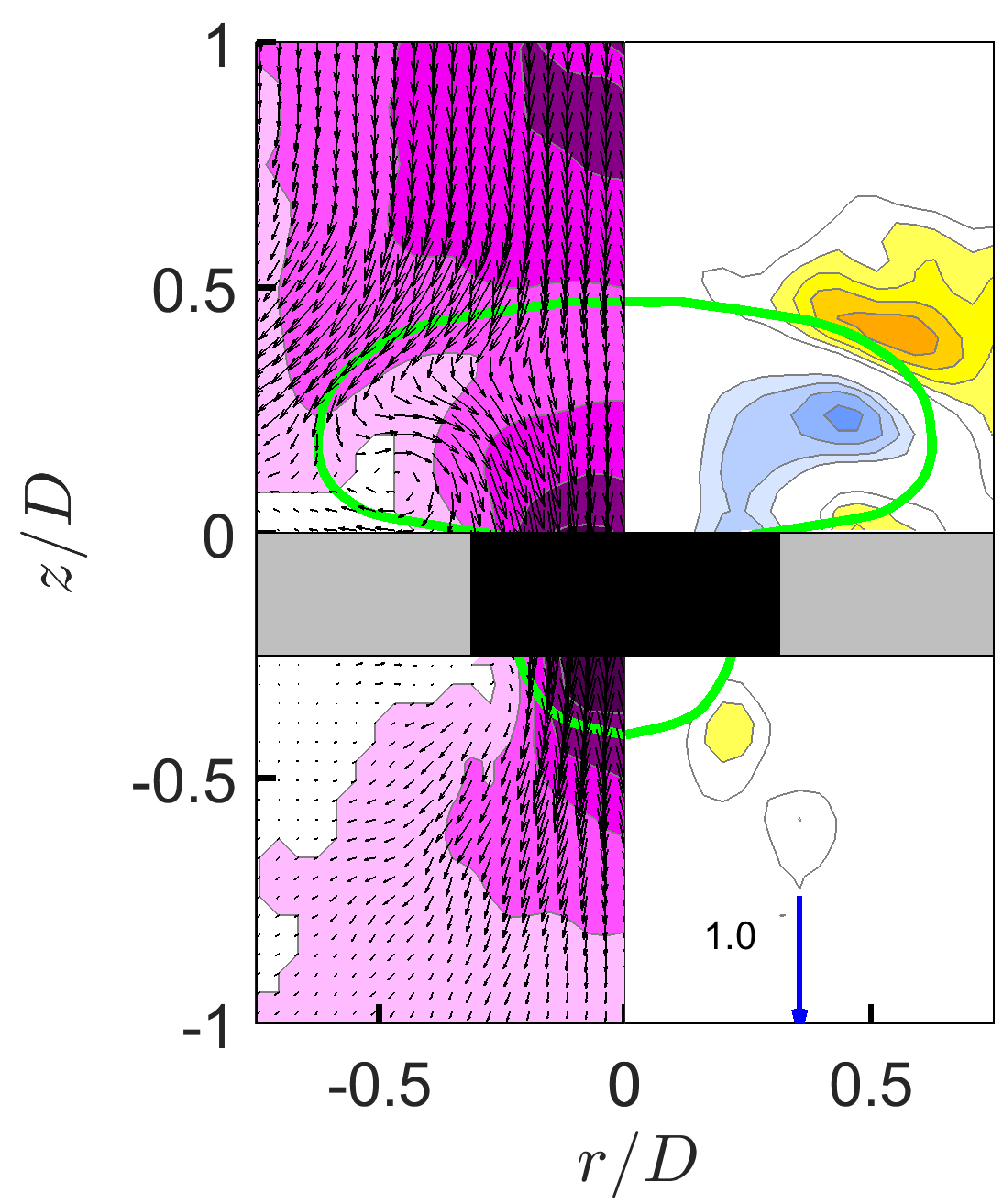}
                \caption{}
                \label{fig13a}
        \end{subfigure}
        \begin{subfigure}[b]{0.27\textwidth}
                \centering
                \includegraphics[width=\textwidth]{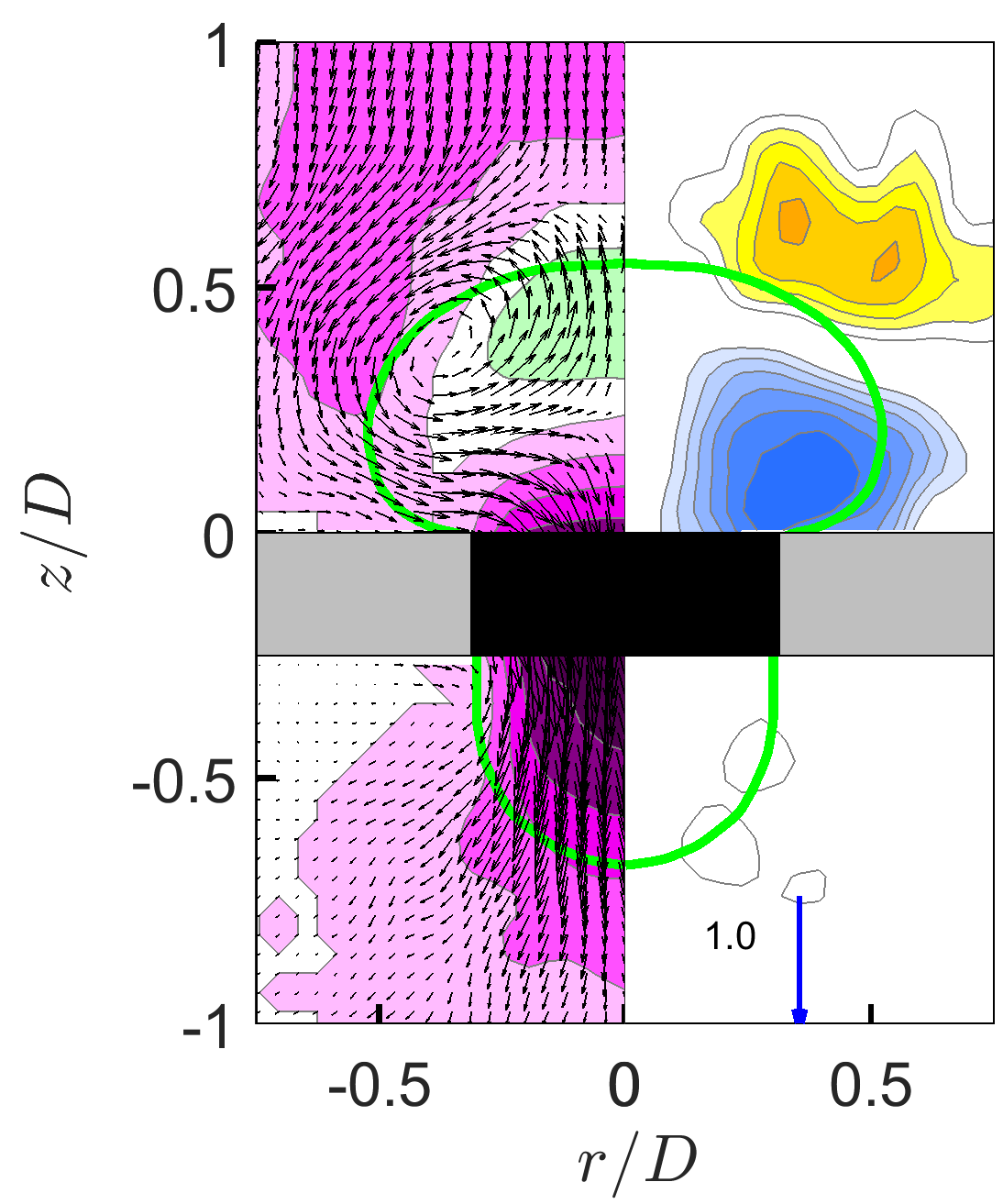}
                \caption{}
                \label{fig13b}
        \end{subfigure}
                \begin{subfigure}[b]{0.35\textwidth}
                \centering
                \includegraphics[width=\textwidth]{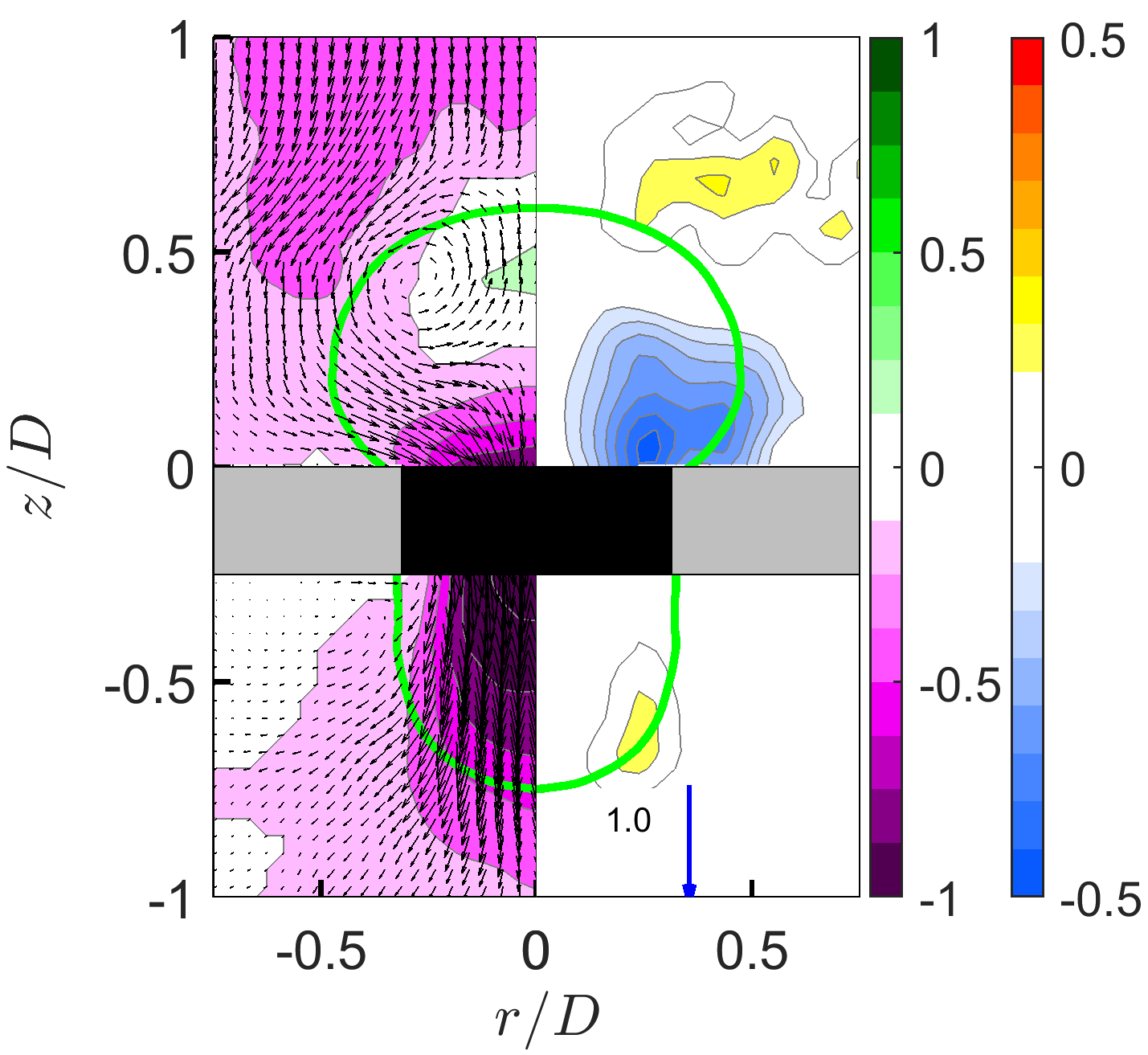}
                \caption{}
                \label{fig13c}
        \end{subfigure}\\
        \begin{subfigure}[b]{0.27\textwidth}
                \centering
                \includegraphics[width=\textwidth]{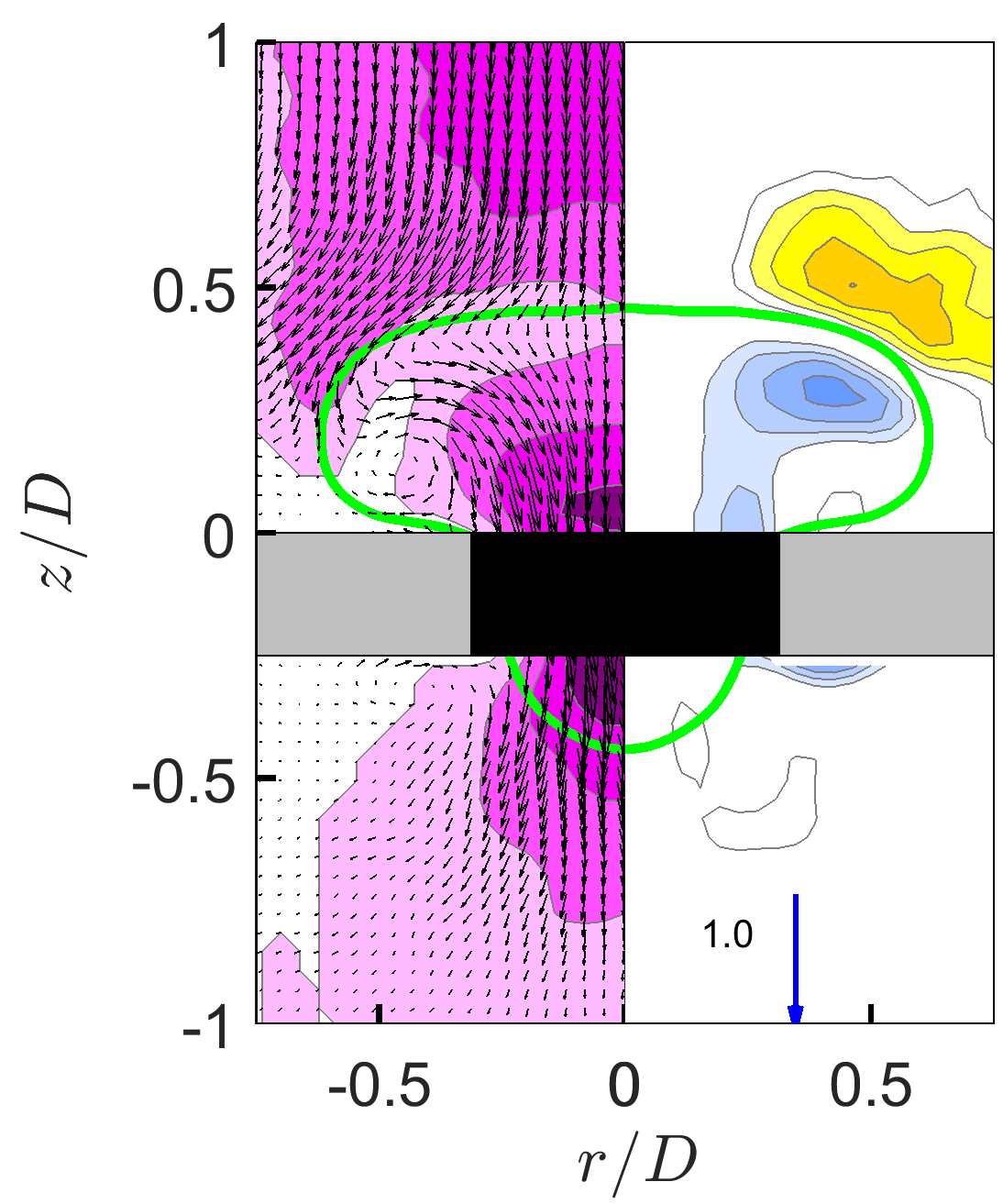}
                \caption{}
                \label{fig13d}
        \end{subfigure}
        \begin{subfigure}[b]{0.27\textwidth}
                \centering
                \includegraphics[width=\textwidth]{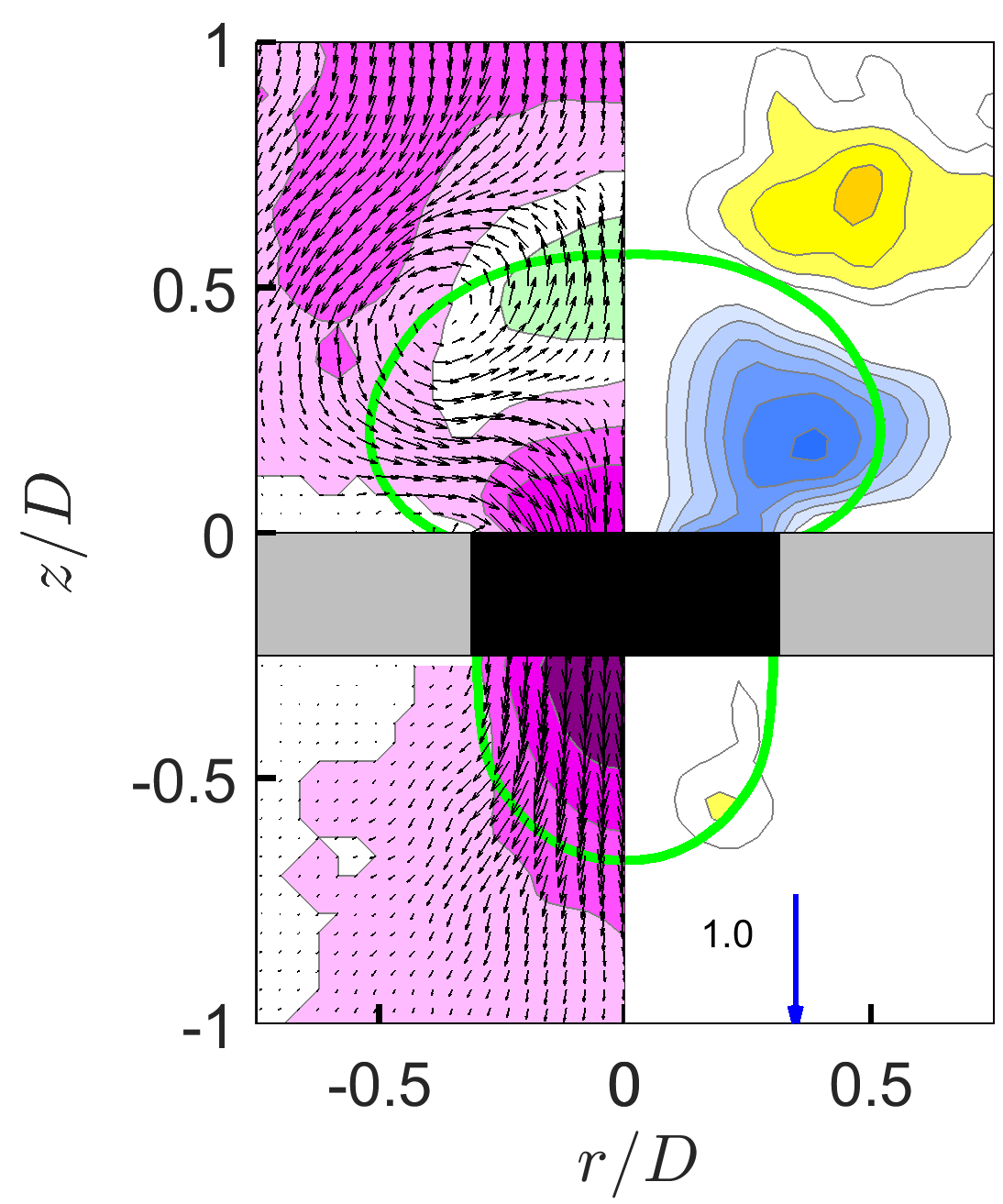}
                \caption{}
                \label{fig13e}
        \end{subfigure}    
        \begin{subfigure}[b]{0.35\textwidth}
                \centering
                \includegraphics[width=\textwidth]{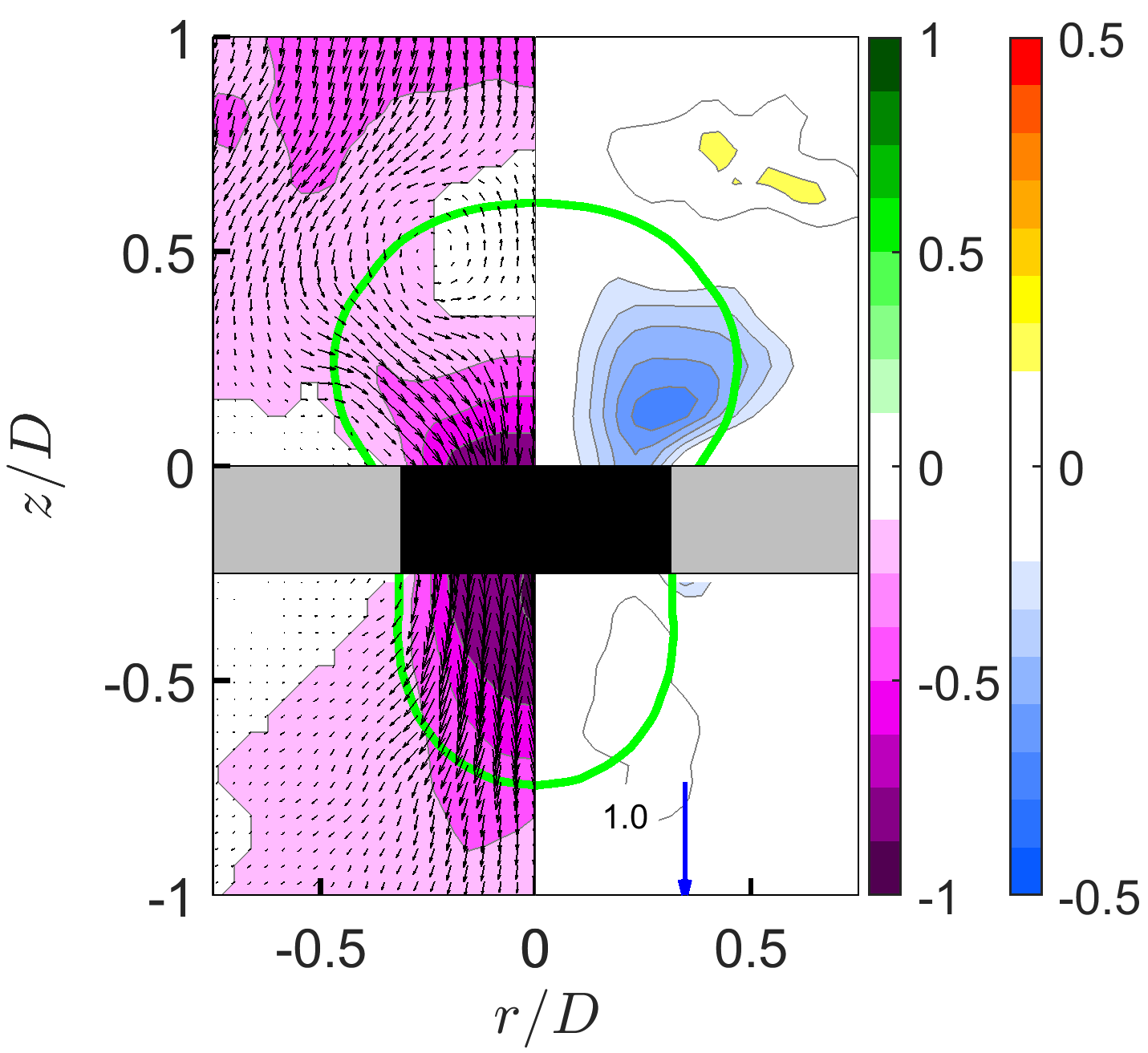}
                \caption{}
                \label{fig13f}
        \end{subfigure}         
        \caption{(Color online) {Contours of axial velocity ($u_z/U_t$: left panel) and radial velocity ($u_r/U_t$: right panel) for drops with $Bo$ = 5.3 and $d/D$ = 0.62 impacting sharp-edged hydrophobic (a, b, c) and sharp-edged hydrophilic (d, e, f) orifices at times $t = 0.17t_i^*$ (a, d), $t = 0.35t_i^*$ (b, e) and $t = 0.42t_i^*$ (c, f).
}}\label{fig13}
\end{figure}

\begin{figure}[h]
 \centering
                \includegraphics[width=0.5\textwidth]{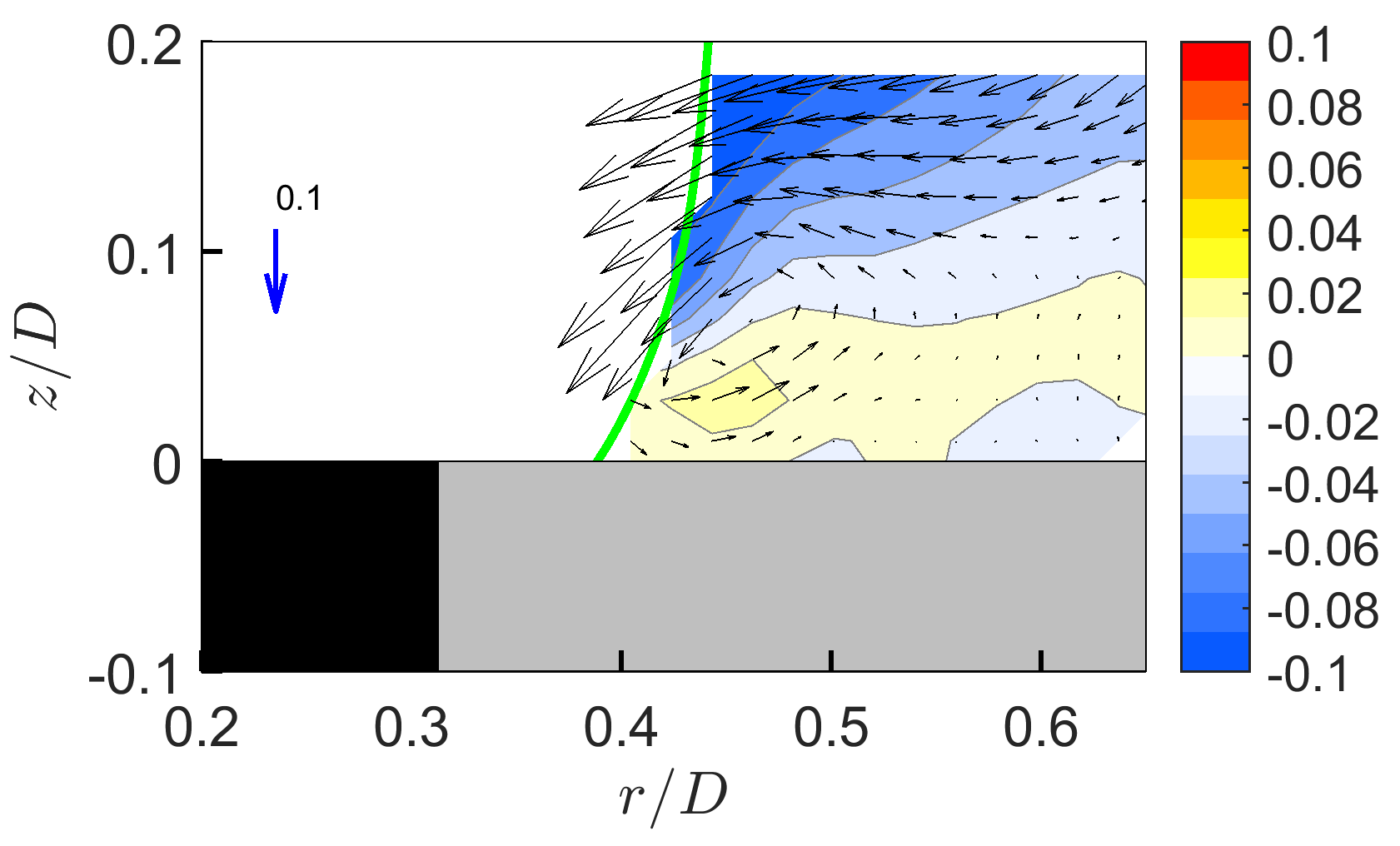}
\caption{(Color online) {High resolution contours of radial velocity ($u_r/U_t$)  superposed with velocity vectors in the surrounding fluid near the contact line for the same velocity field shown in figure} \ref{fig13f} {for the sharp-edged hydrophilic orifice case.} 
} \label{fig_inset}
\end{figure}

\begin{figure}[h]
 \centering
        \begin{subfigure}[b]{0.27\textwidth}
                \centering
                \includegraphics[width=\textwidth]{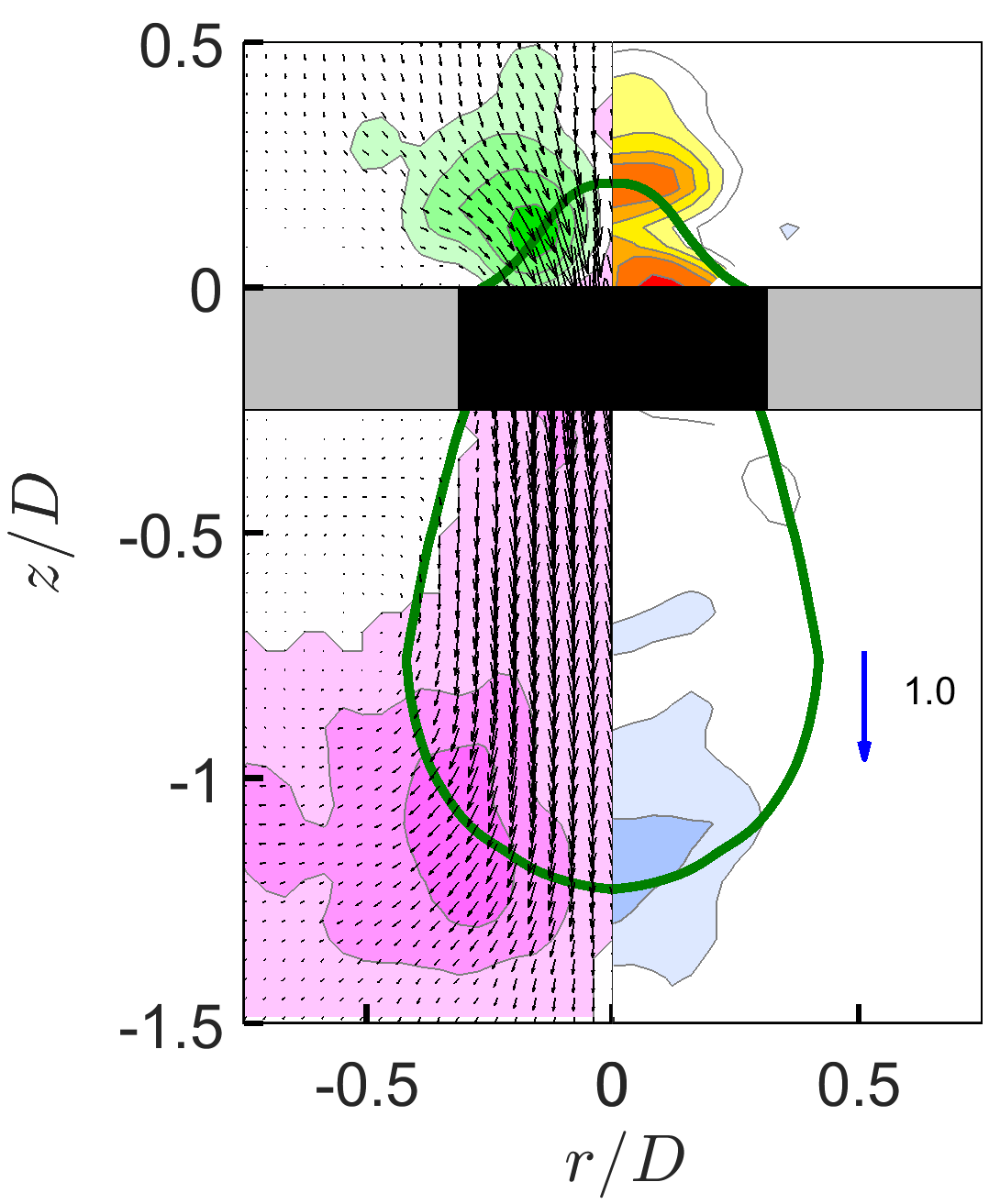}
                \caption{}
                \label{fig14a}
        \end{subfigure}
        \begin{subfigure}[b]{0.35\textwidth}
                \centering
                \includegraphics[width=\textwidth]{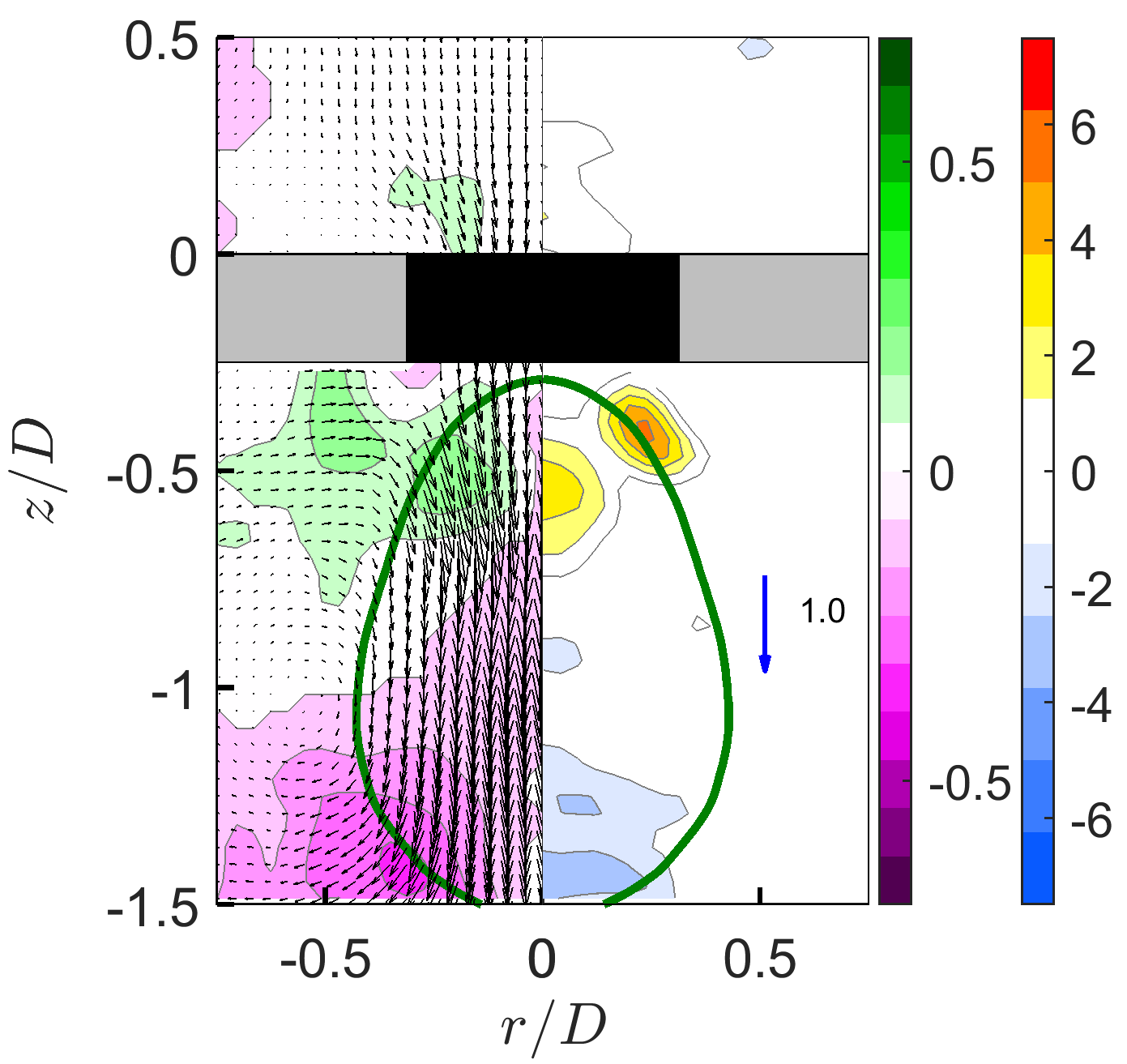}
                \caption{}
                \label{fig14b}
        \end{subfigure}\\
        \begin{subfigure}[b]{0.27\textwidth}
                \centering
                \includegraphics[width=\textwidth]{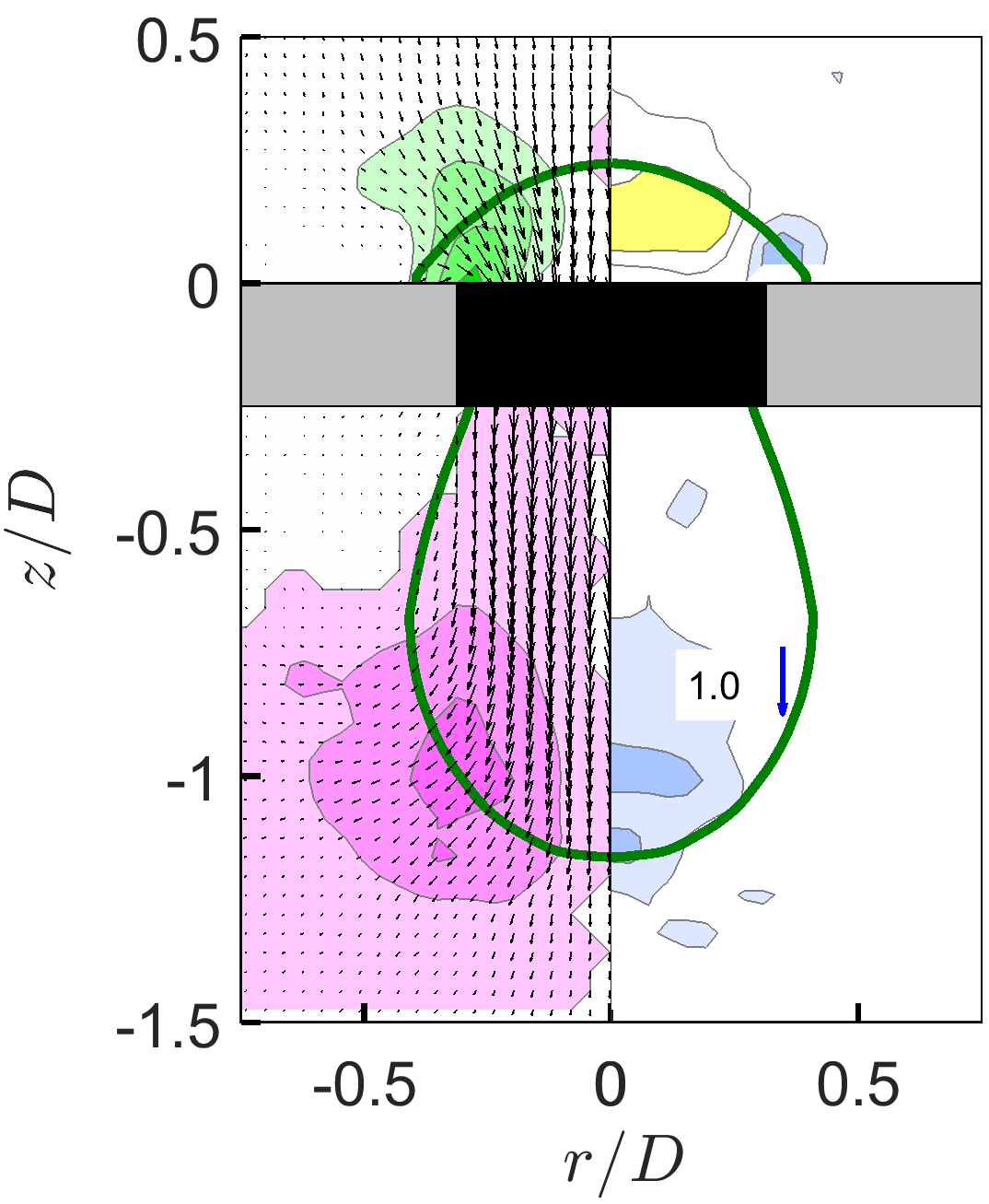}
                \caption{}
                \label{fig14c}
        \end{subfigure}
        \begin{subfigure}[b]{0.35\textwidth}
                \centering
                \includegraphics[width=\textwidth]{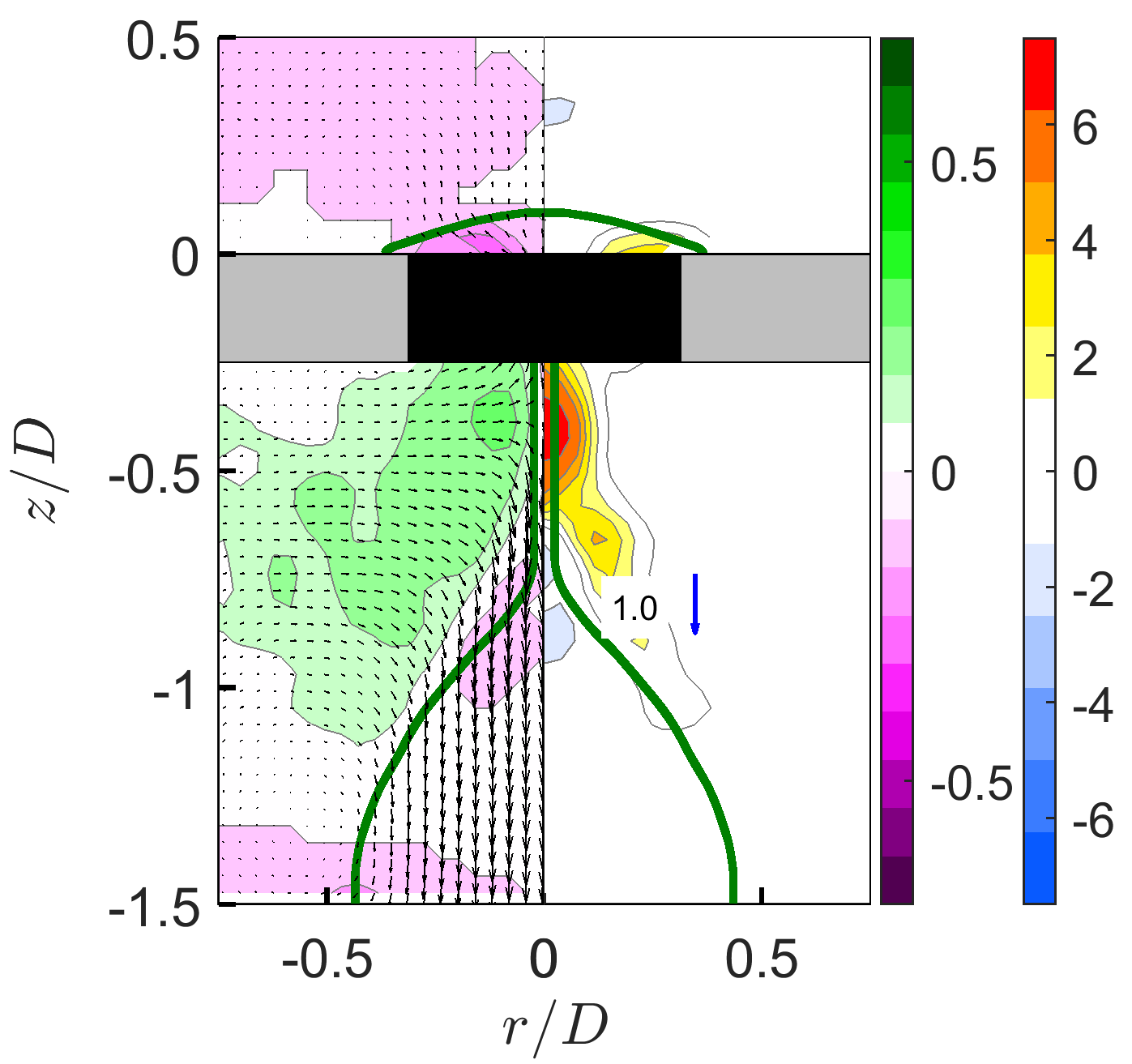}
                \caption{}
                \label{fig14d}
        \end{subfigure}          
        \caption{(Color online) {Contours of axial velocity ($u_z/U_t$: left panel) and the axial strain ($\alpha D/U_t$: right panel) for a drop with $Bo$ = 5.3 and $d/D$ = 0.62 impacting and completely releasing through a sharp-edged hydrophobic orifice at time a) $t=0.75t_i^*$, b) $t=0.92t_i^*$,  and partially releasing through a hydrophilic orifice at time c) $t=0.75t_i^*$, d) $t=1.17t_i^*$}}\label{fig14}
\end{figure}

To examine wettability effects, we first compare velocity fields for somewhat larger drops ($Bo$ = 5.3) impacting on hydrophobic and hydrophilic orifices with $d/D$ = 0.62 (SHPB2 and SHPL1 in Table \ref{tab02}).  Three early post-impact times are shown in figure \ref{fig13}, and two later times are shown in figure \ref{fig14}. Given the larger $Bo$, both drops carry larger inertia ($We$ = 3.4), which is sufficient to overcome the initial surface tension barrier and to penetrate the orifice with significant axial velocity $\approx U_t$. Thus, differences in wettability have little effect on the leading part of the drop. The wettability plays a strong role, however, in the motion of the trailing drop fluid above the plate. In the hydrophobic case with weaker wettability, the trailing contact line remains restricted near the orifice edge shortly after the initial contact  (see figure \ref{fig13a} at t=0.17$t_i^*$). {During the subsequent rebound, drop fluid continues to penetrate through the orifice, causing  radially inward motions in both the drop and surrounding fluids near the orifice edge} (see figures \ref{fig13b} and \ref{fig13c} ). In the hydrophilic case, on the other hand, the stronger surface wettability causes the contact line to propagate radially outward across the upper plate surface at a speed of $\approx 0.1U_t$. This speed is an order of magnitude larger than in the hydrophobic case (SHPB1) discussed in Section \ref{orifice_edge}. The outward motion of the contact line opposes the inward motion of the contracting drop and the  {penetrating} drop fluid. In figure  \ref{fig13f}, the radial velocity contours reveal a different pattern than that in figure \ref{fig13c}:  the inward velocity is confined to a smaller region focused further above the plate surface. To improve resolution near the contact line, the velocity field in this region was processed separately for the hydrophilic  case with a larger interframe time ($\Delta t$ = 15 ms). The results, plotted in figure \ref{fig_inset}, show that the opposing motions near the contact line result in local circulation in the surrounding fluid outside of it. 

\begin{figure}[h]
 \centering
                \includegraphics[width=0.45\textwidth]{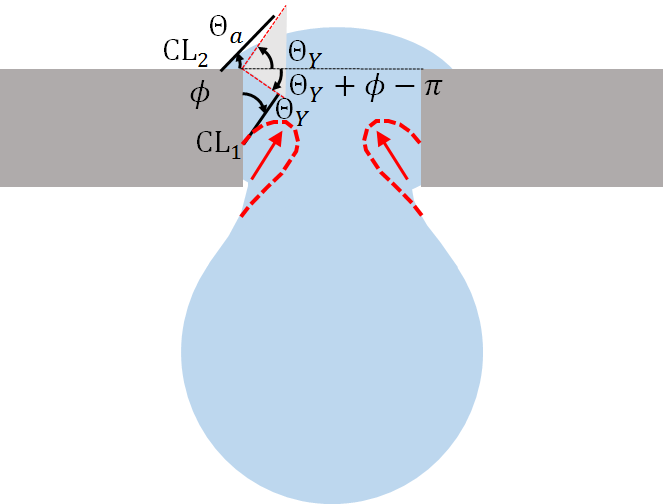}

        \caption{Schematic of drop penetrating a sharp-edged hydrophilic orifice showing various angles of contact: Young\rq{}s angle ($\Theta_Y$) of the leading contact line ($CL_1$), the experimentally observed apparent contact angle ($\Theta_a$) of the trailing contact line ($CL_2$), the wedge angle ($\phi$), and the \emph{canthotaxis regime} [$\Theta_Y+\phi-\pi,\Theta_Y$] marked by dashed lines. {Thick dashed lines with arrows indicate concept of interface deformation before breakup.} }\label{pinning_HPL}
\end{figure}

After most of the drop fluid has penetrated into the orifice, the trailing interface in the hydrophobic case experiences strong curvature near the central axis giving rise to a downward surface tension force on the drop fluid (see figure \ref{fig14a}). This force causes a significant gradient in the axial velocity ($\partial u_z/\partial z$) across the interface as well as strong interface rotation near the orifice edge.  The inertia of the drop fluid above the orifice is strong enough to overcome the pinning force as discussed earlier. {The entire drop has exited the orifice by 0.9$t_i^*$.} In the hydrophilic case, by contrast, the trailing interface maintains a lower curvature, and since the contact line extends approximately 0.1$D$ beyond the orifice edge, part of the downward surface tension force is directed at the plate itself. In this case, the axial velocity gradient across the trailing interface is significantly weaker (see figure \ref{fig14c}). At later times, the trailing interface oscillates vertically before eventually stabilizing to an equilibrium shape. {The apparent contact angle of the trailing contact line is then approximately $10^{\circ}$.  This angle is much less than the Young\rq{}s angle of $55^\circ$, yet within the canthotaxis limit ($-35^\circ \leq \Theta_a \leq 55^\circ$; see figure} \ref{pinning_HPL}).

{The trapped drop fluid above the plate combined with the downward-moving front below it causes strong axial strain in the drop such that a long thin filament develops. The filament breaks in two places, the first being inside the orifice. The second break occurs 0.7D downstream of the lower plate surface producing a satellite that follows the leading drop. The maximum axial strain before the breakup is estimated as 6.5$D/U_t$ which is much larger than the value measured for the smaller drop and shorter filament pinned to the hydrophobic orifice in Section \ref{orifice_edge}. The initial breakup location is also well upstream of that for the drop exiting the hydrophobic orifice.  The earlier breakup in the hydrophilic case may be encouraged by the leading contact line dynamics} (see $CL_1$ in figure \ref{pinning_HPL}). {The Young\rq{}s contact angle ($\Theta_Y = 55^{\circ}$) encourages an inflection in the shape of the leading interface as emphasized by the dashed lines in figure} \ref{pinning_HPL}). {This inflection gives rise to a sharper neck angle as observed in figure} \ref{fig14d} {potentially contributing to the earlier breakup location. The fluid trapped by the hydrophilic orifice is approximately 17\% of the impacting drop volume.}

\subsection{Contact line motion}
\label{dcontact}
\begin{figure}[h]
 \centering
        \begin{subfigure}[b]{0.25\textwidth}
                \centering
                \includegraphics[width=\textwidth]{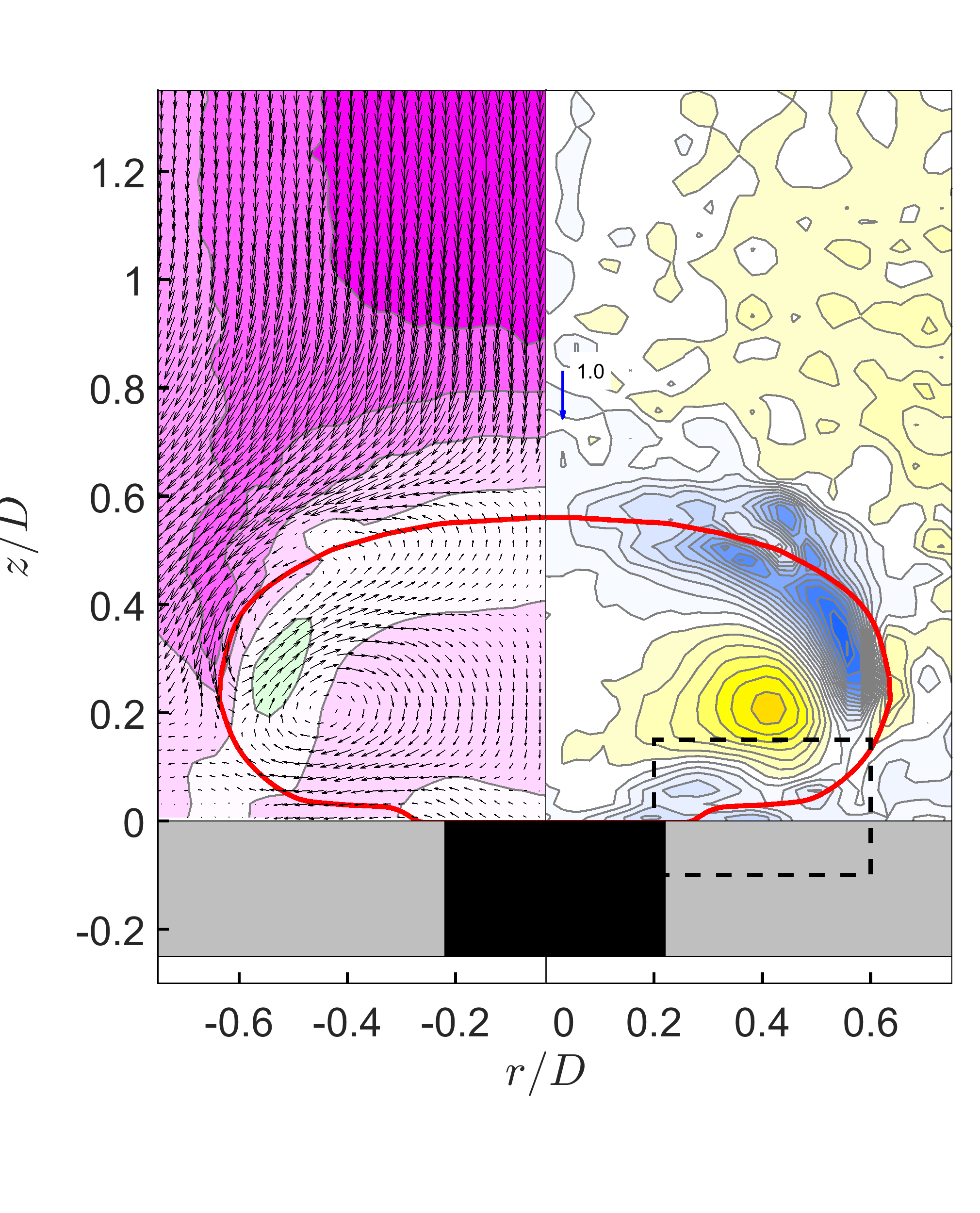}
                \caption{}
                \label{fig15a}
        \end{subfigure}
        \begin{subfigure}[b]{0.25\textwidth}
                \centering
                \includegraphics[width=\textwidth]{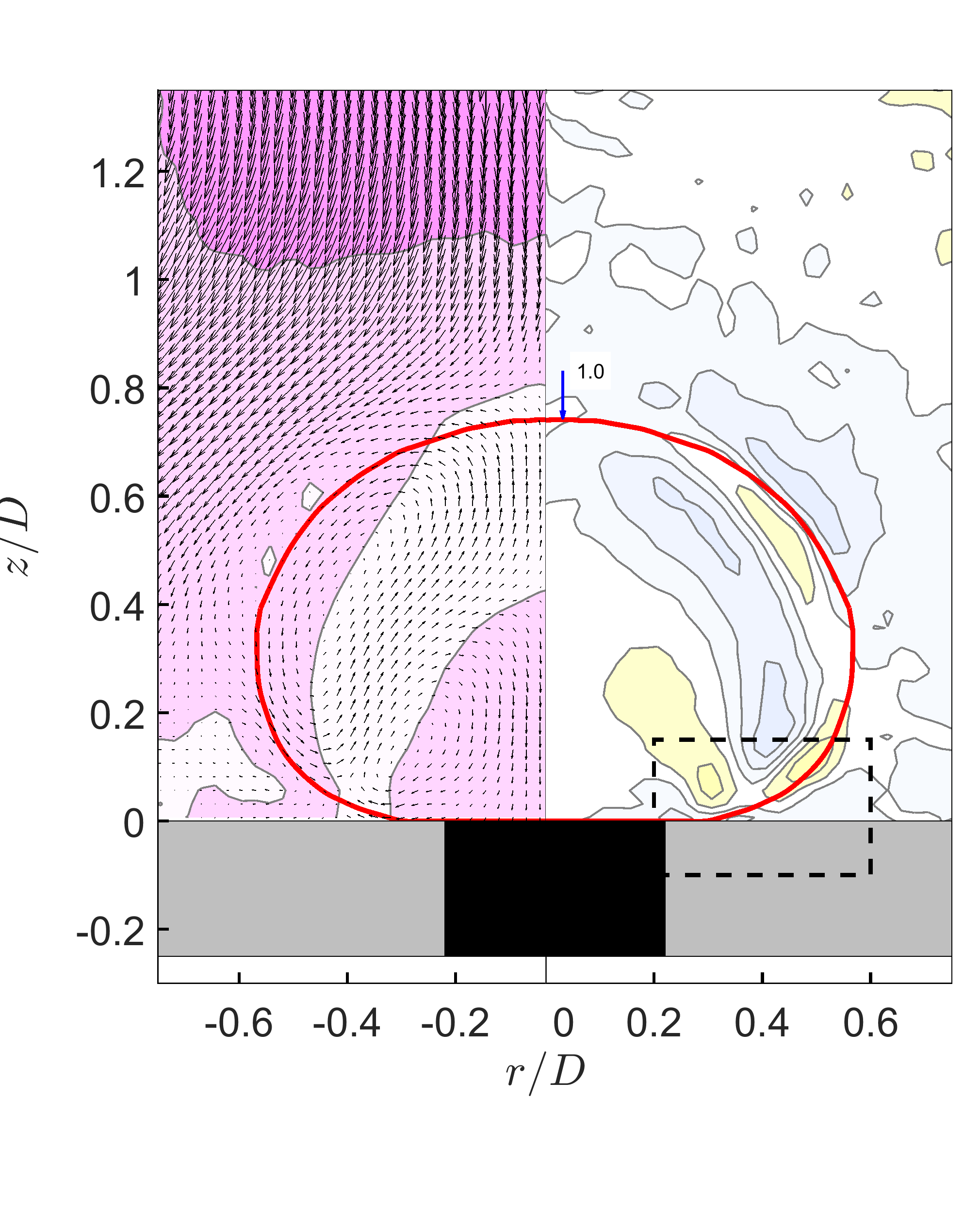}
                \caption{}
                \label{fig15b}
        \end{subfigure}
        \begin{subfigure}[b]{0.31\textwidth}
                \centering
                \includegraphics[width=\textwidth]{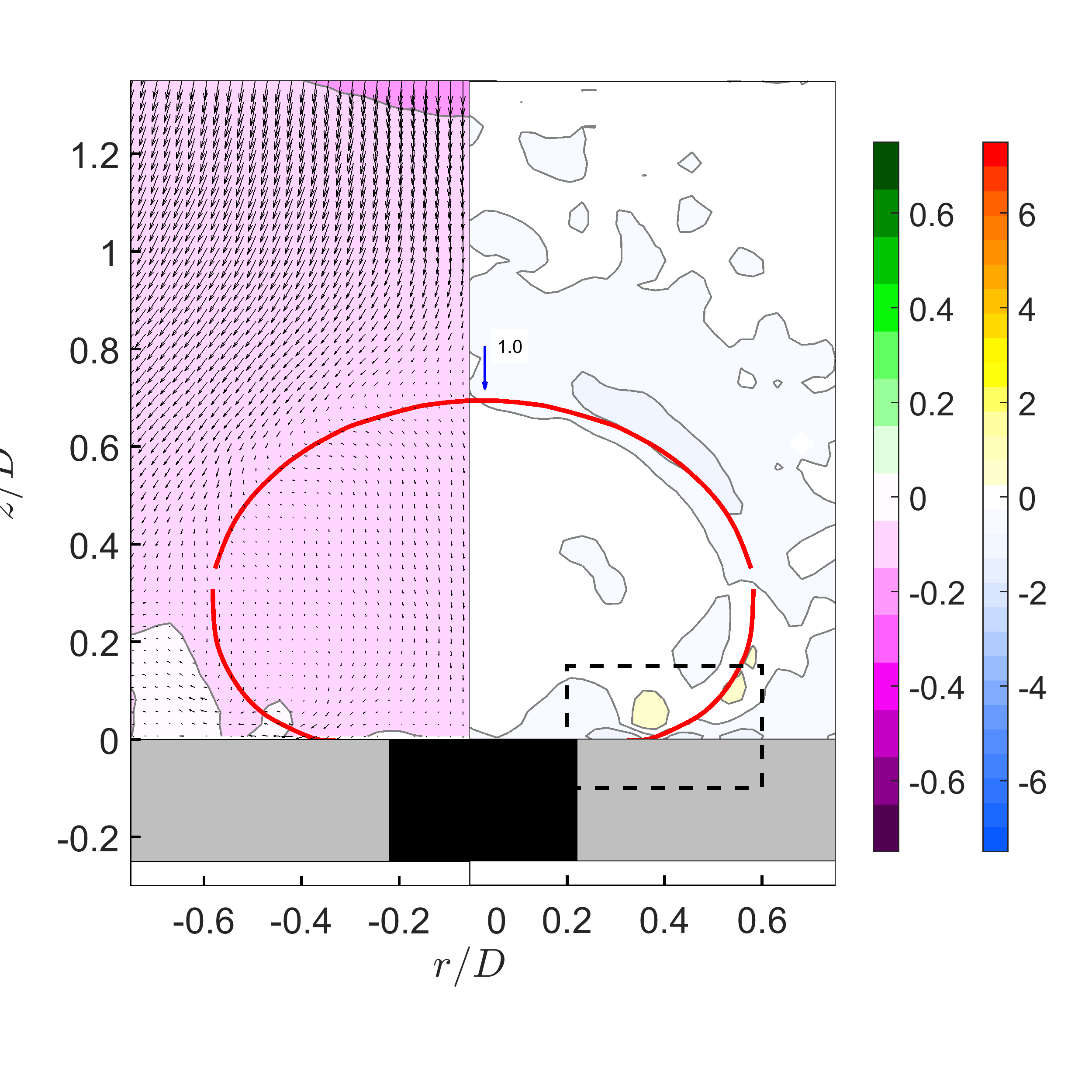}
                \caption{}
                \label{fig15c}
        \end{subfigure}\\
        \begin{subfigure}[b]{0.25\textwidth}
                \centering
                \includegraphics[width=\textwidth]{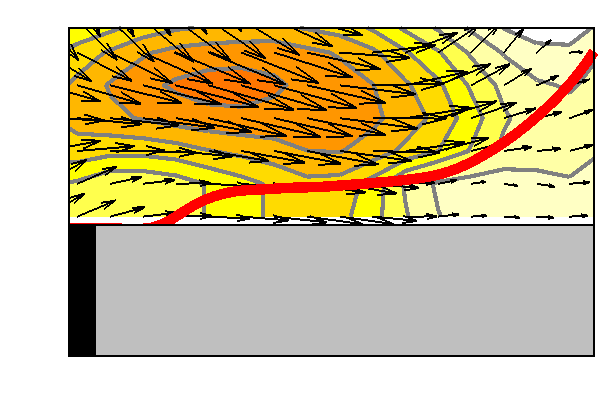}
                \caption{}
                \label{fig15d}
        \end{subfigure}  
	  \begin{subfigure}[b]{0.25\textwidth}
                \centering
                \includegraphics[width=\textwidth]{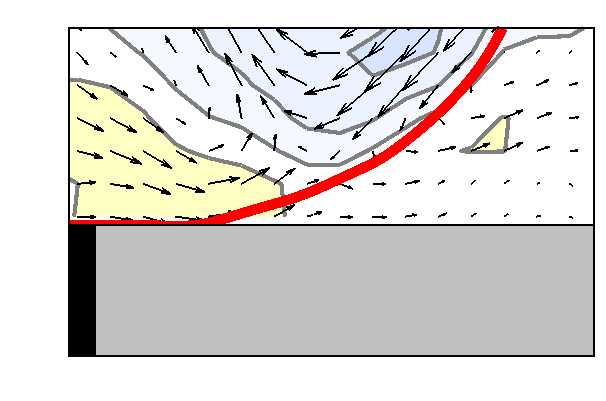}
                \caption{}
                \label{fig15e}
        \end{subfigure}     
        \begin{subfigure}[b]{0.31\textwidth}
                \centering
                \includegraphics[width=\textwidth]{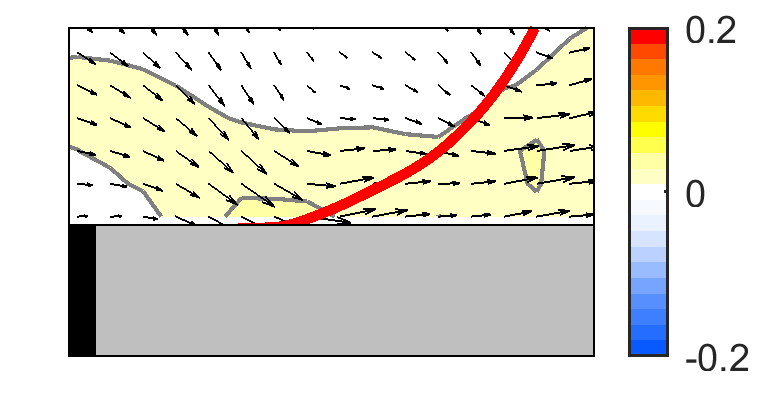}
                \caption{}
                \label{fig15f}
        \end{subfigure}                     
        \caption{(Color online) Contours of axial velocity ($u_z/U_t$: left panel) superposed with velocity vectors and out-of-plane vorticity ($\omega_{\theta}U_t/D$: right panel) at {a) $t=0.07t_i^*$, b) $0.21t_i^*$,  and c) $0.35t_i^*$} for drop with $Bo$ = 4.6 impacting on sharp-edged hydrophilic orifice with $d/D$ = 0.44. Corresponding local radial velocity ($u_r/U_t$) contours superposed with velocity vectors near the contact line (d, e, f).}\label{fig15}
\end{figure}

\begin{figure}[h]
 \centering
        \begin{subfigure}[b]{0.25\textwidth}
                \centering
                \includegraphics[width=\textwidth]{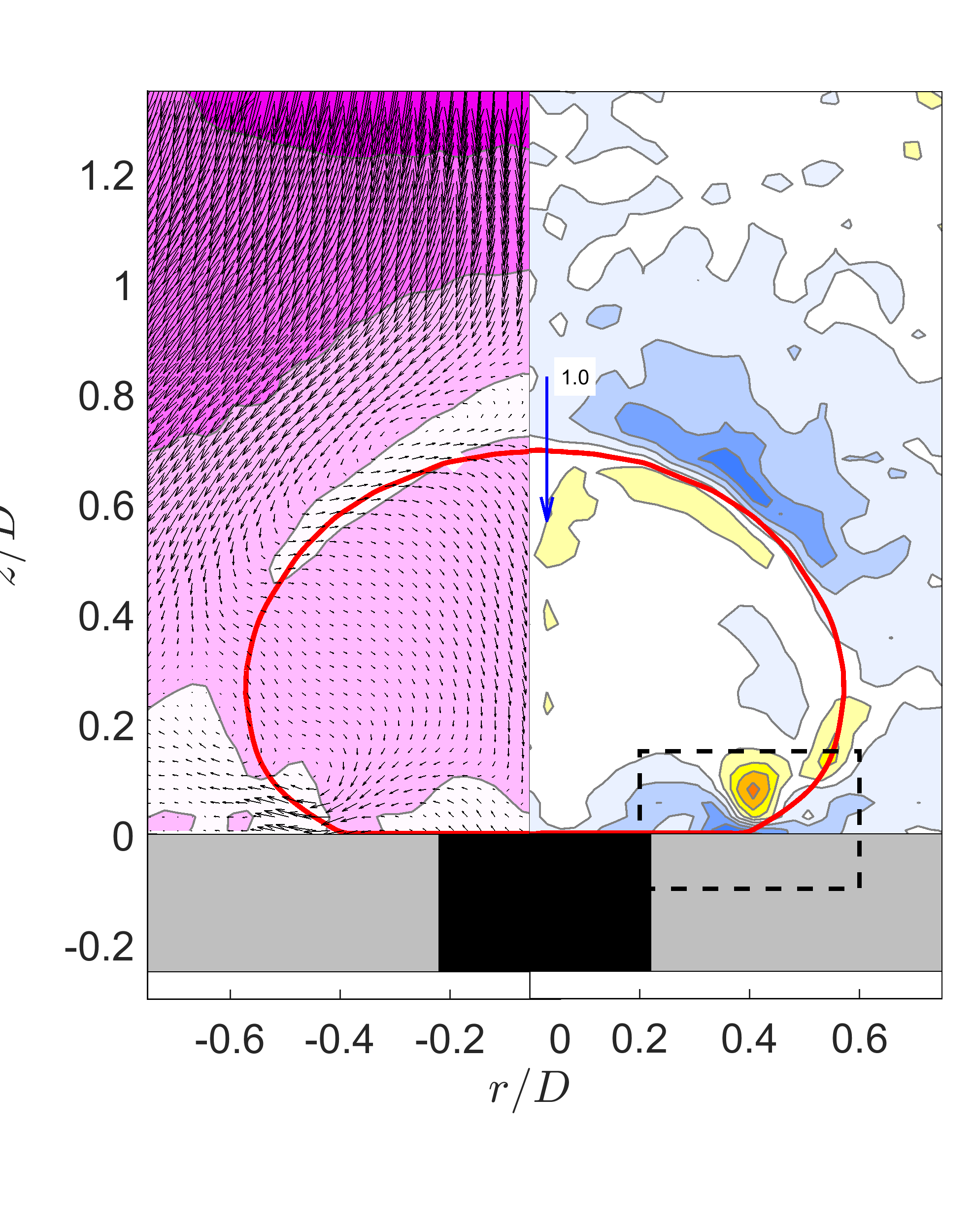}
                \caption{}
                \label{fig16a}
        \end{subfigure}
        \begin{subfigure}[b]{0.25\textwidth}
                \centering
                \includegraphics[width=\textwidth]{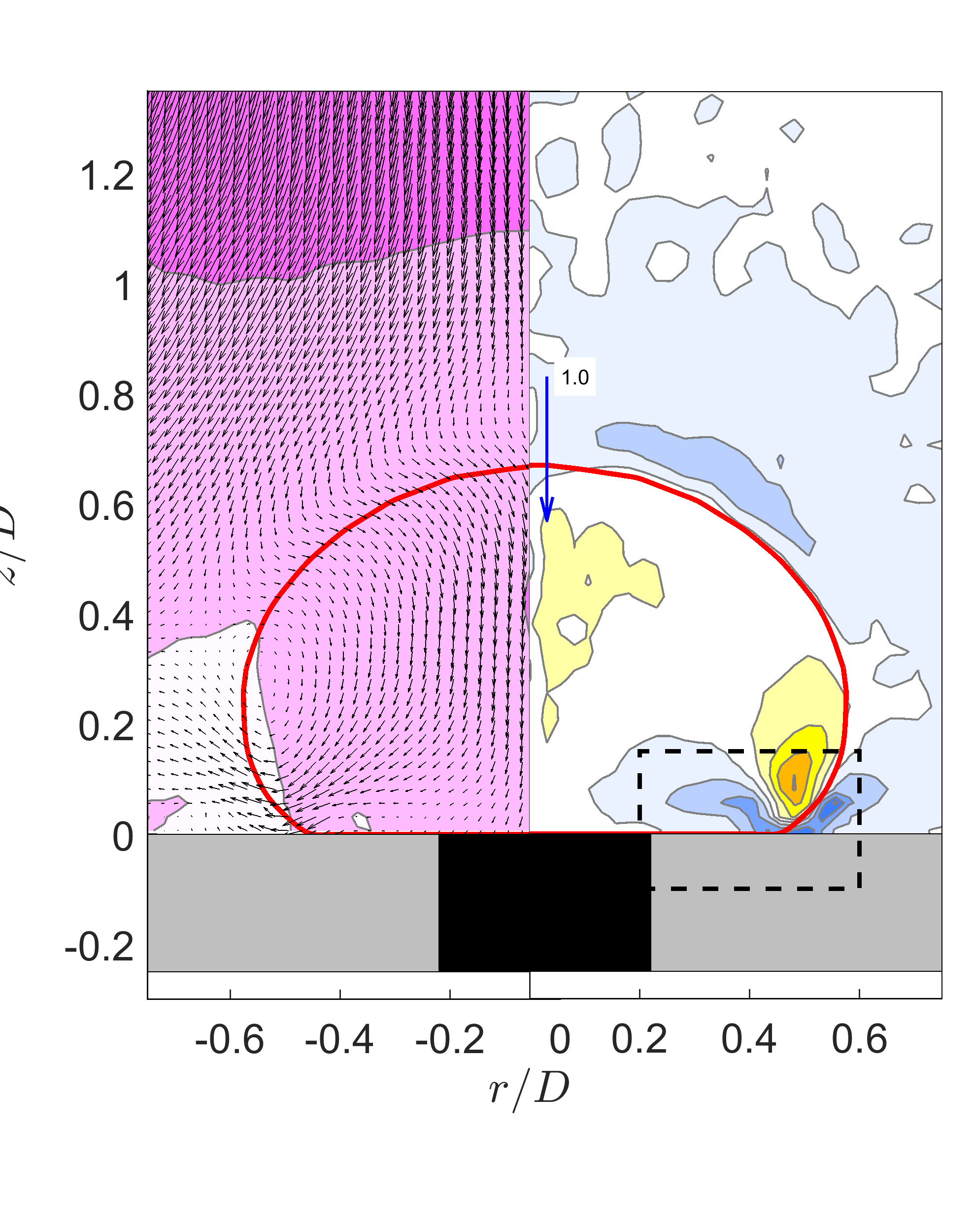}
                \caption{}
                \label{fig16b}
        \end{subfigure}
        \begin{subfigure}[b]{0.31\textwidth}
                \centering
                \includegraphics[width=\textwidth]{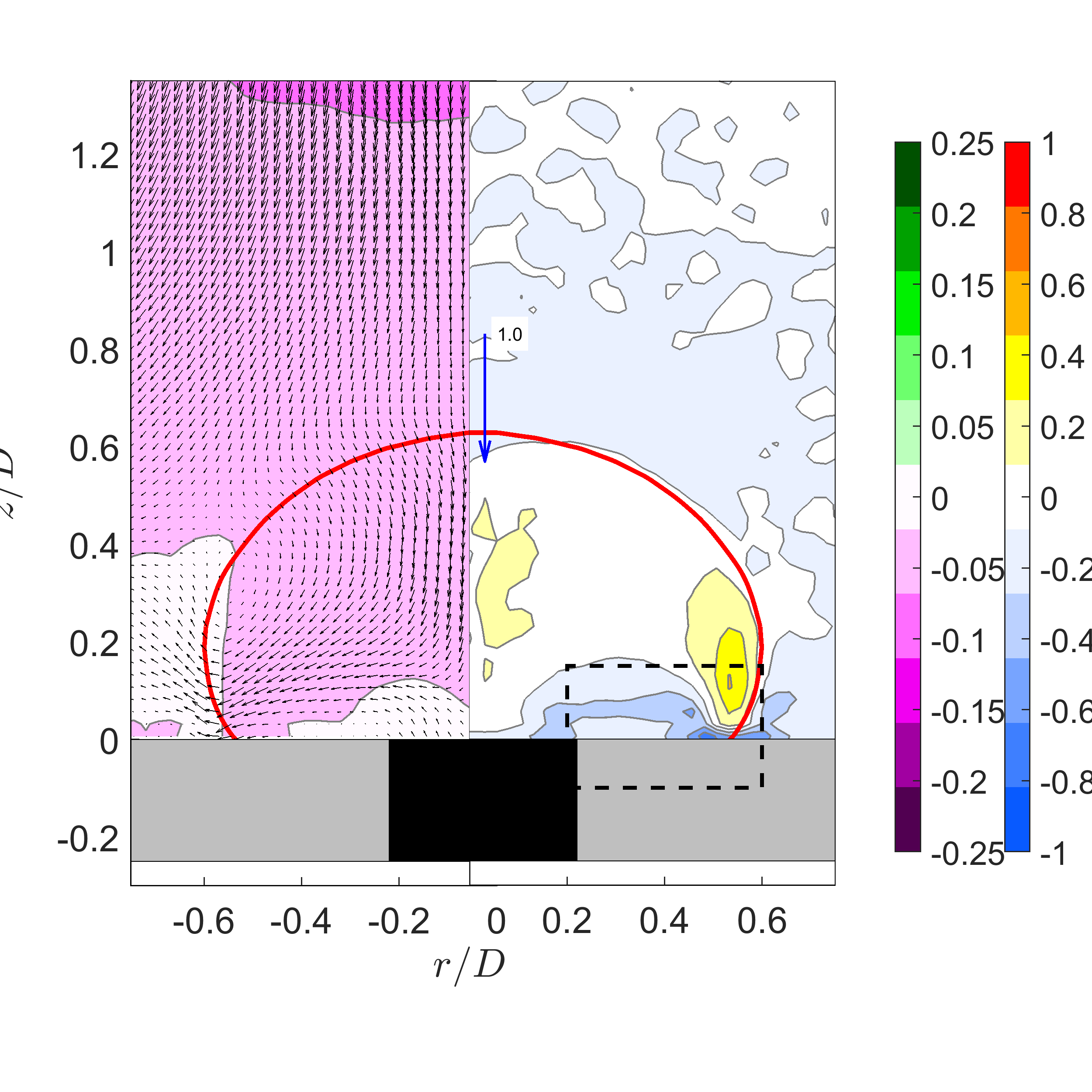}
                \caption{}
                \label{fig16c}
        \end{subfigure}\\
        \begin{subfigure}[b]{0.25\textwidth}
                \centering
                \includegraphics[width=\textwidth]{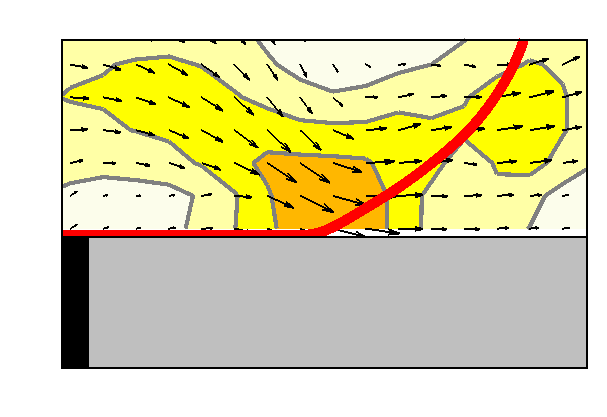}
                \caption{}
                \label{fig16d}
        \end{subfigure}  
	  \begin{subfigure}[b]{0.25\textwidth}
                \centering
                \includegraphics[width=\textwidth]{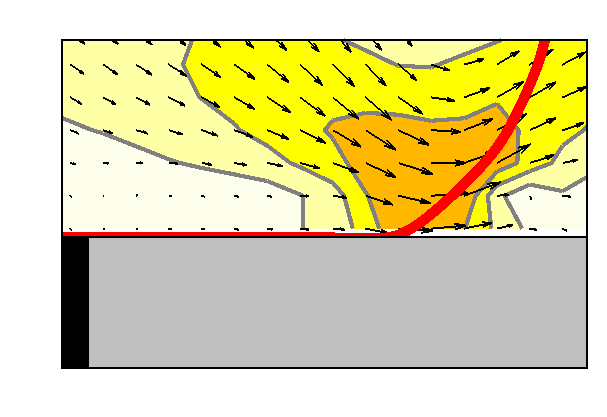}
                \caption{}
                \label{fig16e}
        \end{subfigure}     
        \begin{subfigure}[b]{0.31\textwidth}
                \centering
                \includegraphics[width=\textwidth]{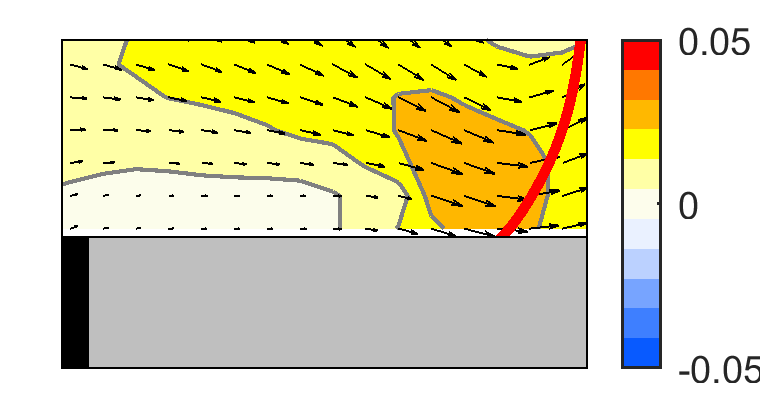}
                \caption{}
                \label{fig16f}
        \end{subfigure}                     
        \caption{(Color online) Contours of axial velocity ($u_z/U_t$: left panel) superposed with velocity vectors and out-of-plane vorticity ($\omega_{\theta}U_t/D$: right panel) at {a) $t=0.5t_i^*$, b) $0.67t_i^*$, and c) $0.95t_i^*$} for drop with $Bo$ = 4.6 impacting on sharp-edged hydrophilic orifice with $d/D$ = 0.44. Corresponding local radial velocity ($u_r/U_t$) contours superposed with velocity vectors near the contact line (d, e, f).}\label{fig16}
\end{figure}

{With the surface wettability effects demonstrated in the Section} \ref{sec3.3}, {it is instructive to examine the local motion pertaining to the contact lines above a hydrophilic orifice. For this purpose, we consider a similar drop ($Bo = 4.8$) impacting on a smaller hydrophilic orifice with $d/D$=0.44 (SHPL2 in Table} \ref{tab02}). In this case, the smaller $d/D$ prevents penetration of the drop fluid, and the drop is captured above the plate. The contact line motion above the plate is resolved at a higher magnification with a vector spacing of 0.24 mm (see Table \ref{tab03}).  Figures \ref{fig15a}-\ref{fig15c} show contours of axial velocity ($u_z/U_t$) and out-of-plane vorticity ($\omega_\theta D/U_t$) at three early times following the initial contact ($t = 0.03t_i^*$) at the orifice edge when the drop impact inertia is significant. The corresponding radial velocity ($u_r/U_t$) contours superposed with  velocity vectors  near the contact line are shown in figures \ref{fig15d}-\ref{fig15f}, respectively. Similar plots at three later times when the moving contact line dominates the global flow field are shown in figure \ref{fig16} {with a different velocity scale}.  

Immediately after the contact, the velocity field near the contact line is dominated by the post-impact inertia that causes the drop to flatten above the plate surface (figure \ref{fig15a}). The overall flow comprises of an axial downflow of the drop fluid above the orifice, an outflow of the precursor oil film squeezed between the drop and the plate diverging away from the orifice, and the wake of surrounding fluid shearing against the drop that initiates a secondary circulation ($\omega_\theta = 5U_t/D$) inside the drop near the trailing interface. {The outwardly moving precursor film drags the contact line across the plate at a speed of about 0.05$U_t$ }(figures \ref{fig15a} and \ref{fig15d}) until the film is completely drained from underneath the drop at $t=0.21t_i^*$(figures \ref{fig15b} and \ref{fig15e}). At this time, the drop rebounding upward exhibits secondary circulation that reduces the radial motion near the contact line. In the next frame at $t=0.35t_i^*$, the drop settles above the plate under gravity allowing the contact line to propagate further across the plate with a speed of approximately 0.03$U_t$ (figure \ref{fig15f}). 

We find that this early stage of contact line spreading is independent of surface wettability based on comparison to an equivalent case with a hydrophobic orifice (not shown) that yielded the same propagation speed. The effect of wettability on contact line speed during underlying film drainage was also found to be insignificant for a drop in air spreading on a flat surface \citep{Eddi2013}. 

After the inertial effects on the drop weaken, the contact line speed decreases steadily. In figure \ref{fig16}, the contact line propagates at about 0.03$U_t$ and is accompanied by rotational motion with regions of opposing vorticity.  The outward moving contact line causes an axial descent of drop fluid at a speed of $u_z = 0.05U_t$ until the drop slowly approaches an equilibrium shape. The bulk motion in the drop fluid resulting from the contact line motion also causes vortical motion in the wake of the drop (see figures \ref{fig16a}-\ref{fig16c}). The contact line motion continues to spread the drop until it reaches an equilibrium shape at $t=5.3t_i^*$ (not shown) with an acute contact angle of 55$^\circ$. 

\begin{figure}[h]
 \centering
        \begin{subfigure}[b]{0.32\textwidth}
                \centering
                \includegraphics[width=\textwidth]{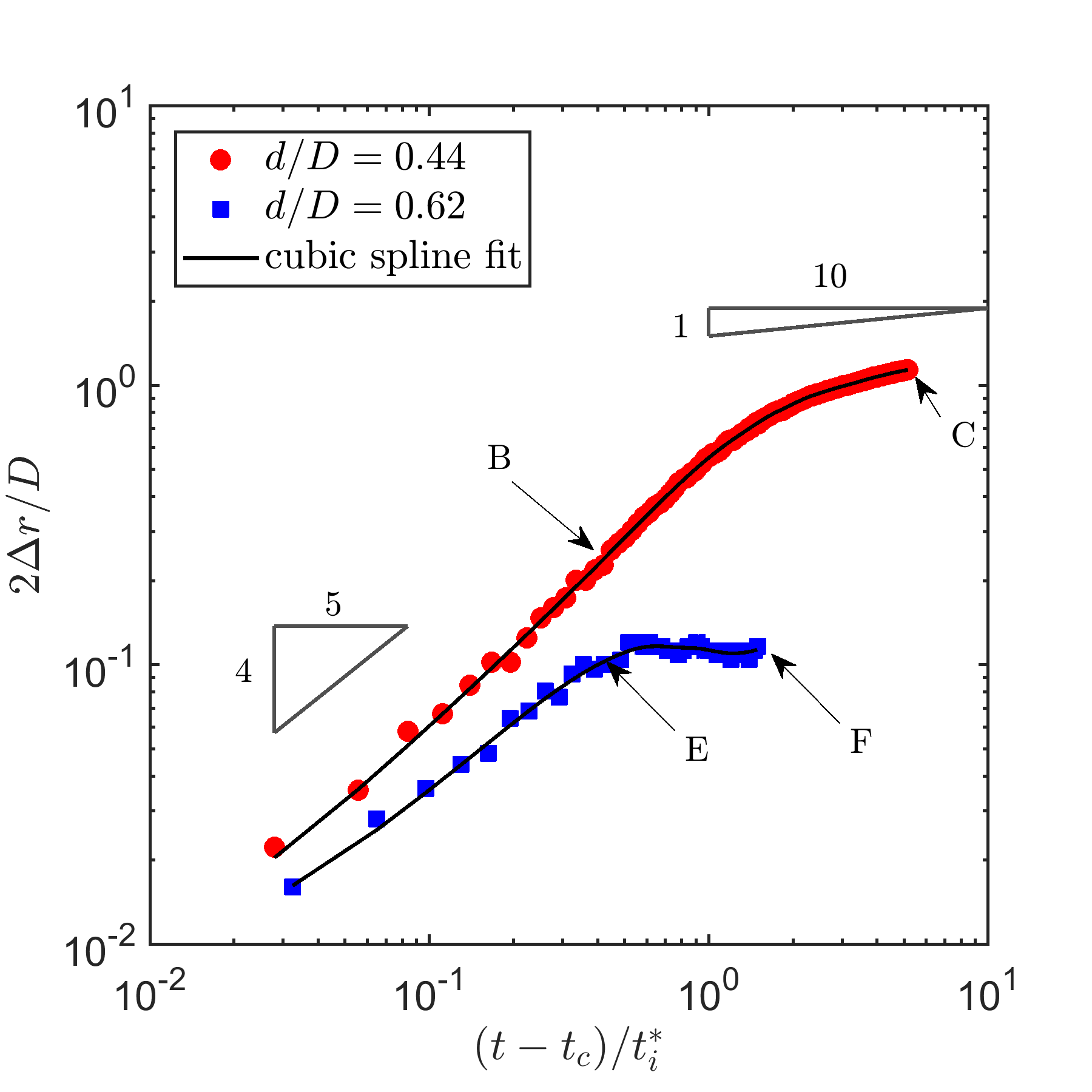}
                \caption{}
                \label{fig17a}
        \end{subfigure}
        \begin{subfigure}[b]{0.32\textwidth}
                \centering
                \includegraphics[width=\textwidth]{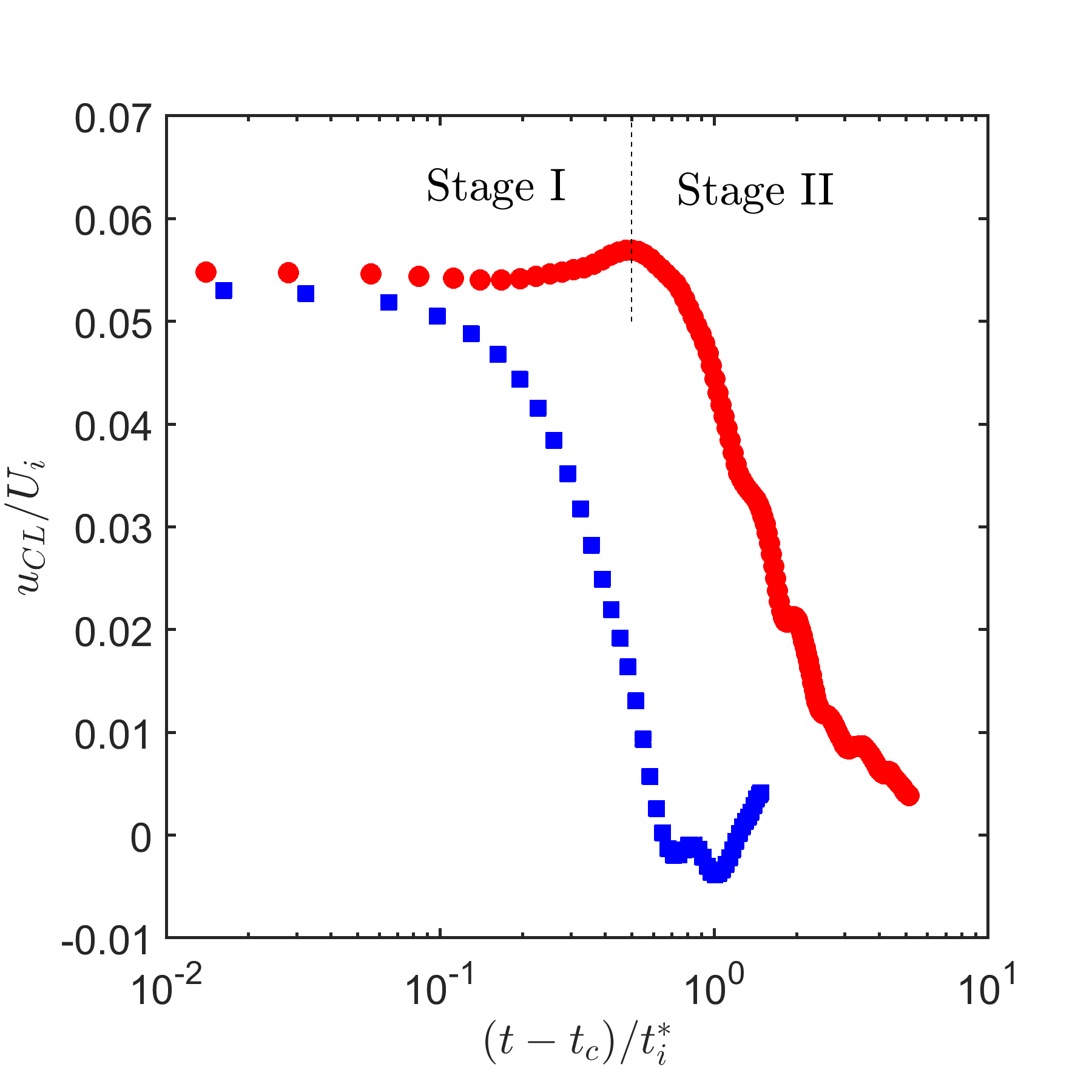}
                \caption{}
                \label{fig17b}
        \end{subfigure}
        \begin{subfigure}[b]{0.32\textwidth}
                \centering
                \includegraphics[width=\textwidth]{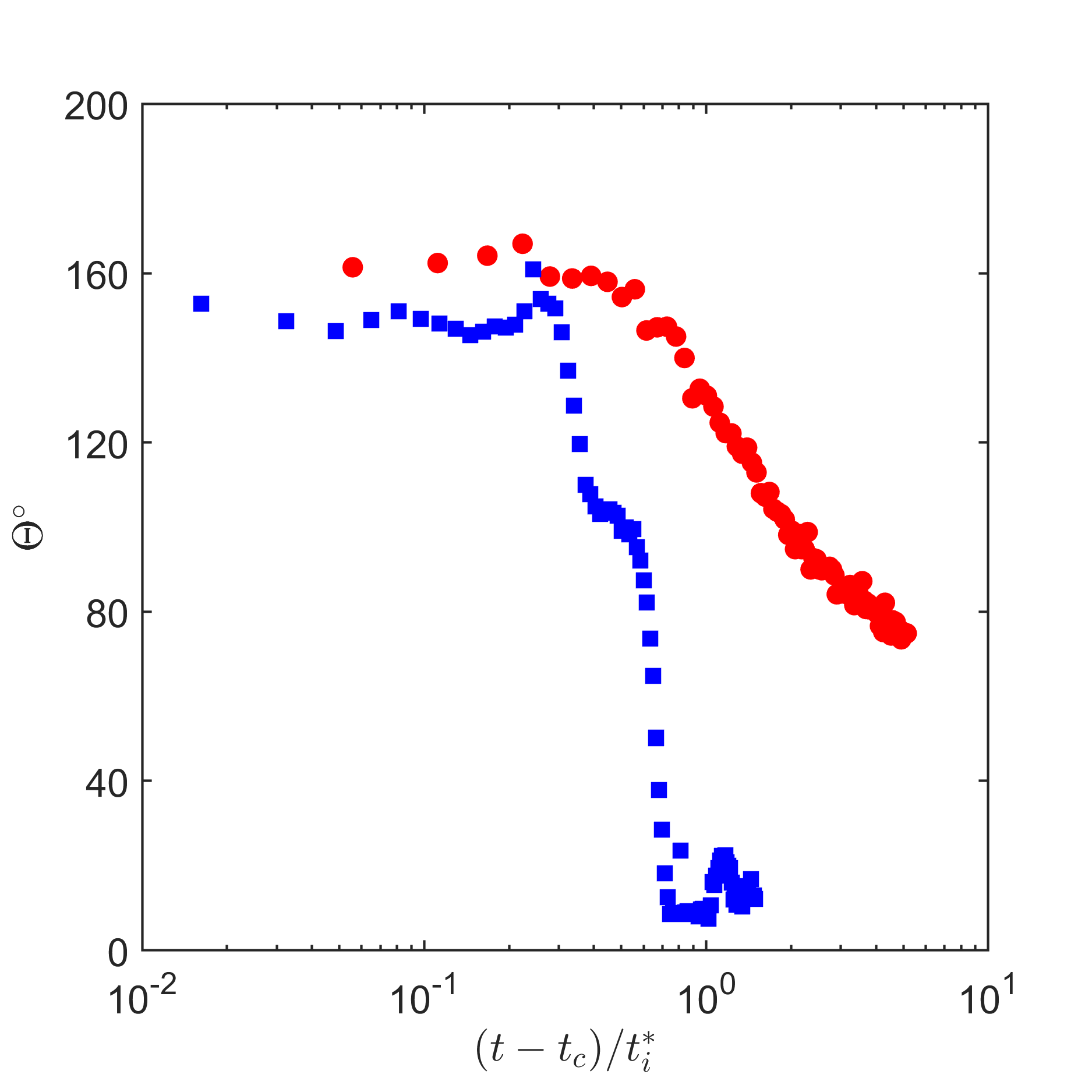}
                \caption{}
                \label{fig17c}
        \end{subfigure}\\
        \begin{subfigure}[b]{0.6\textwidth}
                \centering
                \includegraphics[width=\textwidth]{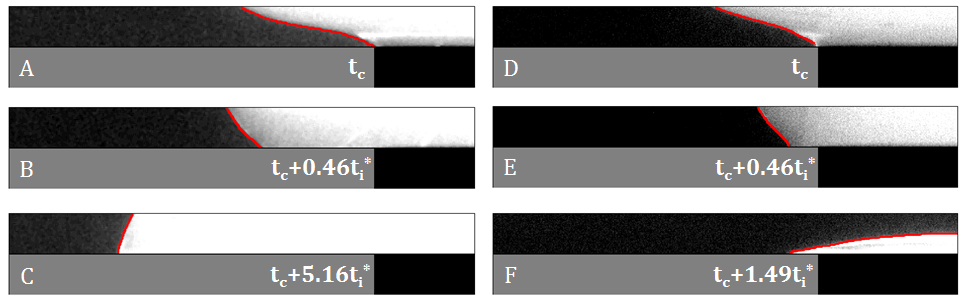}
                \caption{}
                \label{fig17d}
        \end{subfigure}
\caption{a) Variation of normalized spreading distance ($2\Delta r/D$) from orifice edge, b) contact line velocity ($u_{CL}/U_i$) based on cubic-spline fit with respect to time ($(t-t_c)/t_i^*$) measured from the time of initial contact ($t_c$), c) dynamic angle of contact ($\Theta_d$) with respect to $2\Delta r/D$, and  d) example snapshots of contact line location and orientation above the plate surface for a drop with $Bo \approx$ 5 impacting a hydrophilic orifice with $d/D$ = 0.44 (A, B, C) and 0.62 (D, E, F). The triangles in \ref{fig17a} demonstrate power-law variations during early inertia dominated ($\kappa=4/5$) and later viscous dominated ($\kappa=1/10$) spreading.}\label{fig17}
\end{figure}

Using the image processing method discussed in section \ref{sec2.2}, we explore the stages of contact line propagation in further detail. The contact line location measured from the orifice edge ($\Delta r$) and the dynamic contact angle ($\Theta_d$) are resolved every 10 ms from back-lit visualization images. The temporal variation of $\Delta r$, contact line speed ($u_{CL}$) and $\Theta_d$ for the two hydrophilic examples (SHPL1 and SHPL2 in Table \ref{tab02}) are presented in figures \ref{fig16a}, \ref{fig16b} and \ref{fig16c}. The contact line speed, $u_{CL}$ was estimated by first fitting the radial location ($\Delta r$) on a cubic smoothing spline with smoothing parameter $p=1/(1+1000\epsilon)$ with $\epsilon = \Delta t^{1/16}$ (see csaps() in MATLAB), and then using a central difference scheme. The contact line velocity was normalized with the drop impact velocity ($U_i$) instead of $U_t$ because of the slight difference in $Bo$ between the two cases. For convenience, these two cases with $d/D$ = 0.44 and 0.62 will be referred to as small-orifice and large-orifice, respectively. {The impact velocity ($U_i$) for the small and the large orifices are 0.67$U_t$ and 0.8$U_t$ , respectively.}

The first stage of contact line propagation {($t<0.5t_i^*$)} lasts until the post-impact shape oscillations in the bulk of the drop diminish (see I in figure \ref{fig17b}).  The limits of this stage are depicted in snapshots A and B in figure \ref{fig17d}. Because of minimal drop penetration into the orifice, the small-orifice case exhibits distinct effects of impact inertia, surface tension and gravity through this stage.  Immediately after the contact, as shown in figure \ref{fig15}, the contact line is pushed outward along with the precursor film due to a) downward inertia squeezing the film, and b) the surface tension near the contact point causing the film to retract.  The contact line speed begins at $u_{CL} \approx 0.06U_i$ (frame A), remains nearly constant until the precursor film drains completely, and then increases slightly while the rebounding drop settles above the plate under gravity (frame B). Since most of the drop fluid remains above the plate, the gravitational settling enhances the spreading and somewhat increases the contact line speed. For the large-orifice case, which has similar impact inertia and film thickness (frame D), the contact line propagation begins at nearly the same speed as in the small-orifice case. However, the continuous penetration of drop fluid into the large orifice opposes the outward motion of the contact line, and the propagation speed $u_{CL}$ decreases rapidly toward zero. 

After the precursor film drains and the inertial effects subside, the contact line propagation enters the typical viscous stage where its motion is driven primarily by surface wettability and opposed by the drop viscosity (see II in figure \ref{fig17b}) . For the small-orifice case, this stage corresponds with decelerating contact line speed. For the large-orifice case, the drop fluid beneath the plate continues to pull the contact line inward, and the contact line speed oscillates near $u_{CL} =0$ until part of the drop fluid breaks away from the orifice (see figure \ref{fig14} and the related discussion). 

The evolution of the dynamic contact angle ($\Theta_d$) is captured in figure \ref{fig17c}. During the inertial stage, the contact angle is affected by the precursor film and the local curvature near the contact point (see A, D in figure \ref{fig17d}). In both cases considered, $\Theta_d$ begins at $\approx$150$^\circ$. The value oscillates somewhat as the drop shape oscillates above the plate. In the large-orifice case, after the film drains (frame E), $\Theta_d$ decreases rapidly as the remaining drop fluid above the plate continues to flow into the orifice. At the end of the sequence, the contact line is located close to but outside of the orifice edge, and $\Theta_d$ oscillates between 10$^\circ$ and 20$^\circ$. In the small orifice case, $\Theta_d$ also begins decreasing during the viscous stage, although more gradually. It approaches {$\approx 75^\circ$} at the end of the sequence.

The contact line propagation in figure \ref{fig17a} appears to follow a power law $\Delta r \approx t^{\kappa}$ with variable exponent. For a liquid drop impacting on a flat surface in air, \citet{Eddi2013} showed that the exponent $\kappa$ decreases continuously from 4/5 to 1/10 during the viscous stage. The small-orifice results demonstrate a similarity, with $\kappa$=4/5 during the inertial stage and decreasing during the viscous stage.  Over the range shown, $\kappa$ decreases to 0.18 but does not reach 1/10.

{Upon impact, the sharp edge orifice promotes spontaneous rupture of the oil film around the orifice periphery resulting in symmetric spreading of drop fluid across the plate surface. Due to the quick film rupture, the early stage of spreading is strongly affected by impact inertia. This behavior can be contrasted with that occurring beneath a drop falling through liquid onto a solid surface. In the latter scenario, the underlying film typically ruptures away from the central axis after a prolonged drainage} \citep{Foister1987, Foister1990}. {The resulting contact line motion is decoupled from the impact inertia and the early stage of spreading is driven exclusively by the surface tension in the precursor film near the contact location. This mode of spreading leads to a power-law relationship with $\kappa$=1/2} \citep{Biance2004, Bird2008}, which is clearly lower than 4/5. 
\section{Discussion and Conclusions}

Using refractive index matching and time-resolved PIV, we examined details of the motion within a gravity driven drop impacting on and moving through an orifice plate and within the surrounding immiscible liquid.  The initial gravitational potential of the drop was redistributed into kinetic energy within both fluids, and the surface deformation energy of the drop was tracked through during various stages of the drop's motion through a round-edged orifice (without surface contact). We quantified the kinetic energy resulting from local circulation and strain inside the drop ($KE_{d,D}$), {which was not accounted for in earlier studies} \citep{Clanet04, Attane07}. {Compared to the translational kinetic energy $KE_{d,T}$ of the drop, $KE_{d,D}$  increased from 4\% of the total at free-fall to nearly 100\% during the post-impact oscillations of the drop. When the drop reaccelerated through the orifice, $KE_{d,D}$ was of similar order to $KE_{d,T}$.}

For the current liquid combination, a significant amount of drop energy ($\approx$ 50\%) was transferred to and dissipated within the surrounding liquid during its fall and as it approached the orifice plate. This effect is typically parameterized by the drop-to-surrounding fluid viscosity ratio $\lambda$ \citep{Mohamed03}. A lower value of $\lambda$ leads to greater momentum transfer and stronger dissipation and consequently weaker drop inertia at the time of impact. The liquid wake carried significant energy and inertia unlike liquid/air flows with $\lambda \gg 1$ \citep{Lorenceau03, Delbos10,Clanet04, Attane07,  Kumar17}. The momentum carried by the surrounding fluid influences the drop\rq{}s local motion, shape and the final outcome in a few ways.  First, the downward moving liquid wake opposed the upward motion of the drop fluid post-impact, limiting the drop's rebound. The velocity field in the wake produces both shear and normal stresses on the trailing interface of the drop. Shear stresses from the wake act as a source of secondary circulation within the drop. The normal stress ($\tau_{zz} = \mu\frac{du_z}{dz}$), on the other hand, pushes the drop downward against the resistance of the plate and the capillary pressure at the orifice ($P_{\sigma}\approx \sigma/d$). Based on the PIV results, the maximum $\tau_{zz}$ scales as $\mu_sU_t/D$, which when compared with $P_{\sigma}$ provides a modified Capillary number: $Ca^{*} = \frac{\mu_s U_t}{\sigma}\frac{d}{D}$. For the cases considered in this study $Ca^*$=0.1, so that the integrated wake normal stress was insufficient to push a drop through the orifice. However, increasing $\mu_s$ and decreasing $\sigma$ each by an order of magnitude could increase $Ca^*$ by two orders of magnitude and provide a scenario for which the wake stress would enhance penetration of the drop fluid through the orifice. 

{In a liquid/air system, a drop typically makes contact with an underlying solid surface immediately after impact, irrespective of whether the surface is impermeable} \citep{Josserand16,Kumar17}, {or permeable} \citep{Lorenceau03, Delbos10}). {Therefore, the post-impact drop shape and subsequent dynamics can depend strongly on the surface wettability. In a liquid/liquid system with round-edged orifice, contact is prevented by a film of surrounding oil, and wettability effects are absent.} The simpler scenario of a round-edged orifice was contrasted with that of a sharp-edged orifice, where contact-upon-impact and wettability complicated the dynamics. 

In the sharp-edged case with hydrophobic surface,  both leading contact line \emph{slipping} and trailing contact line \emph{pinning} influenced local velocity variations in both liquids during the drop\rq{}s passage. Although the contact at the orifice edge delayed the onset of drop penetration, eventual \emph{slipping} of the leading contact line re-initiated downward motion within the orifice. The drop fluid then accelerated downward and passed through the orifice with a rate and an internal velocity field similar to those in the round-edged orifice case. This similar behavior implied that the contact location remained close to the upper orifice edge so that a film of surrounding fluid persisted over most of the orifice interior. {The pinning force of the trailing contact line was significant.  Unless the downward inertia of the drop fluid could overcome the pinning force, the drop pinched off so that a fraction remained captured at the orifice. {Based on the ratio between the inertia ($F_i$) and the pinning force  ($F_p$), this effect can be parameterized  by a modified Weber number, $We_d^* = \frac{1}{f_c}We\frac{d}{D}$, with $f_c = cos\Theta_Y - cos(\Theta_Y+\phi-\pi)$ being the canthotaxis factor.} Although depinning occurred for cases with sufficient inertia, the depinning could not be observed in detail because of limitations in optical access. Therefore, it was not possible to correlate the depinning process with measurements of the dynamic contact angle and whether or how it exceeded the canthotaxis limit. 

When the plate surface was hydrophilic, the trailing contact line moved quickly outward and away from the upper orifice edge so that \emph{pinning} was never observed. Instead, the portion of drop fluid above the plate formed a cap shape that promoted drop breakup and fractional drop capture at the orifice. For a case with minimal penetration, the outward spreading of drop fluid along the hydrophilic surface could be characterized by two stages. Initially, the spreading rate was affected by the impact inertia and gravity, and followed a power law dependence for distance vs. time with an exponent of 4/5. Later, after the inertial effects subsided, the contact line propagated mainly due  to wettability of the surface, and the rate of spreading decreased steadily. The power law exponent approached that of 1/10 previously observed for viscous spreading of liquids in air \citep{Tanner1979}. 


An aspect that requires attention is the location and orientation of the leading contact line interior to a sharp-edged orifice. Although the leading contact line orientation and motion could not be resolved optically in the current experiments, it clearly plays a significant role in determining the outcome of a drop impact. In the hydrophobic case, the motion of this contact line was sufficient to re-initiate downward motion in the stagnated drop fluid. In the hydrophilic case, the motion and orientation of the leading contact line appeared to influence the neck development and breakup location (see figures  \ref{fig14} and \ref{pinning_HPL}). It would be interesting to observe how the bulk fluid motion inside the orifice influences the contact line and vice versa. This behavior could potentially be characterized in future experiments with a refractive index matched liquid/solid system.

 \clearpage
 \section{References}
 \bibliography{EXIF2017}

\end{document}